%% file: main.tex
\documentclass[fleqn,usenatbib]{mnras}

\usepackage{newtxtext,newtxmath}
\usepackage[T1]{fontenc}

\DeclareRobustCommand{\VAN}[3]{#2}
\let\VANthebibliography\thebibliography
\def\thebibliography{\DeclareRobustCommand{\VAN}[3]{##3}\VANthebibliography}

\usepackage{graphicx}	% Including figure files
\usepackage{amsmath}	% Advanced maths commands
\usepackage[dvipsnames]{xcolor}	% Advanced maths commands
\usepackage{comment}
\usepackage{textgreek}
\usepackage{xspace}
\usepackage{booktabs}
\usepackage{multirow}
\usepackage{rotating}
\usepackage{longtable,lscape}

\newcommand{\sersic}{S\'ersic\xspace}
\newcommand{\eazy}{\textsc{EAZY}}
\newcommand{\sig}{\ensuremath{\sigma}\xspace}
\newcommand{\JWST}{\textit{JWST}\xspace}
\newcommand{\Mstar}{\ensuremath{M_\star}\xspace}
\newcommand{\Msun}{\ensuremath{\mathrm{M}_\odot}\xspace}
\newcommand{\Zsun}{\ensuremath{\mathrm{Z}_\odot}\xspace}
\let\oldAA\AA
\renewcommand{\AA}{\text{\oldAA}\xspace}

\newcommand{\tng}{\textsc{Illustris-TNG300}\xspace}
\newcommand{\shark}{\textsc{SHARK}\xspace}

\title[High-$z$ quiescent galaxies]{The abundance and nature of high-redshift quiescent galaxies from JADES spectroscopy and the FLAMINGO simulations}

\author[W.M. Baker et al.]{
William M. Baker$^{1,2,3}$\thanks{E-mail: wb308@cam.ac.uk},
Seunghwan Lim$^{1,2}$,
Francesco D'Eugenio$^{1,2}$,
Roberto Maiolino$^{1,2,4}$,
\newauthor{}
Zhiyuan Ji$^{5}$,
Santiago Arribas$^{6}$,
Andrew J. Bunker$^{7}$,
Stefano Carniani$^{8}$,
Stephane Charlot$^{9}$,
\newauthor{}
Anna de Graaff$^{10}$,
Kevin Hainline$^{5}$,
Tobias J. Looser$^{1,2}$,
Jianwei Lyu$^{5}$,
Pierluigi Rinaldi$^{5}$,
\newauthor{}
Brant Robertson$^{11}$,
Matthieu Schaller$^{12,13}$,
Joop Schaye$^{12}$,
Jan Scholtz$^{1,2}$,
Hannah \"Ubler$^{1,2}$,
\newauthor{}
Christina C. Williams$^{14}$,
Christopher N. A. Willmer$^{5}$,
Chris Willott$^{15}$,
Yongda Zhu$^{5}$\\
$^{1}$Kavli Institute for Cosmology, University of Cambridge, Madingley Road, Cambridge, CB3 OHA, UK\\
$^{2}$Cavendish Laboratory - Astrophysics Group, University of Cambridge, 19 JJ Thomson Avenue, Cambridge, CB3 OHE, UK\\
$^{3}$DARK, Niels Bohr Institute, University of Copenhagen, Jagtvej 128, DK-2200 Copenhagen, Denmark\\
$^{4}$Department of Physics and Astronomy, University College London, Gower Street, London WC1E 6BT, UK\\
$^{5}$Steward Observatory, University of Arizona, 933 N. Cherry Avenue, Tucson, AZ 85721, USA\\
$^{6}$Centro de Astrobiolog\'ia (CAB), CSIC–INTA, Cra. de Ajalvir Km.~4, 28850- Torrej\'on de Ardoz, Madrid, Spain\\
$^{7}$Department of Physics, University of Oxford, Keble Road, Oxford, OX1 3RH, UK\\
$^{8}$Scuola Normale Superiore, Piazza dei Cavalieri 7, 56126 Pisa, Italy\\
$^{9}$Sorbonne Universit\'e, CNRS, UMR 7095, Institut d'Astrophysique de Paris, 98 bis bd Arago, 75014 Paris, France\\
$^{10}$Max-Planck-Institut f\"ur Astrophysik, Karl-Schwarzschild-Str. 1, 85741 Garching, Germany\\
$^{11}$Department of Astronomy and Astrophysics, University of California, Santa Cruz, 1156 High Street, Santa Cruz, CA 95064, USAA\\
$^{12}$Leiden Observatory, Leiden University, PO Box 9513, 2300 RA Leiden, The Netherlands\\
$^{13}$Lorentz Institute for Theoretical Physics, Leiden University, PO Box 9506, 2300 RA Leiden, the Netherlands\\
$^{14}$NSF’s National Optical-Infrared Astronomy Research Laboratory, 950 North Cherry Avenue, Tucson, AZ 85719, USA\\
$^{15}$NRC Herzberg Astronomy and Astrophysics, 5071 West Saanich Road, Victoria, BC V9E 2E7, Canada\\
}

\date{Accepted XXX. Received YYY; in original form ZZZ}

\pubyear{\the\year{}}

\begin{document}
\label{firstpage}
\pagerange{\pageref{firstpage}--\pageref{lastpage}}
\maketitle

\begin{abstract}

We use NIRSpec/MSA spectroscopy and NIRCam imaging to study a sample of 18 massive ($\log \Mstar/\Msun > 10$~dex), central quiescent galaxies at $2\leq z \leq 5$ in the GOODS fields, to investigate their number density, star-formation histories, quenching timescales, and incidence of AGN.
The data depth reaches $\log \Mstar/\Msun \approx 9$~dex, yet the least-massive central quiescent galaxy found has $\log \Mstar/\Msun > 10$~dex, suggesting that quenching is regulated by a physical quantity that scales with \Mstar. With spectroscopy, we assess the completeness and purity of photometric samples, finding number densities 10 times higher than predicted by galaxy formation models, confirming earlier photometric studies.
We compare our number densities to predictions from FLAMINGO, the largest-box full-hydro-simulation suite to date. We rule-out cosmic variance at the 3-$\sigma$ level, providing spectroscopic confirmation that galaxy formation models do not match observations at $z>3$. Using FLAMINGO, we find that the vast majority of quiescent galaxies' stars formed \textit{in situ}, with these galaxies not having undergone multiple major dry-mergers. This is in agreement with the compact observed size of these systems and suggests that major mergers are not a viable channel for quenching most massive galaxies. Several of our observed galaxies are old, with four displaying 4000-\AA breaks with formation and quenching redshifts of $z\geq8$ and $z\geq6$. Using tracers, we find that 8 galaxies host AGN, including old systems, suggesting a high AGN duty-cycle with a continuing trickle of gas to fuel accretion.

\end{abstract}


%%%%%%%%%%%%%%%%%%%%%%%%%%%%%%%%%%%%%%%%%%%%%%%%%%

%%%%%%%%%%%%%%%%% BODY OF PAPER %%%%%%%%%%%%%%%%%%

\section{Introduction}

While precise boundaries may vary \citep[e.g.,][]{Marchesini2023}, massive quiescent galaxies are generally defined as having stellar masses $\Mstar>10^{10}~\Msun$ \citep[e.g.,][]{Glazebrook2017,Carnall2023abundances} and negligible star formation in the last 100~Myr \citep[$\mathrm{sSFR_{100\,Myr}}\,{\leq}\,0.2/t_{\rm obs}$, where $t_\mathrm{obs}$ is the age of the Universe at the redshift of the galaxy; e.g.][]{Gallazzi2014,Carnall2024}.
These systems are one of the key puzzling results of the first two years of \JWST operations \citep[e.g.][]{Carnall2023abundances,Valentino2023,Carnall2024,Long2024}.
Their number density and physical properties are a crucial test for our models of galaxy formation \citep[e.g.,][]{Beckmann2017,Donnari2021a,Donnari2021b,Remus2023,Hartley2023}, while the earliest and most massive systems may present a challenge even for cosmological models \citep[e.g.,][]{Boylan_kolchin2023,Glazebrook2024}.

On the cosmological side, the most massive and earliest quiescent galaxies are important tests of $\Lambda$CDM cosmology.
The possible tension arises due to the early Universe lacking dark-matter haloes sufficiently massive to host these galaxies.

This appears to be the case even with the maximum possible baryon conversion efficiency of 100~percent \citep{Glazebrook2024,Park2024,Carnall2024,deGraaff2024} \citep[but see also][for the effect of the chosen priors]{Turner2024}.
The advantage of studying quiescent galaxies is that
their stellar masses are more accurate than for star-forming galaxies, as shown by the reduced scatter of quiescent \textit{vs} star-forming galaxies in scaling relations \citep[e.g., the stellar mass plane;][]{Hyde2009,deGraaff2021}.
However, there are still some possible explanations for these massive quiescent galaxies that could resolve the apparent tension, including small systematic errors in the stellar mass measurements due, e.g., to non-solar abundance ratios \citep[e.g.,][]{Beverage2021,Beverage2024}, varying IMFs \citep[e.g.,][]{vanDokkum2024}, possible contributions from AGN to the continuum \citep{Inayoshi2024}, and cosmic variance, making these single objects exceptions and not the rule \citep[e.g.,][]{Carnall2023Nature,Valentino2023,deGraaff2024,Kragh2024}.

As for galaxy formation models, massive quiescent galaxies are fascinating because some of them seem to have formed most of their stars extremely rapidly \citep[e.g.,][]{Belli2019,Park2023,Park2024,Beverage2024}. Furthermore, these galaxies have ceased forming new stars despite their high accretion rates of cosmic gas (implied by the connection between high stellar mass and high halo mass). This disconnect between the halo accretion rate and the galaxy star-formation rate requires some physical mechanism `quenching' star formation. Whatever processes have quenched massive quiescent galaxies, they appear to operate not only in the 
very early Universe \citep[with galaxies quenched as early as z$\approx7\text{--}9$;][]{Glazebrook2024,Weibel2024}, but also in incredibly little time \citep[even accounting for the short timescale set by the star-forming main sequence at $z=5\text{--}7$; e.g.,][]{Koprowski2024, Clarke2024}.

Many mechanisms have been proposed to quench massive central galaxies for 100's of Myr \citep[e.g.,][]{Man2018,Curtis-Lake2023}, but theoretical models and energy arguments seem to require some form of feedback from accreting supermassive black holes (SMBHs), through episodes of Active Galactic Nuclei \citep[AGN;][]{Silk1998,Binney2004,DiMatteo2005,Bower2006,Croton2006}.

This AGN feedback could act either by removing the star-forming gas through outflows \citep[ejective feedback;][]{DiMatteo2005,Maiolino2012,Carniani2016}, or by stopping the accretion of cold gas by heating the circum-galactic medium \citep[preventive feedback;][]{Bower2006,Croton2006}.

In support of preventive feedback, evidence from redshifts $z=0\text{--}2$ suggests that the best predictor of quiescence is SMBH mass \citep{Terrazas2016,Piotrowska2022} or, in the absence of direct measurements of this mass, its closest proxies, such as central velocity dispersion or bulge-to-total mass ratio \citep{Bluck2022,Bluck2023}. Following these findings, quenching has been attributed to time-integrated AGN feedback because (most of the) SMBH mass arises from the black-hole accretion history. This time-integrated feedback would act to prevent cosmic gas accretion onto the galaxy, which is then `starved' of fuel for star formation \citep{Peng2015,Trussler2020, Baker2024}. Most notably, these observational results match the predictions of both semi-analytical models \citep{Bluck2022} and cosmological simulations \citep{Piotrowska2022}, which in turn correctly predict the local number density of quiescent galaxies.

However, \JWST has challenged this successful picture; the discovery of a surprising abundance of massive quiescent galaxies at redshifts $z=3\text{--}4$ \citep{Carnall2023abundances}, seems to defy the predictions of cosmological simulations \citep[and semi-analytic models, particularly at $z=4\text{--}5$,][]{Lagos2025}, even accounting for cosmic variance in the observations \citep{Valentino2023}.
One limitation of past studies, however, has been the size of the cosmological simulations used to obtain the predicted number densities. Observed densities of order $10^{-5}$~cMpc$^{-3}$ \citep{Valentino2023} mean that only a few objects are found per 100-cMpc simulation box\footnote{We use the notation of cMpc (or cGpc) to refer to the comoving size.}, meaning that cosmic variance could still explain the observed discrepancy between observations and theory, without any change to our models.

Alternatively, assuming the mismatch between observations and theory to be representative, explaining the overabundance of observed quiescent galaxies would require a stronger or more efficient quenching mechanism at the earliest epochs, possibly some form of AGN-driven ejective feedback \citep[e.g.,][]{Xie2024}. Evidence in this last direction seems to come from the discovery of AGN-driven neutral-gas outflows in at least one third of massive quiescent galaxies at $z=1.5\text{--}2.5$ \citep{Davies2024}. These outflows have high mass loading, comparable to or exceeding their galaxy star-formation rate \citep{Belli2024,D'Eugenio2023a}, and their high incidence may point to a high duty cycle of AGN feedback.

One key difficulty in establishing the causes of quenching is that quenching mechanisms may have short timescales relative to the duration of quiescence. This is particularly true for AGN feedback, because SMBHs are believed to quench galaxies over many AGN episodes, each lasting shorter than 1--10~Myr \citep[e.g.,][]{Harrison2018,Harrison2024,Scholtz2018,Scholtz2021}. In contrast, galaxies can stay quiescent for 100's of Myr \citep[e.g.,][]{Thomas2005,McDermid2015}.
There has been a lot of work in studying massive quiescent galaxies at $z<1\text{--}2$, to investigate the causes of quenching \citep[e.g.,][]{Thomas2005,Cappellari2011,McDermid2015,Newman2018,Belli2019,Barone2022,Bluck2022,Piotrowska2022,Woodrum2024}.
However, by studying quiescent galaxies at increasingly higher redshifts, the younger age of the Universe necessarily leaves less room for supermassive black holes to be in inactive phases because the multiple AGN episodes necessary to quench the galaxy must occur more frequently in the limited time frame available. Therefore, an interesting question regarding quenching mechanisms is how many high-z quiescent galaxies are found to host AGN.

Indeed, a surprising result of the last two years has been the large number of high-z AGN candidates detected, including both Type 1 \citep{Maiolino2023, Maiolino2024, Kokorev2023, Harikane2023, Matthee2024, Greene2024} and Type 2 \citep{Scholtz2023, Tacchella2024}, many of which appear to be overmassive compared to the host galaxy \citep{Ubler2023,Furtak2024,Juodzbalis2024, Mezcua2024, Maiolino2024} based on local scaling relations \citetext{but see \citealp{Sun2024} for a different view}.
Their overmassive nature, if confirmed, almost certainly implies that a higher-than-expected amount of energy has already been released into the host galaxy and/or halo, and this release happened in a short amount of time.
This suggests that these black holes have had significant accretion rates within their history, meaning that AGN feedback is likely to have affected the host galaxy.

However, establishing whether a galaxy is an AGN host can prove challenging \citep[e.g.,][]{Hickox2018,Scholtz2023}, and may require a combination of different tracers \citep[e.g.,][]{Calabro2023}, including X-rays, emission-line diagnostics, mid-infrared and radio emission.

A different but related problem is the lower-mass boundary of massive quiescent galaxies. The discovery of low-mass, `mini' or rapidly-quenched galaxies at $z=5\text{--}7$ \citep[$\Mstar\sim10^7\text{--}10^9~\Msun$;][]{Strait2023,Looser2024,Looser2023, Baker2025} raises the question of what is the mass threshold where `bursty' star formation and short-lived quenching are supplanted by long-term quenching, i.e., quiescence.

Many recent studies have focused on photometric samples \citep{Merlin2019,Carnall2023abundances,Valentino2023,Alberts2023,Ji2024,Long2024}, deep spectroscopic samples at redshifts $z=1\text{--}3$ \citep{Beverage2024,Park2024}, small samples (or even individual objects) confirmed by spectroscopy at $z>4$ \citep{Carnall2023Nature,deGraaff2024,Weibel2024b, Wu2025}, and shallow spectroscopic samples at $z>3$ \citep{Nanayakkara2024, Nanayakkara2025}. Each of these approaches has some disadvantages. Photometric studies may suffer from inconsistent filter coverage and are insensitive to the presence of optical AGN; deep spectroscopy at $z=1\text{--}3$ targets the epoch when the mismatch between theory and observations is relatively small. Single-object studies may not be representative and shallow spectroscopy is not sensitive to the lower-mass end of the population.

In this work, we use deep low-resolution \JWST spectroscopic data to estimate an accurate number density of quiescent galaxies and to measure their lower-mass envelope. We complement this sample with medium-resolution rest-frame optical spectroscopy and X-ray to radio observations to measure the fraction of AGN hosts. Finally, we combine our number density measurements with theoretical predictions from FLAMINGO, the largest-box cosmological simulations to date, to provide a comparison of observations and theory over a meaningful survey volume.

Throughout this work, we assume the standard flat $\Lambda$CDM cosmology from \cite{PlanckCollaboration2020}. Emission lines are given using vacuum (air) wavelengths in the UV (optical--NIR) rest-frames (but in the analysis we use vacuum wavelengths as appropriate). We assume a \citet{Chabrier2003} initial mass function (IMF).

\section{Data}

\begin{figure*}
    \centering
    \includegraphics[width=1.\columnwidth]{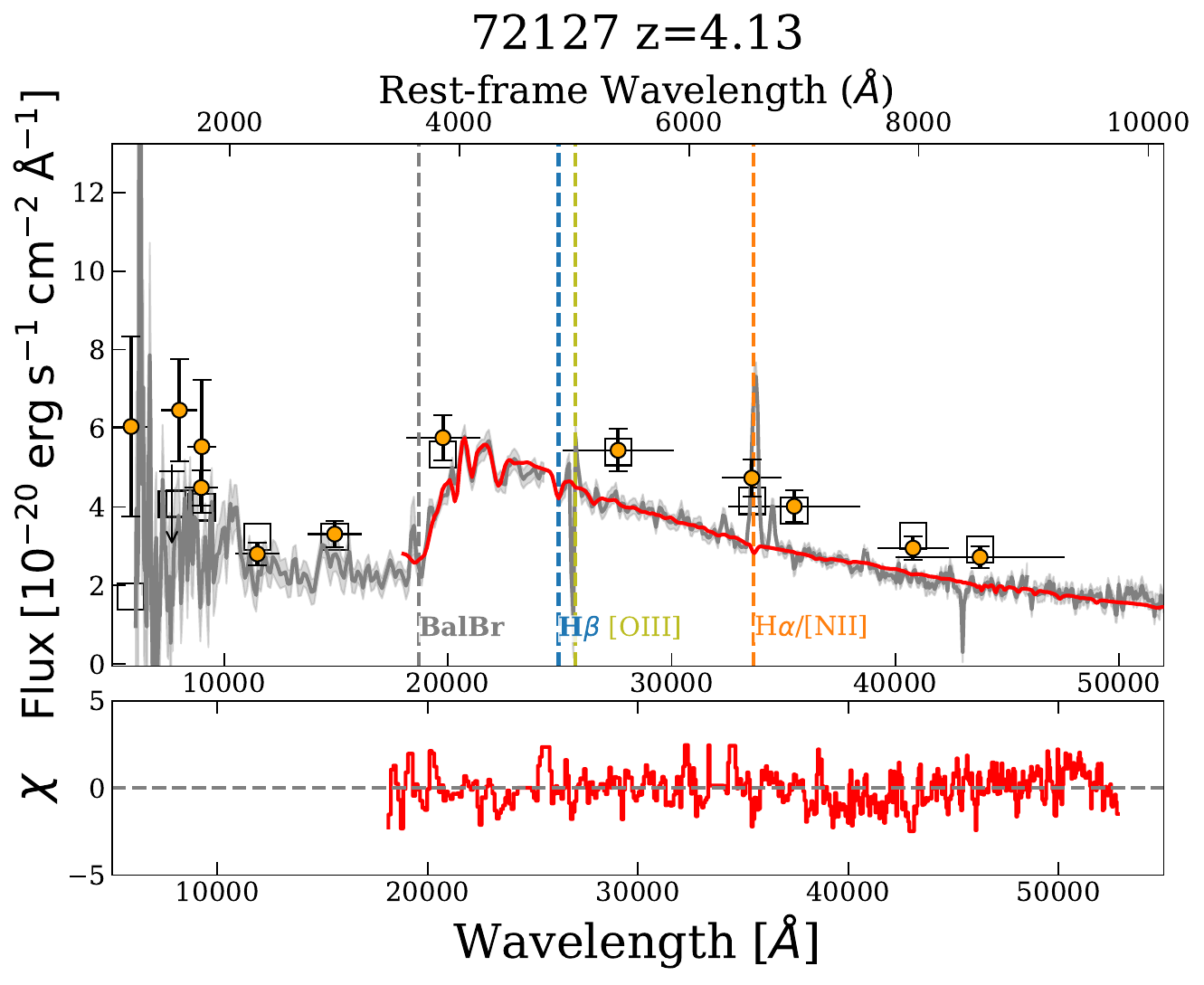}  
    \includegraphics[width=1.\columnwidth]{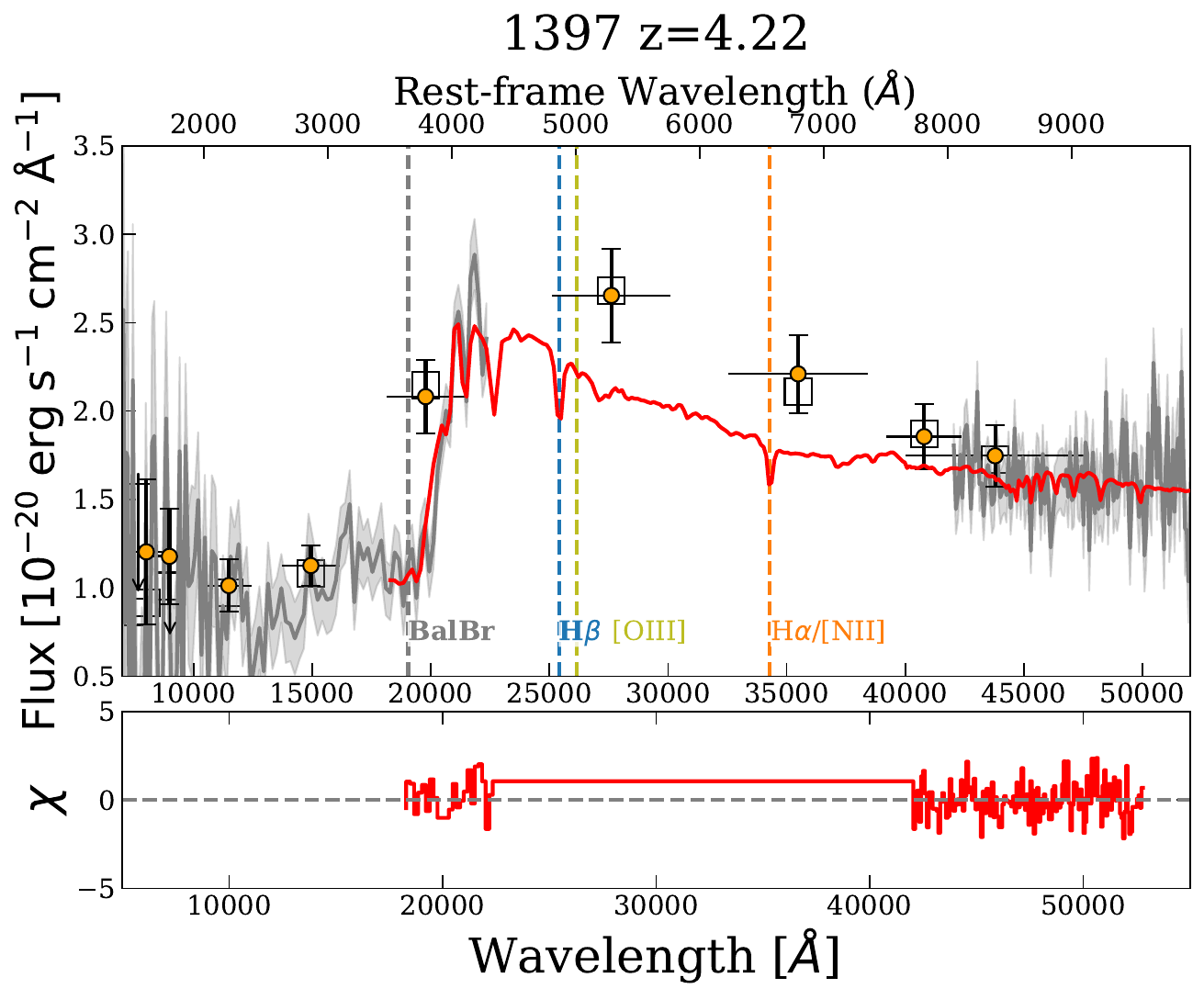}
    \includegraphics[width=1.\columnwidth]{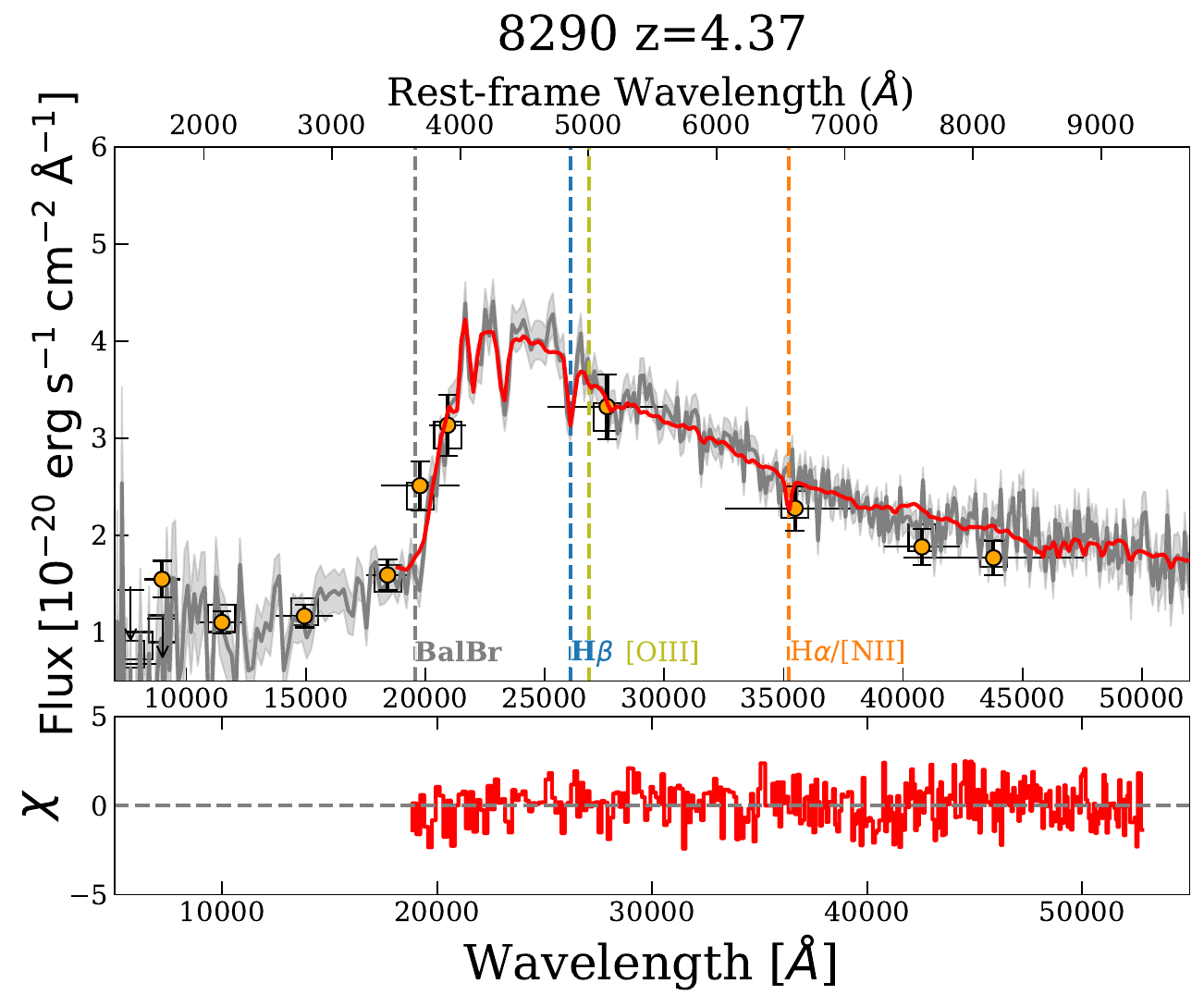}
    \includegraphics[width=1.\columnwidth]{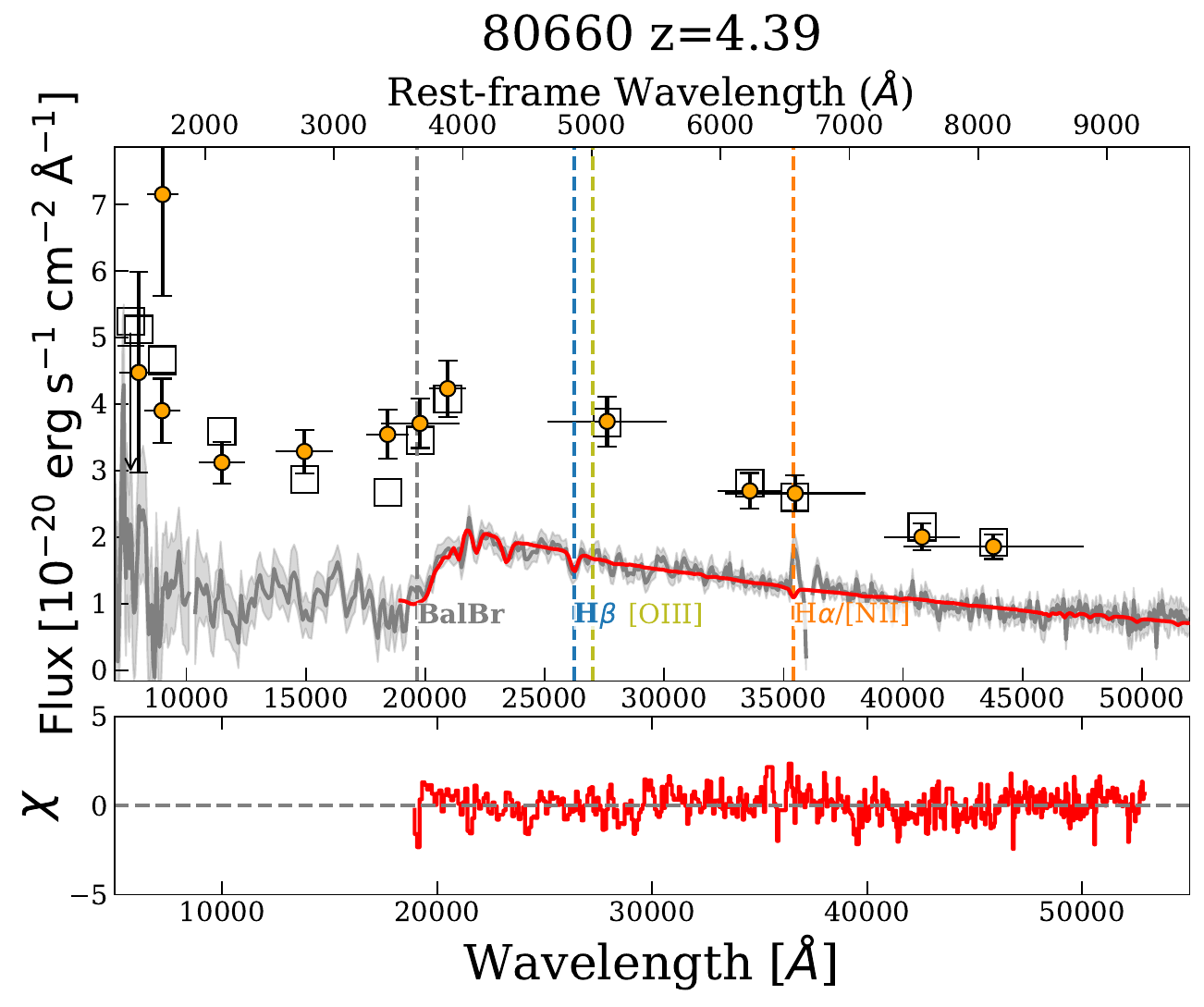}
    \includegraphics[width=1.\columnwidth]
    {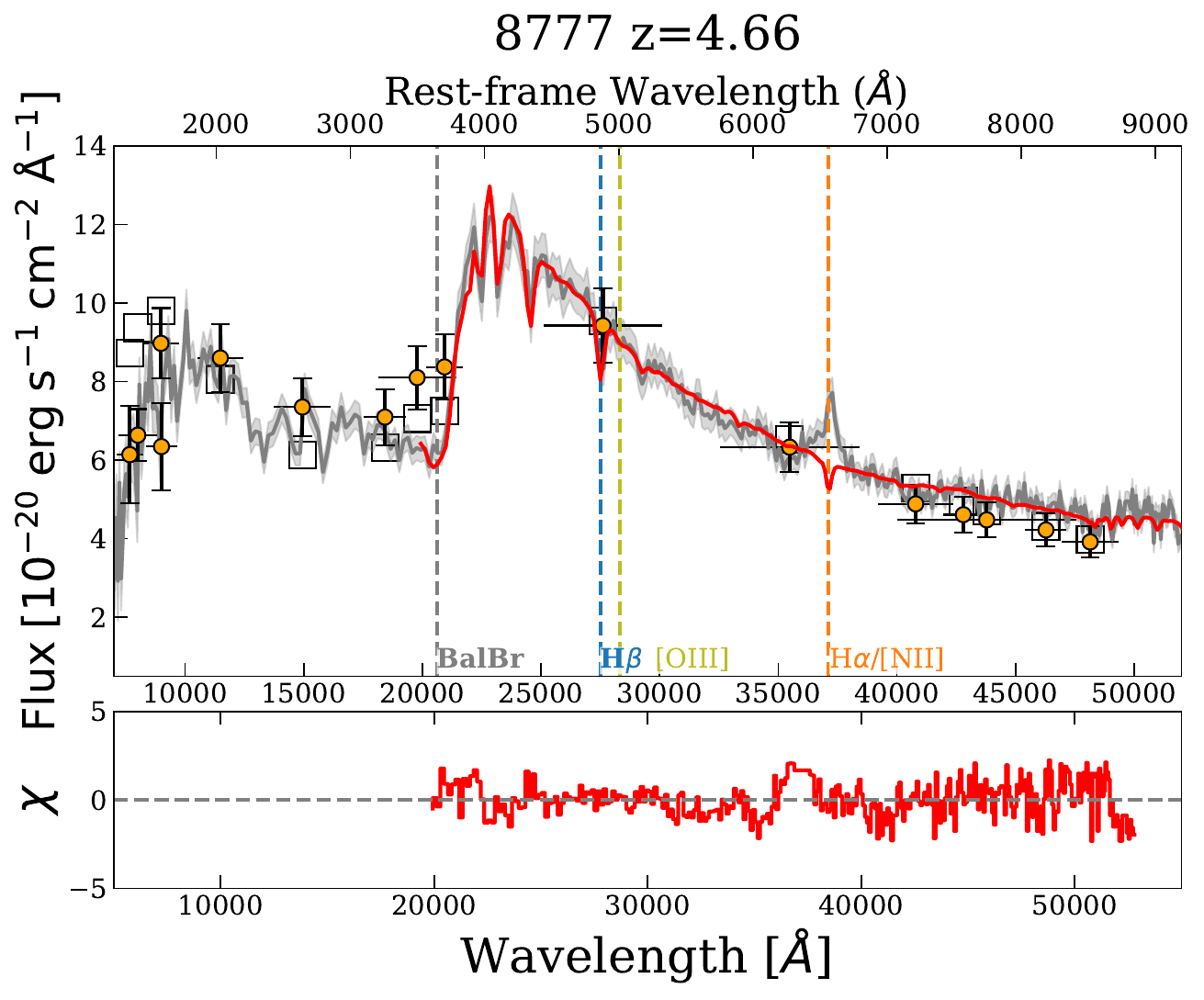}
    \caption{NIRSpec prism spectra for the five highest redshift quiescent galaxies in the sample ($z>4$). We fit both the spectrum and the photometry simultaneously.
    The grey represents the intrinsic spectrum and errors, while the red corresponds to the best-fit spectrum. The yellow points are the photometry, and the black squares the best-fit model photometry. The $\chi$ values in the bottom panel correspond to the best-fit spectrum. The spectra are upscaled to the photometry with a second-order polynomial (to account for extended sources and other effects).
    We marginalise over the emission lines because we cannot fit these for AGN - this is not taken into account in the $\chi$ figure. The spectra of Galaxies 1397, 8290 and 8777 come from GO programme 2198 \citet{Barrufet2021}. All other spectra and all photometry come from JADES \citet{Eisenstein2023} or JEMS \citet{Williams2023}.}
    \label{fig:spectra}
\end{figure*}

\subsection{NIRSpec Data}\label{subsec:nirspec}

In this work, we use \JWST/NIRSpec multi-object spectroscopy \citep{Jakobsen2022} from the GOODS fields \citep[Great Observatories Origins Deep Survey;][]{Giavalisco2004}. These data were obtained as part of three programmes: \JWST Programme ID (PID) GO 2198 \citep{Barrufet2021}; PID~1207 (PI~Rieke); and JADES \citep[the \JWST Advanced Deep Extragalactic Survey; PIDs~1210, 1180, 1181, 1286, 1287, and 3215][]{Eisenstein2023, Eisenstein2023b}.
In the GOODS fields, we can combine the accurate total fluxes of deep NIRCam photometry with the superior precision of MSA NIRSpec spectroscopy \citep[Micro Shutter Assembly;][]{Ferruit2022}. 
An extensive description of the NIRSpec observations and data reduction process has been provided in \cite{Bunker2024, D'eugenio2024}. Here we provide only a short summary, highlighting the differences compared to the publicly available data products.

All JADES NIRSpec observations use slitlets consisting of three shutters. Each observation uses a three-nod observing pattern and a variable number of dithers, as described in \citet{D'eugenio2024}.
The three-nod pattern is normally used for accurate background subtraction, but can cause wavelength-dependent self-subtraction of extended sources \citep{D'eugenio2024}. To avoid this problem, where possible, in this work, we always use the `two-nod' background subtraction, which only subtracts observations that are 2 micro shutters apart. This costs one third of the exposure time. An exception to this is the four galaxies drawn from PID~2198; these spectra were obtained from the DAWN JWST Archive\footnote{\href{https://dawn-cph.github.io/dja/index.html}{https://dawn-cph.github.io/dja/index.html}.} \citep[DJA;][]{Heintz2024}, and were reduced using a combination of the \textsc{jwst} pipeline and \textsc{msaexp}\footnote{\href{https://github.com/gbrammer/msaexp}{https://doi.org/10.5281/zenodo.7299500}}. While these four sources do not use the two-nod background subtraction, they are all very compact \citep[as expected from $z>4$ quiescent galaxies;][]{Ji2024}.
The integration times range from 2~h (PIDs 1180 and 1181) to 50~h (PID~3215), with the minimum duration considered sufficient to measure reliable star-formation histories \citep[SFHs; e.g.,][]{Nanayakkara2024,Glazebrook2024}.

The NIRSpec data undergo an initial processing through the \JWST Calibration Pipeline applying standard detector-level corrections such as dark subtraction, linearity calibration, and cosmic ray identification. The pipeline also performs flat-fielding, background subtraction, and wavelength calibration.

For each NIRSpec exposure, we utilise the \JWST grating equation models to determine the trace and wavelength solution for each source in the field of view. We then performed optimised spectral extraction, accounting for the curved traces and varying spatial profiles of the sources along the dispersion direction.
Our extraction routines are optimised to handle differential slit-losses for point-like sources. To correct for these issues in the slit-loss correction and, possibly, in the overall flux calibration, we recalibrate the spectra using NIRCam photometry.

A critical component of our NIRSpec reduction process is the determination of accurate spectroscopic redshifts for the observed sources. We use an initial redshift estimate using \textsc{bagpipes} \citep{Carnall2018}, then refine this estimate with a combination of visual inspection, full spectrum fitting with \textsc{ppxf} \citep{Cappellari2023}, and further emission-line fitting in the medium-resolution data (when available, and when emission lines are detected). The procedure is described in \citet{D'eugenio2024}.

All galaxies that we explore spectroscopically have spectroscopy obtained with the prism disperser, from PID~2198 and from JADES. These data have a spectral resolution $R=30\text{--}300$ in the wavelength range 0.6--5.3~$\mu$m \citep{Jakobsen2022}. For a subset of %11 
galaxies, we also have medium-resolution spectroscopy. 
%2 
A fraction of these are from PID~1207, which uses a combination of G140M/F100LP and G235M/F170LP, spanning 1--1.6~$\mu$m with $R=700\text{--}1,200$. The remainder 
are from JADES which uses G140M/F070LP, G235M/F170LP and G395M/F290LP, covering the same spectral range as the prism.
Most of the medium-resolution observations are too shallow to constrain the stellar continuum (2~h). Moreover, the mask design of the JADES survey allows for spectral overlaps in the medium-grating observations \citetext{\citealp{Bunker2024,D'eugenio2024}; except for the highest-priority targets; e.g., \citealp{Bunker2024}}. Although overlaps are easy to identify and remove for emission line science \citep{D'eugenio2024}, they may severely affect the continuum. For this reason, in this work we focus on prism spectroscopy.
However, medium-resolution data are still crucial for detecting optical AGN via emission line diagnostics.
For this reason, we use these data to measure the flux of nebular emission lines, given that the spectral resolution of the gratings enables us to spectrally resolve the [NII]$\lambda\lambda$6543,6584--H$\alpha$ complex; Section~\ref{subsec:ppxf}).
For one galaxy from PID~1207, we also use the grating data to measure the SFH and to assess the reliability of the prism-based inference (Appendix~\ref{app:grating.sfh}). 
We find that the two SFHS are are consistent to within 1\sig 1. For this comparison we do not use the JADES gratings observations, because the JADES masks are designed by allowing spectral overlaps, which affect the continuum and can therefore bias the SFH.

\subsection{NIRCam Data}

The Near-Infrared Camera (NIRCam) imaging data from the James Webb Space Telescope (\JWST) obtained by JADES \citep[PIDs: 1180, 1181, 1210, 1286, 3215 ][]{Eisenstein2023, Eisenstein2023b, Rieke2023} and JEMS \citep[PID: 1963][]{Williams2023} are processed through the JADES custom reduction pipeline.

We start with the \JWST Calibration Pipeline version 1.9.6, incorporating the latest Calibration Reference Data including dark frames, distortion maps, bad pixel masks, read noise, superbias, and flat-field files \citep{Rieke2023}.

At the detector level, we perform group-by-group corrections that include dark subtraction, linearity calibration, and cosmic ray identification, followed by ramp fitting to recover count-rate images \citep{Rieke2023}.
Further calibration steps apply flat-fielding, photometric calibration, and background subtraction to ramp-fitted images \citep{Rieke2023}.
We use a custom version of TweakReg to align exposures, matching sources across images to calculate relative and absolute astrometric solutions. The calibrated aligned exposures are then combined into mosaics using the \JWST Stage 3 pipeline \citep{Rieke2023}.

We used the standard photometric JADES catalogues for the photometry \citep{Rieke2023,Eisenstein2023b,D'eugenio2024}. A brief outline of these are as follows. We use the latest version of \textsc{SExtractor} \citep{Bertin1996} to detect sources in mosaicked NIRCam images. The detection parameters are tuned to optimise sensitivity to faint sources while minimising spurious detections. Isophotal photometry with careful deblending allows us to measure accurate colours for blended objects.
Astrometric alignment is performed using common sources across datasets to enable consistent multi-wavelength photometry. 

 We used a combination of Kron convolved and CIRC5 photometry (which corresponds to an aperture radius of 0.35~arcsec.). We checked both and found no significant difference in the derived photometry, just differences in the goodness of fit due to extended sources and near neighbours. In each case, we chose the better fitting photometry.

\subsection{FLAMINGO simulation}

For comparisons with models and to help interpret our observational results, we adopt predictions from the Full-Hydro Large-scale Structure Simulations with All-sky Mapping for the Interpretation of Next Generation Observations (FLAMINGO; \citealt{Schaye2023, Kugel2023}) project. The FLAMINGO suites are state-of-the-art cosmological hydrodynamical simulations with large volumes of up to $(2.8\,\rm cGpc)^3$. The simulations are run using the code \textsc{SWIFT} \citep[Smoothed-particle-hydrodynamics With Inter-dependent Fine-grained Tasking,][]{Schaller2024}.

The main advantage of using FLAMINGO for this kind of study is clear: the massive objects of interest are very rare, and combined with the relatively narrow area probed by deep surveys including \JWST, the cosmic variance is expected to be significant, going up to or beyond an order of magnitude \citep{Xiao2023, Lim2024}. Given the comoving volume probed by our data or other similar studies of $10^5{-}10^6\,{\rm cMpc}^3$, only simulations larger than a cGpc size allow the probe of cosmic variance safely up to 3~$\sigma$, which is the minimum for evidencing tensions. 

The suites have variations in volume and resolution as well as physics models and cosmology. For our analysis, we use its $(1\,\rm cGpc)^3$ flagship run, L1\_m8, where the mass resolution of the initial baryonic elements is $m_{\rm gas}\,{=}\,1.34\times10^8\,{\rm \Msun}$, and the Plummer-equivalent comoving gravitational softening length is 11.2\,ckpc. The reason we chose the smaller-box higher-resolution simulation is that the $1\,{\rm cGpc}$ box is already large enough, thousands of times bigger than the survey volume (see Sect.~\ref{s.results.ss.numberdens}). This large box size enables us to explore the cosmic variance up to 3$\sigma$ and achieve our science goals. Also, the SFH predicted from the simulations is found to be largely dependent on the numerical resolution, with the lower-resolution FLAMINGO yielding a much younger SFH (and essentially zero, falling below the resolution, at $z\,{\gtrsim}\,7$). Although the relatively poor resolution of even the highest resolution FLAMINGO run is potentially a limitation for comparisons, increasing resolution typically leads to a lower quenched fraction (e.g. \citealt{Schaye2023}), because more star-forming small-scale gas is resolved, and because fewer galaxies fall below the resolution to be assigned zero SFR. Therefore, our results from FLAMINGO can be rather taken as a conservative limit for comparisons of e.g. the number densities, concerning the resolution. 

The free parameters for the subgrid models employed in FLAMINGO were calibrated by machine learning to both the stellar mass function and the cluster gas fraction from $z\,{\lesssim}\,0.3$ observations. The simulation employs the cosmological models from the Dark Energy Survey year three results \citep{Abbott2022}, which is a flat universe constrained by 3$\times$2pt correlation functions. Although this is different from the Planck cosmology assumed elsewhere in this paper, we checked, with a lower-resolution FLAMINGO that adopted the Planck cosmology, that the difference is insignificant. Finally, the simulation assumes a \citet{Chabrier2003} IMF, the same as for the observations.

\section{Analysis}\label{s.analysis}

\subsection{SED Fitting}\label{s.analysis.ss.sedfit}

We use the flexible \textsc{prospector} code \citep{Johnson2021} to perform a simultaneous SED modelling of our multi-wavelength photometric and spectroscopic data sets from \JWST.
Prospector is a Bayesian SED modelling framework designed to generate realistic galaxy SEDs from flexible stellar population synthesis models and constrain the corresponding model parameters given observational data. Source-frame galaxy SEDs are constructed using the \textsc{fsps} software \citep{Conroy2009, Conroy2010}, allowing for user-specified parametric forms or non-parametric representations of the SFH, metallicity evolution, dust attenuation, and other components.

We used the MILES spectral library \citep{Sanchez-Blazquez2006,Falcon-Barroso2011} and MIST isochrones \citep{Choi2016}. Note that these assume fixed (solar) abundances, which are not representative of the observed properties of massive quiescent galaxies at $z\lesssim 2$ \citep{Kriek2016,Beverage2021,Beverage2024}, and may result in over-predicted stellar ages.
We assume a \citet{Chabrier2003} initial mass function and a non-parametric star-formation history, described by a constant star-formation rate (SFR) over a set of predetermined intervals in look-back time. The model is parametrised by the logarithm of the SFR ratio between adjacent time intervals. We adopt the `continuity' probability prior on this SFH model \citep{Leja2019}, which consists of a Student's $t$ distribution with mean 0, and standard deviation 0.3. The mean of 0 centres the prior on a constant SFR with cosmic time, while the broad wings of the $t$ distribution allow for variations in the SFR over time, albeit with a modest penalisation.

Dust attenuation assumes the \citet{Calzetti2000} attenuation law. 

We perform a full spectro-photometric fitting by combining the prism spectra with multi-band photometry, enabling us to incorporate information from both of these sources.  We used a 2\textsuperscript{nd}-order polynomial to account for offsets between the photometry and the spectra due to wavelength-dependent slit losses. This has the effect of modelling both the spectrum and the photometry, particularly important in the case of larger extended sources where the spectrum only probes a region of the overall galaxy. We note, however, that in this process we are still fitting the photometry combined with the spectrum (so we include the shape and normalisation of the photometry etc.) as this is crucial for determining the total galaxy properties. We are not simply upscaling the spectrum. This is shown in Appendix \ref{app.s.offset} and Fig. \ref{fig:141147}. In the case of galaxies 12619, 171147, and 72396 we found we required a 10\textsuperscript{th}-order polynomial to better fit the observed photometry. We also test using a zero order polynomial, i.e. not accounting for the wavelength dependent slit losses, and find it leads to significantly poorer fits.

We test the effects of using a bursty continuity prior on the SFH \citep[as in][]{Tacchella2022} - this is enacted by increasing the width of the Student's $t$ distribution prior between the SFH bins. This has the effect of allowing greater variation in the SFR between adjacent bins, i.e. more `burstiness' in the SFH. We find that this does not significantly change our results. 

We also explore different priors on the stellar metallicity. This is important for two reasons. In general, for the youngest quiescent galaxies in our sample \citep[i.e., spectral `post-starburst' galaxies, e.g.][]{Goto2005,Wu2020,Suess2022,D'Eugenio2020}, metal absorption lines have low equivalent width, thus requiring high continuum SNR to reliably measure metallicity. In addition, because the spectral resolution of our prism data is insufficient to capture variations in metal-absorption features, we do not expect to constrain the stellar metallicity \citep{deGraaff2024}. We start with a uniform prior in $\log Z$, with $\rm \log(Z/\Zsun)\in (-2,0)$. We then vary this to a higher metallicity prior to $\rm log(Z/\Zsun)\in(-1,0.19)$, and we also test a uniform prior in linear $Z$. We find that whilst obtaining a multitude of different, relatively unconstrained metallicities for these galaxies, this does not alter the other stellar population properties such as formation and quenching times.

In our fiducial run, we also mask the NIRSpec spectrum for rest-frame wavelengths bluer than 3500~\AA, to avoid deriving metallicity information from the relatively poorly known rest-frame UV spectrum, which may additionally suffer from strong non-stellar features, such as ISM absorption \citep[e.g.,][]{Calzetti2000} or AGN continuum emission \citep[e.g.,][]{Carnall2023Nature}.

The high-quality spectroscopic redshifts from NIRSpec are incorporated as priors, reducing the number of degeneracies in the SED modelling. Stellar-population properties, including ages, metallicities, and SFHs, are derived from the combined spectrophotometric constraints, 
leveraging the continuity SFH to obviate the lack of constraining power in the data for the earliest stages of star formation.
We marginalise over the emission lines because we cannot accurately model those galaxies that contain AGN. 

Fig. \ref{fig:spectra} shows an example of our \textsc{Prospector} fits (in this case to the  highest redshift quiescent galaxies found in this work, see Section \ref{s.sample}). The observed spectrum and errors are given in grey and the observed photometry is represented by the yellow points. The best fit spectrum is in red and the best-fit photometry are the black squares. The lower panel shows the $\chi$ values for the fit of the spectra. We note that the $\chi$ values do not take into account the marginalisation of the emission lines hence the divergence seen in those regions. In some galaxies there is an offset between the spectrum and the photometry, but the spectrum has been scaled to the photometry with a second-order polynomial (see above).

\subsection{Emission-line fluxes}\label{subsec:ppxf}

The emission line fluxes for H$\beta$, [OIII]$\lambda$5007, H$\alpha$ and [NII]$\lambda$6583 were measured by simultaneously fitting the continuum and emission lines with \textsc{ppxf} \citep{Cappellari2023}. We used the same setup as in \citet{D'eugenio2024}, for both the prism and medium-resolution spectra.

When measuring H$\beta$ and [OIII]$\lambda$5007 we always use the prism data, because prism spectra are available for all galaxies, and because they have a higher signal-to-noise ratio (SNR) than the shorter-exposure grating spectra.
We use the latter to measure H$\alpha$ and [NII]$\lambda$6583, given that the medium-resolution data has sufficient spectral resolution to separate H$\alpha$ from the [NII]$\lambda\lambda$6548,6583 doublet.
Measurement uncertainties are obtained by adding random noise to the observed spectra and repeating the fit. Random noise is derived by using the pipeline uncertainties. We tested sampling from the best-fit residuals in a neighbourhood of each spectral pixel \citep[e.g.,][]{Looser2023}, and found similar results. This procedure also varies the shape of the underlying stellar continuum, but artificially increases noise (by adding random noise to already noisy data). We reduce the uncertainties derived by a factor $\sqrt{2}$ to take into account the additional noise introduced by this procedure.

\subsection{Morphological Fitting}\label{s.analysis.ss.morphology}

To fit the morphology of the quiescent galaxy sample, we use the code \textsc{PySersic} \citep{Pasha2023}. This software forward models and fits the light distribution of the galaxy assuming a parametric model. We use a standard, axisymmetric \citep{Sersic1968} profile. For the PSF, we use model PSFs from the JADES collaboration as in \cite{Ji2023,Ji2024}. As input, we use the F444W images because this is the only wide-band filter available for all targets, whilst also not being contaminated by strong emission lines from AGN (we also test using F200W to make better use of the SW PSF and find it makes no significant difference to our results). For our quiescent galaxy sample, this corresponds to rest-frame optical continuum emission from the stellar populations, given that at $z<5$ we can rule out strong emission line in this filter and a major contribution from hot-dust emission due to AGN. We create 3\arcsec $\times$ 3\arcsec square cutouts centred on each target galaxy; this size is large enough to provide source-free regions to estimate any residual background, but small enough to save computational time. We use \textsc{Source Extractor} to model any remaining background in the cutouts and to perform source detection. 
We fit a single \sersic profile \citep{Sersic1968} to every source in the cutout, except those with fluxes below 10~nJy, which we model as point-source components due to their small size and low SNR. 
In almost all cases, the primary source contains almost the entire flux in the image, but we still fit the remaining sources to avoid biasing the model by trying to account for excess light away from the central source. 
We use NUTS (No U Turn sampling) to sample the posterior in order to determine the best-fit model and to obtain more realistic uncertainties.
The best-fit half-light radii and \sersic indices are reported in Table \ref{tab:values_errors}.

\section{Sample}\label{s.sample}

\subsection{JADES Sample}\label{s.sample.ss.jades}

 \begin{figure}
    \centering
    \includegraphics[width=\columnwidth]{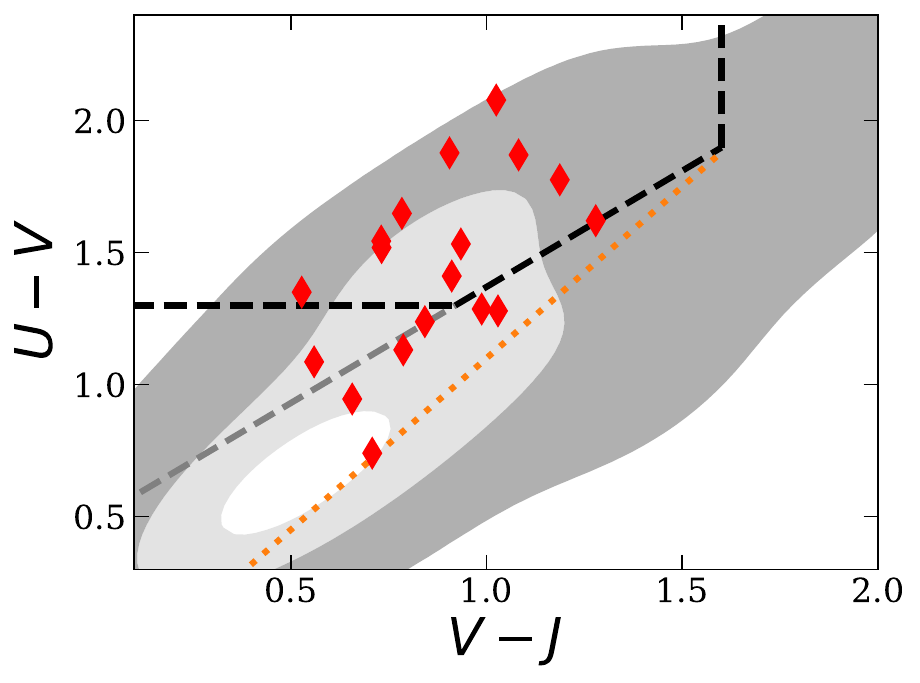}
    \caption{A UVJ quiescent galaxy selection diagram with the UV colours plotted against VJ colours of our final sample of 18 massive quiescent galaxies (diamonds).  Overplotted in black is the typical UVJ selection criteria used traditionally based upon \citet{Schreiber2015}. We expand this (grey line) to include fast quenching as in the case of \citet{Belli2019}.  We treat any galaxy that falls within this criteria within 1 $\sigma$ of the colours as a possible quiescent galaxy to maximise completeness. The sample that fufill this criteria are denoted by the grey contours with the outer contour containing 90\% of the population. Overplotted as an orange dashed line would be the required selection to include all spectroscopically found quiescent galaxies in our sample. This motivates a comprehensive study into UVJ colours as selection at high-redshift.}
    \label{fig:uvj}
\end{figure}

\begin{figure*}
  \includegraphics[width=\textwidth]{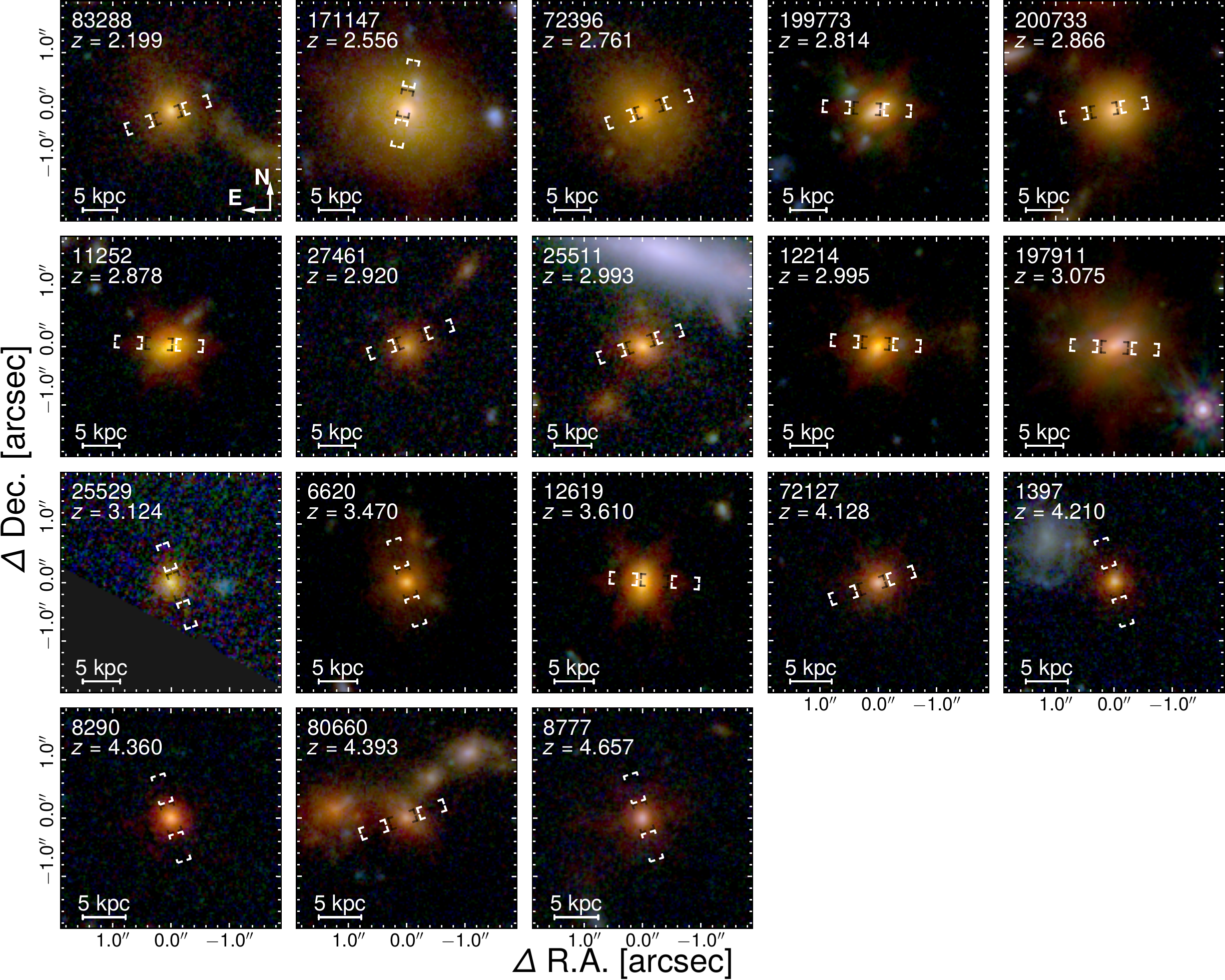}
  \caption{False-colour images of the sample, sorted by redshift. Red, green, and blue are from F444W, F200W, and F090W NIRCam imaging. The NIRSpec shutters are overlaid (black and white colours to enhance contrast).
  There is a clear trend of increasing galaxy size with cosmic time.
  Several sources present low-mass satellites. The three galaxies that look near in projection to 80660 are at different photometric redshifts. Three quiescent galaxies are spectroscopically confirmed satellites; these are not part of the sample (Appendix~\ref{app:satellites}; Fig.~\ref{fig:rgb.satellites}).}
  \label{fig:rgb}
\end{figure*}

\begin{figure*}
    \centering
    \includegraphics[width=2\columnwidth]{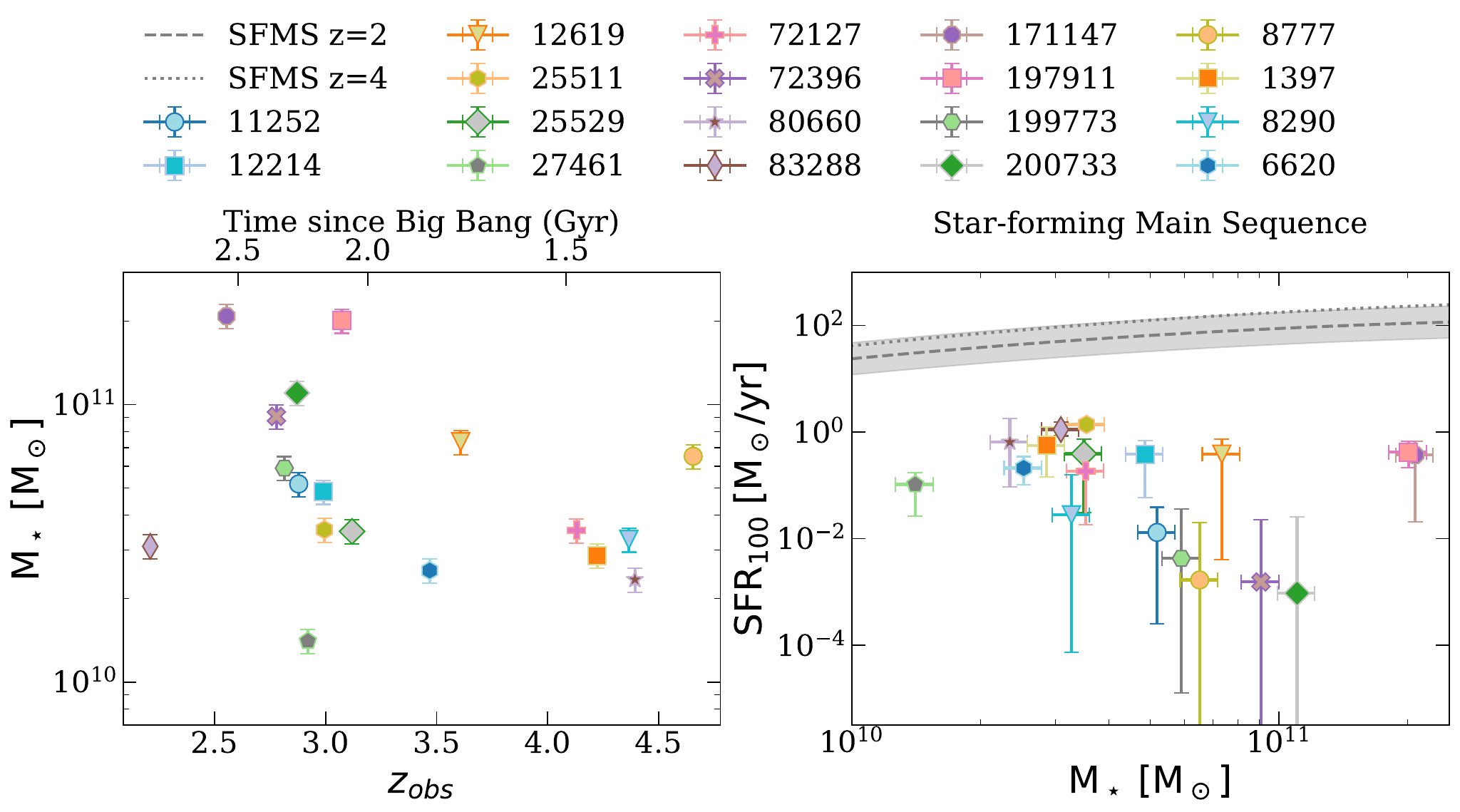}
    \caption{The distribution of our sample of massive quiescent galaxies on the stellar mass vs redshift and SFR vs stellar mass planes. The points correspond to the galaxies. Left: stellar mass vs spectroscopic redshift. Right: Star-formation rate (averaged over 100~Myr) vs stellar mass. The dashed and dotted line corresponds to the SFMS at redshifts 2 and 4 respectively \citep{Popesso2023}. The grey shaded region corresponds to a 0.3dex uncertainty on the SFMS at redshift 2.}
    
    \label{fig:sample}
\end{figure*}

We selected our samples from spectroscopic data in the GOODS-S and GOODS-N fields. We limit the selection to galaxies with prism data, and to two programmes: JADES and PID: 2198 (see Section~\ref{subsec:nirspec}). As we are interested in higher redshift quiescent galaxies we imposed a redshift cut of z$\geq$2. 

First, we select candidate quiescent galaxies based on UVJ colours obtained by template fitting photometry with the fitting code \eazy  \;\citep{Brammer2008}, as described in \cite{Hainline2024}. \eazy\xspace uses galaxy templates to perform a grid search based upon redshift. The templates used are a combination of the \eazy "v1.3" templates \citep{Brammer2008, Erb2010}, alongside mock templates from the JAGUAR simulations \citep{Williams2018}. These are supplemented by new templates designed using \textsc{FSPS} \citep{Conroy2009} which were added to those of \citet{Coe2006}.\footnote{The templates used are all publicly available \citep{hainline_2023_7996500}. }

We use the UVJ criteria of \cite{Schreiber2015} with the fast quenching addition of \cite{Belli2019}. Specifically, the dividing line given by
\begin{equation}
    U-V>0.88\;(V-J)\;+\;0.49,
\end{equation}
where U, V and J are rest-frame magnitudes in the same filters as \citet{Williams2009}. However, as we are looking for rare, high-redshift quiescent galaxies, we want to ensure that we recover all quiescent galaxies with spectra. Therefore, we include the one-\sig uncertainties, thereby selecting any candidate galaxy that falls into the extended \cite{Schreiber2015} criteria \citep{Belli2019} if the U,V,J colours are perturbed by one \sig. Also, as we are looking for massive quiescent galaxies, we impose a stellar mass cut of $\Mstar>10^{10}\;\Msun$ based on the best-fitting \eazy\; masses. We test our best-fit Prospector masses vs the template derived \eazy masses finding no systematic offset and a scatter of within 0.2dex.

Next, to select the final sample from these candidates, we imposed the time-dependent criterion based on the specific star-formation rate (sSFR) obtained via SED fitting (Section~\ref{s.analysis.ss.sedfit}).\footnote{Note that, because our SED fitting approach does not model AGN-dominated nebular emission, we do not model emission lines, which are fitted and removed from the spectrum during optimisation.
For this reason, our modelling approach is not sensitive to star formation on very short timescales. We rely on the BPT diagram to assess the origin of the emission lines in our sample. } This criterion is given by
\begin{equation}
    \mathrm{sSFR_{100Myr}}\leq \frac{0.2}{t_\mathrm{obs}}
    \label{eq:quenching}
\end{equation}
where $\rm sSFR_{100Myr}$ is the SFR averaged over 100~Myr divided by the total formed stellar mass by the time of observation, and $t_\mathrm{obs}$ is the age of the Universe at the redshift of the galaxy. This criterion has been used previously in many studies and is generally considered equivalent to a colour cut in the UVJ colour-colour diagram \citep[e.g.,][]{Gallazzi2014, Carnall2024}. Interestingly, in our case we find that quiescent galaxies that fall outside of the standard UVJ quiescent region are still clearly quiescent based on the sSFR cut. We also, as a double check, visually inspect the spectra and see clear evidence for strong Balmer breaks and a lack of emission lines. This motivates further study to explore how many high-z quiescent galaxies are lost by relying on colour selection alone at high redshift where significant extrapolation can be required \citep[e.g. ][]{Valentino2023, Gould2023, Antwi-Danso2023}.
 
The sSFR selection results in a sample of 21 quiescent galaxies, 17 from JADES and 4 from PID2198 \citep[which includes CANDELS GS-9209;][]{Carnall2023Nature}. 
After visual inspection, we reject three galaxies from this sample; one due to a spectrum contaminated by short circuits in NIRSpec \citep{Rawle2022}, and two that exhibit a combination of a Balmer break and dust-obscured star formation -- our treatment of emission lines in the spectrum are unable to disentangle this kind of system.

This gives us a final sample of 18 massive quiescent galaxies.

The UVJ colours of the final sample are shown in Fig. \ref{fig:uvj}. The \cite{Schreiber2015} criteria is given by the dashed black line and the fast quenched extension \citep{Belli2019} is given by the grey dashed line. 

We find that in order to include all our quiescent galaxies based on their 50th percentile \eazy\; colours, we would require a selection criteria of
\begin{equation}
    U-V>1.30\;(V-J)\;-\;0.20,
\end{equation}
shown as the orange-dotted line in Fig.~\ref{fig:uvj}.
This motivates the exploration of the systematics involved in the colour selection at higher redshift.

We note that during the selection process we also identified two low-mass quiescent satellite galaxies below the mass cut. Moreover, we identified another galaxy above the mass cut that is clearly a satellite. Because we are interested only in \textit{internal} quenching, we reject these three galaxies, which are clearly satellites, based on their proximity to higher-mass galaxies at the same redshift \citetext{one of which has been already reported by \citealp{Sandles2023}}. Nevertheless, the fact that our methods correctly identify even the two low-mass satellite galaxies as quiescent means that the deepest spectroscopic data we use are sensitive enough to probe the lowest-mass envelope of the quiescent population. We explore these galaxies' spectra and SFHs in more detail in the Appendix Section~\ref{app:satellites}.

For our final sample of 18 massive quiescent galaxies, we report galaxy NIRSpec IDs, coordinates, Balmer-break strengths, and D$_n$4000 index as well as other derived properties in the Appendix Table \ref{tab:values_errors}.

Fig.~\ref{fig:rgb} shows RGB images of all 18 massive quiescent galaxies in our sample. All appear classically quiescent (i.e. red, with no blue clumps or other signs of recent star-formation) on the basis of the NIRCam imaging. We model all the quiescent galaxies and other sources in the imaging with a single-component \sersic profile with a freely varying \sersic index (see Sections \ref{s.analysis.ss.morphology} and \ref{sec:morphology}). 

Fig. \ref{fig:sample} shows stellar mass vs spectroscopic redshift (left panel) and star-formation rate (SFR, averaged over 100~Myr) vs stellar mass (right panel).
From the left-hand panel we see that we have massive quiescent galaxies from redshift 2 (our lower cut) to redshift 4.7. 

We find five galaxies above redshift four - two of which (1397, and 80660) have not previously been published, one of which was previously identified as a little red-dot candidate \citep{Kokorev2024}, one of which (8290) had its spectra previously published in \cite{Barrufet2024b} but not explored (i.e. fit), and one (8777) whose R100 spectra were published in \cite{Barrufet2024b} and whose R1000 spectra was explored in detail in \cite{Carnall2023Nature}. This galaxy (ID~1297, CANDELS~ID GS-9209) is the highest-redshift quiescent galaxy in our sample at z=4.7 and hosts a broad-line region (BLR) AGN \citep{Carnall2023Nature}. 
As mentioned above, 1397 has not been published previously, probably because of the detector gap affecting part of the prism spectrum. However, we still see clear evidence of a Balmer break, which, complemented by JADES photometry, suggests a lack of high-equivalent-width (EW) emission lines (estimating EWs with the photometry and model spectrum gives EW(H$\beta$+[OIII])=$7\pm0.3$\AA and EW(H$\alpha$+[NII])=$25\pm 1.3$). 
The two quiescent galaxies from JADES both reside in GOODS-N, at redshifts 4.1 and 4.4 respectively.

The observed spectra and photometry for these five (z>4) galaxies can be seen in Fig. \ref{fig:spectra} along with the best-fit models. We show the remaining spectra in Fig. \ref{fig:D4000spectra} and in Appendix Figs. \ref{fig:spectra_1} and \ref{fig:spectra2}.
We see clear evidence for strong Balmer breaks and absorption features, confirming these galaxies' quiescent nature. 

We find that our galaxies lie within the stellar mass range of $10^{10}-10^{11.3}\Msun$, with the majority having a lower mass than other comparable samples at this redshift \citep[e.g.][]{Carnall2024,Nanayakkara2024}.

We next explore the galaxies' position relative to the star-forming main sequence \citep[SFMS][]{Brinchmann2004, Noeske2007, Speagle2014, Popesso2023,Baker2023a}. This relation describes the strong correlation between the SFR of a galaxy and its stellar mass, which stems from a combination of the molecular gas main sequence, which describes the amount of gas within a galaxy relative to its stellar mass, and the Schmidt-Kennicutt relation, which encodes information on the SFR of a galaxy relative to its amount of gas \citep[][]{Baker2022, Baker2023a}. Although the SFMS does not directly encode a causal relation \citep[e.g.,][]{Baker2023a}, it is still a useful benchmark for studying the evolutionary stage of galaxies, even when the gas content is known \citep{Tacconi2020, Scholtz2024}.

Fig. \ref{fig:sample} right panel shows the galaxies' positions relative to the star-forming main sequence at redshifts 2 and 4 \citep[grey dashed and dotted lines from][]{Popesso2023}. We see that all our quiescent galaxies reside far below the SFMS. This is to be expected, as they are all defined as quiescent on the basis of several different criteria. This shows that over a timescale of the past 100~Myr there has been no considerable star-formation activity.

\subsection{FLAMINGO sample}

\label{ssec_FLAMINGO_sample}

Based on the characteristics of the observational data, we select galaxies from the FLAMINGO simulation. Specifically, we select massive quenched galaxies from FLAMINGO with ${M_\ast\,{\geq}\,10^{10}\,{\rm M}_\odot}$ and ${\rm sSFR_{100Myr}}\,{\leq}\,0.2\,/\,t_{\rm obs}$, mimicking the observational selection. Although the stellar mass of the JADES samples extends down to ${\simeq}\,10^{10}\,{\rm M}_\odot$, we apply this conservative cut because the completeness of the samples is not clearly known. We tested with several other selections, such as the mass cut of ${M_\ast\,{\geq}\,10^{10.4}\,{\rm M}_\odot}$, and found that the results almost always vary within a factor of 2. The sSFR is measured by summing up the mass of star particles newly born during the last 100\,Myr, and by dividing it by the total formed stellar mass as well as 100\,Myr. The selection of simulated galaxies is made independently of the two snapshots at $z\,{=}\,2.5$ and 3.5. This is because the observational data cover a wide range of redshifts, during which galaxies undergo substantial evolution in both physical properties and number density. We also add a log-normal scatter of 0.3\,dex to the stellar mass and SFR originally from the simulation before the selection, to represent the observational errors in the derived quantities. 

\section{Results}\label{s.results}

\subsection{Number densities of quiescent galaxies}\label{s.results.ss.numberdens}

\begin{table}
    \caption{Quiescent Galaxy Number Densities. The errors indicate the 1\,$\sigma$ range, obtained by dividing the sky area into 25 sub-areas, and drawing 10,000 bootstrap sample thereby.}
    \centering
    \label{table:number_densities}
    \begin{tabular}{lccc} % Adjusted column alignment
        \toprule % Top horizontal line
        Redshift Range & Number Density (cMpc\textsuperscript{-3}) 
        \\ 
        \midrule % Middle horizontal line
         2 $\leq$ $z$ $\leq$ 3
         & $1.28^{+0.77}_{-0.26}\times10^{-4}$
         \vspace{1.5mm} \\ 
         3 $\leq$ $z$ $\leq$ 4.5 
         & $3.96^{+2.64}_{-0.66}\times10^{-5}$
         \vspace{1.5mm} \\
         4 $\leq$ $z$ $\leq$ 5 
         & $1.28^{+0.85}_{-0.43}\times10^{-5}$
         \vspace{1.5mm} \\
         2 $\leq$ $z$ $\leq$ 4.5  
         & $8.01^{+4.37}_{-1.46}\times10^{-5}$ 
        \\ 
        \bottomrule % Bottom horizontal line
    \end{tabular}
\end{table}

\begin{figure*}
    \centering
    \includegraphics[width=2\columnwidth]{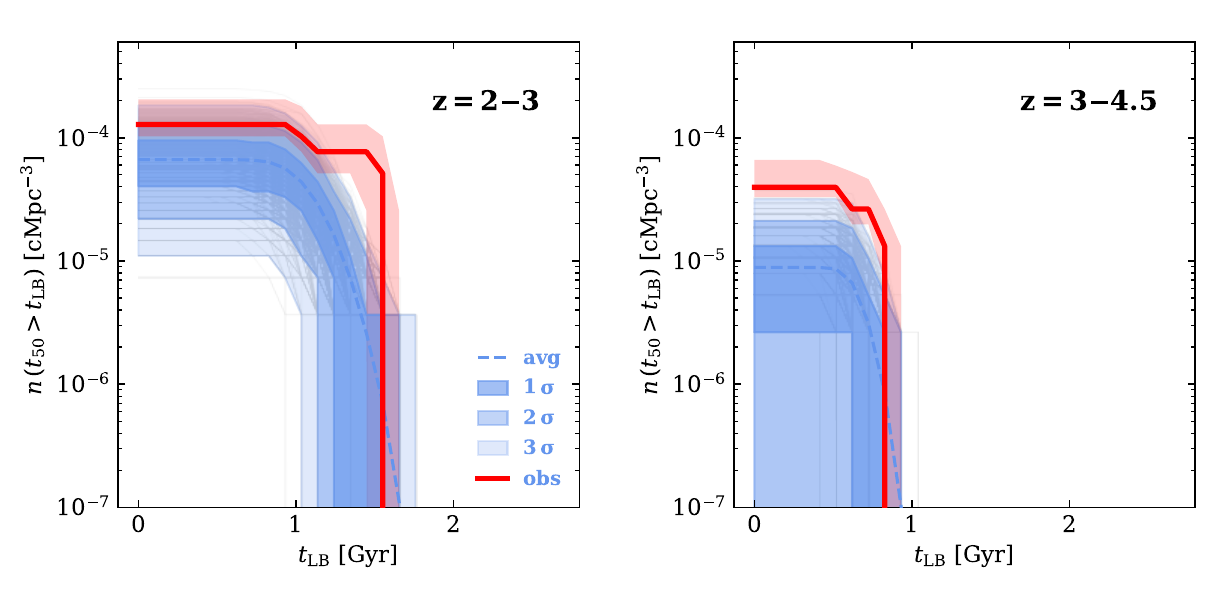}
    \caption{Comoving number densities of massive quiescent galaxies for the (spectroscopically corrected) photometric sample compared to the FLAMINGO simulation, in two redshift bins. The number density is cumulative as a function of the formation time, $t_{50}$, defined as the lookback time by when a galaxy formed 50 percent of its total formed stellar mass. The simulation results were obtained by dividing the entire volume into sub-boxes with the same volume as the observation, and then computing the number densities from each sub-box to assess the cosmic variance. It results in a total of 3,375 and 2,197 sub-boxes for the $z=2\text{--}3$ and $3\text{--}4.5$ bins, respectively. This allows us to probe the cosmic variance safely up to 3$\sigma$, as indicated by the blue bands. For the observational results, we only show the 1$\sigma$ range, which is obtained by dividing the sky area into 25 sub-areas, and thereby drawing 10,000 bootstrap samples, while also altering the age within the derived uncertainty. We see that we observe far more quiescent galaxies compared to the comparable volume in the simulation, especially in the z=3-4.5 bin.}
    \label{fig:Number densities}
\end{figure*}

One of the key questions we want to address is whether the high reported number density of massive quiescent galaxies could still be consistent with models, given a sufficiently large box, and given the higher accuracy of spectroscopic SFH measurements.
This enables us to better understand the population of these galaxies - providing a stronger comparison to theoretical modelling and galaxy formation and evolutionary physics. In order to properly measure the comoving number density, we need an accurate estimate of the survey volume.

However, due to the complicated selection function of JADES spectroscopy \citep[e.g.,][]{Bunker2024,D'eugenio2024}, the actual volume surveyed by this spectroscopic study is difficult to calculate directly. 

To avoid this issue, we used a photometrically selected sample \citep{Ji2024} obtained from products based on the NIRCam UVJ colours of galaxies in the GOODS-S JADES+FRESCO footprint \citetext{Ji, Z., priv.~comm.}. 
These were taken from a survey area of 77.1 arcmin$^2$ \citep{Eisenstein2023,Eisenstein2023b}. 

We will use the volume probed by this photometric sample to compare with the simulations, after applying completeness and purity corrections derived from spectroscopy. 
To compute the volume we simply take the survey area ($\Omega$=77.1 arcmin$^2$), then the differences between the volume of two cones corresponding to the redshifts probed. This corresponds to the equation
\begin{equation}
    V=\frac{\Omega}{3} (d_{z_2}^3-d_{z_1}^3), 
\end{equation}
where $d_{z_2}$ is the comoving distance at the upper redshift range and $d_{z_1}$ is the comoving distance at the lower redshift range.
We divided the volume into two redshift ranges, z = 2--3, which corresponds to $2.73\times 10^5\,{\rm cMpc}^3$ and z = 3--4.5, which corresponds to $3.78\times 10^5\,{\rm cMpc}^3$. One of our galaxies, ID 8777, the highest-redshift one, is beyond z = 4.5, thus outside the volume. The volume completeness of the samples, however, is not clearly known while it is likely to drop at the edge. We therefore just adopt the volume as described for the calculation of the number density, but found that the conclusion does not change when expanding the volume to the redshift of 8777 or even to z = 5. 

To compute the photometric number densities we divide the number of quiescent galaxies found within the volume by the volume itself.

The key advantage we have compared to other photometric number density approaches is that we have spectra, so we can determine false positives and false negatives for these photometric sample. Just by exploring the UVJ colours we expect to find galaxies that may be missed by UVJ colour selection alone (see Fig. \ref{fig:uvj}).

We verified that we recovered this photometric sample with our spectroscopic approach. To do this, we match our spectroscopic sample to the photometric sample. We find that there are nine targets in common within the given areas. Eight of these are selected in our spectroscopic sample (6620, 8290, 8777, 11252, 12619, 171147, 199773, and 200733), whilst one (6887) we manually removed as the spectrum was contaminated by shorts. This means that our spectroscopic sample finds 100~percent of the photometric sample for which we have spectra.

We next look through our spectroscopic sample to find any galaxies missing in the photometric sample.
This is key as we are then able to correct our number densities for quiescent galaxies missed by photometric selection alone.
We find three spectroscopically confirmed quiescent galaxies that are missing from the photometric sample (12214, 1397, and 197911). 
This means we apply a correction factor to our photometric number densities based on the number of quiescent galaxies we expect the photometric sample to have missed. We multiply our number densities by the fraction of the actual number of quiescent galaxies (photometric plus those missed) to the photometrically selected number.
For the $z = 2\text{--}3$ bin, this correction factor is 5/4. For the $z= 3\text{--}4.5$ bin, this correction factor is 6/4. For the combined bin, this corresponds to 11/8.

Our observed number densities are reported in Table \ref{table:number_densities}.
We also report an observed number density of 1.28$\rm\times10^{-5}\;[cMpc^{-3}]$ for the z=4-5 range, but caution that this is based on photometrically identifying two galaxies in this region and spectroscopically finding three. 

We compute the 1$\sigma$ uncertainty range for the observational number densities by dividing the survey area into 25 sub-areas, and thereby drawing 10,000 bootstrap samples. For each bootstrap sample, the age of each galaxy is also altered within the derived uncertainty. 

To compute the model predictions of the number densities, we divide the simulation box of $(1\,\rm cGpc)^3$ into several sub-boxes each of which has the same volume as the observation. This allows us to take into account cosmic variance, i.e. the random chance to find a volume with the observed number densities, thus helping to quantify the significance of the discrepancy with the model. The procedure results in 3,375 and 2,197 sub-boxes for the $z\,{=}\,2{-}3$ and $z\,{=}\,3{-}4.5$ volumes, respectively. This is a large enough number of sub-boxes to explore the cosmic variance up to about $3.5\,\sigma$. 

Fig.~\ref{fig:Number densities} shows the number density of massive quiescent galaxies in these observed volumes compared to the number found in the FLAMINGO simulation. The cumulative number density is presented as a function of $t_{50}$, which is defined as the lookback time by when a galaxy formed 50 percent of its total formed stellar mass.
This means that $t_{LB}$=0 corresponds to our observed number densities, whilst increasing $t_{LB}$, decreases the number of quiescent galaxies observed with that formation time. In both redshift ranges, the simulation predictions disagree with the observed number densities at about or greater than 3$\sigma$. At higher redshifts $z=3\text{--}4.5$, our number densities are an order of magnitude higher than the average predicted by FLAMINGO. 
Although comparisons with FLAMINGO have not been investigated before, a discrepancy with predictions from theory has already been reported by other photometric studies \citep[e.g.,][]{Carnall2023abundances,Valentino2023}, and is also seen in other numerical simulations (see following Sec. \ref{s.results.ss.comp_num_den} and Fig. \ref{fig:num_den_comp}). However, using the large volume of FLAMINGO, it is clear that cosmic variance, by itself, is unlikely to explain the observations, with the number density at $z=3\text{--}4.5$ close to a 4-\sig fluctuation.
The use of spectroscopy rules out significant contamination of our sample as a possible explanation. The use of prism spectroscopy enables us to measure much more precise stellar ages compared to photometry alone \citep[e.g.,][]{Nanayakkara2024}. 

\subsection{Comparison of number densities}
\label{s.results.ss.comp_num_den}
\begin{figure}
    \centering
    \includegraphics[width=1\linewidth]{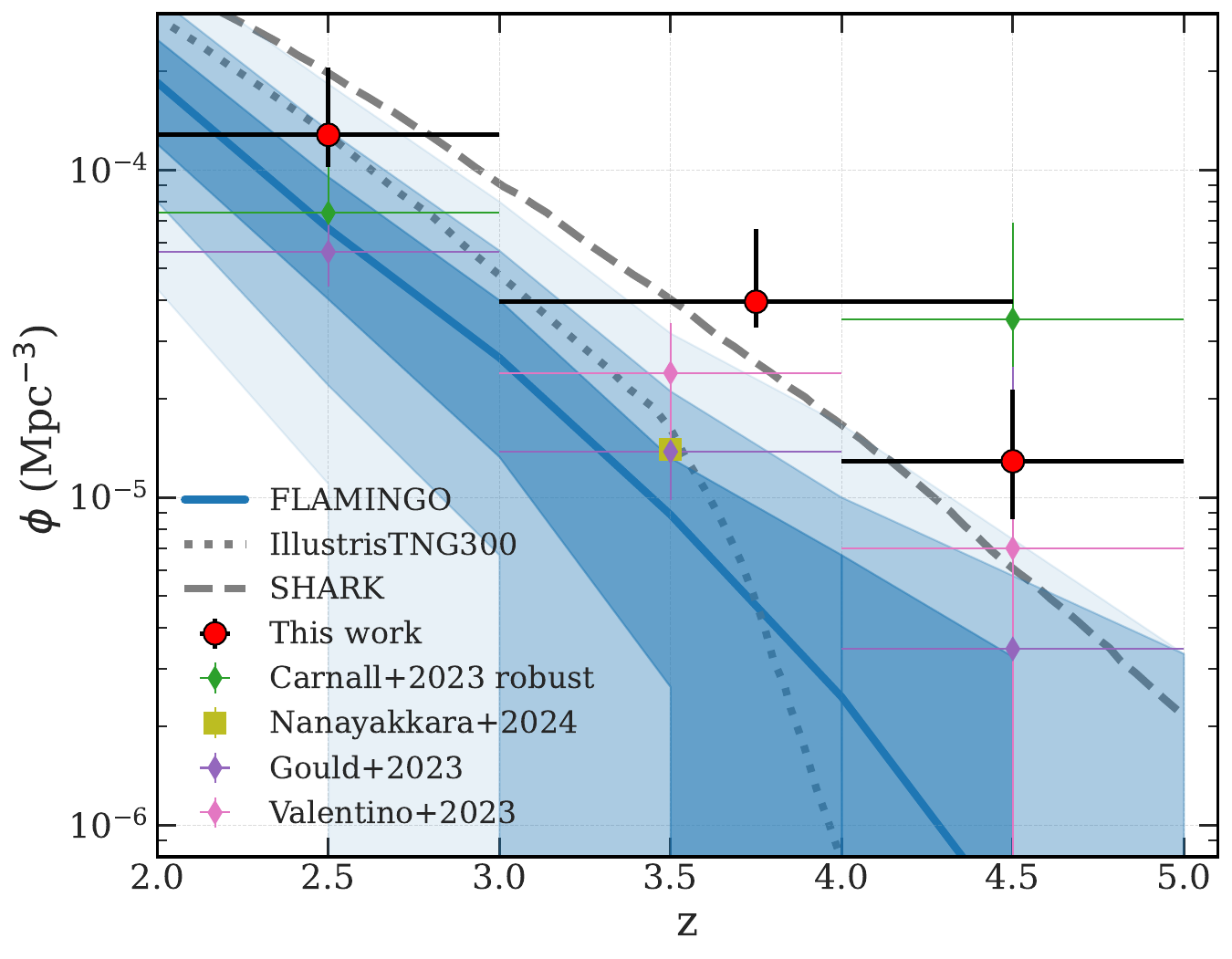}
    \caption{Comoving number densities of massive quiescent galaxies based on observational surveys, cosmological simulations and semi-analytic models. The red points show the number densities computed in this work with our spectroscopically corrected photometric sample. The solid blue line corresponds to number densities from the FLAMINGO simulations with the colour fill corresponding to the 1\sig, 2\sig and 3\sig cosmic variances for a volume of $3\times 10^5\,{\rm cMpc}$, roughly an average volume for our data. Other points correspond to photometric surveys including \citet{Carnall2023abundances}, \citet{Gould2023}, \citet{Valentino2023}, and the single (spectroscopically based) point from \citet{Nanayakkara2024}. The dotted grey line corresponds to the IllustrisTNG300 cosmological simulation and the dashed grey line corresponds to prediction from the semi-analytic model \textsc{SHARK} both of which are obtained from \citet{Lagos2025}.}
    \label{fig:num_den_comp}
\end{figure}

It is worth considering how our computed comoving number densities from the observations and the FLAMINGO simulations compare to those in the literature. Extensive comparison to large scale cosmological simulations \citep[e.g. \textsc{EAGLE}, \textsc{Illustris-TNG}, \textsc{SIMBA}][]{Schaye2015, Crain2015, Pillepich2018, Springel2018, Dave2019} and semi-analytic models \citep[SAMs, e.g. \textsc{SHARK}, \textsc{GAEA}, \textsc{GALFORM}][]{Lagos2018, Lagos2024shark, DeLucia2024, Lacey2016}, have already been analysed in-depth in \citet{Lagos2025}. However, in that case they were limited on the observational side by the lack of spectroscopic correction \citep[the single point from][not withstanding]{Nanayakkara2025} to their photometric samples so were unable to adjust for missed quiescent galaxies or interlopers. They were also limited on the simulations side by the box-size, therefore, being unable to probe or rule out cosmic variance within the simulations themselves. With the addition of our 18 strong spectroscopic sample being able to correct our photometric sample, alongside the large scale 1cGpc$^3$ FLAMINGO box, we have the ability to probe these two effects.

Fig. \ref{fig:num_den_comp} shows comoving number densities versus redshifts for massive quiescent galaxies based on observational surveys, cosmological simulations and semi-analytic models. The red points show the number densities computed in this work with our spectroscopically corrected photometric sample. This includes the z=2--3 bin, z=3--4.5 bin and the z=4--5 bin (see Sec. \ref{s.results.ss.numberdens} for more details on this computation.) 
The solid blue line corresponds to the number densities from FLAMINGO with the shaded regions corresponding to 1\sig, 2\sig and 3\sig errors for a volume of $3\times 10^5\,{\rm cMpc}^3$ (an average volume for our data), again testing cosmic variance. This is similar to Fig. \ref{fig:Number densities}, but the focus here is on the redshift evolution not the stellar ages.

We can again straight away see the discrepancy between our observed number densities and the simulated number densities. There is a 2\sig difference for the z=2--3 bin, but greater than 3\sig for the higher redshift bins. This is highlighting that we likely need both earlier and stronger feedback within the simulations. We also see that for the highest redshift bins, cosmic variance within the simulations cannot account for this discrepancy.

We can also compare these number densities to results from the literature.
The other points in Fig. \ref{fig:num_den_comp} correspond to other photometric surveys including \citet{Carnall2023abundances}, \citet{Gould2023}, \citet{Valentino2023}, and the single point based on spectroscopy from \citet{Nanayakkara2024}. The dotted grey line corresponds to the Illustris-TNG300 cosmological simulation \citep{Pillepich2018, Springel2018} and the dashed grey line corresponds to the prediction from the semi-analytic model \textsc{SHARK} \citep{Lagos2018, Lagos2024shark}, both of which are obtained from \citet{Lagos2025}.

It is immediately apparent from the figure that we see lots of variations between the different measures. Some of these are easily explainable due to simple selection effects. In our work we are selecting massive quiescent galaxies as those with $M_\star>10^{10}M_\odot$. \citet{Gould2023} for example select massive quiescent galaxies with $M_\star>10^{10.6}M_\odot$ and in \citet{Valentino2023} they select two samples, one with $M_\star>10^{10.6}M_\odot$ and the other lower mass bin with $9.5<M_\star<10^{10.6}M_\odot$. In Fig. \ref{fig:num_den_comp} we plot the $M_\star>10^{10.6}M_\odot$ reported values as this most closely replicates our sample (our lowest mass spectroscopically confirmed massive central galaxy in GOODSS has $M_\star=10^{10.4}M_\odot$). Despite this, it is natural to assume the numbers in \citet{Gould2023} and \citet{Valentino2023} would be increased by switching to a $M_\star>10^{10}M_\odot$ selection. If we combine the number densities for both of the bins from \citet{Valentino2023} it becomes more consistent with our estimate from this work. Even with just the $M_\star>10^{10.6}M_\odot$ bin we are within around 1\sig. This is important as \citet{Valentino2023} have a larger area (145.1 arcmin$^2$ compared to our 77.1 arcmin$^2$ ) due to combining photometry from multiple different surveys. 
This is an important check that we are not being too strongly affected by cosmic variance within our observations and remain consistent with number densities calculated in larger volumes. 

There appear to be larger tensions with the number densities computed in \citet{Carnall2023abundances}, particularly in the highest redshift bin of z=4--5. However, as was noted in \citet{Valentino2023}, the CEERS field appears to have an overdensity of massive quiescent galaxies compared to other comparable fields, showing the role cosmic variance can play and this is likely to account for the excess seen in \citet{Carnall2023abundances}.

For the simulations and SAMs, of upmost importance are the various feedback prescriptions for galaxy quenching \citep[e.g.][]{Vogelsberger2020,Lagos2025}. The exact implementation and methodology for this feedback directly relates to the number of massive quiescent galaxies found at high-z. In other words, the observed number densities of massive high-z quiescent galaxies are a key test of the strength and timescales of feedback prescriptions within cosmological simulations and SAMs. \citet{Lagos2025} found that just among simulations and SAMs there are significant differences for the predicted number densities of high-z relating to these feedback implementations, with some predicting better at certain mass scales and redshift ranges. 

In Fig. \ref{fig:num_den_comp} we overplot number densities computed in \citet{Lagos2025} for the \textsc{Illustris-TNG300} \citep{Pillepich2018, Springel2018} cosmological simulations and the \textsc{SHARK} \citep{Lagos2018,Lagos2024shark} SAM. We use the number densities corresponding to quiescent galaxies with $M_\star>10^{10}M_\odot$. We also use the most generous sSFR threshold of sSFR${\lesssim}\rm10^{-10}yr^{-1}$, although note that this makes little difference. 

One of the first aspects that becomes apparent is that \tng dramatically underestimates the number of quiescent galaxies above z$\sim$3.5, with zero quiescent galaxies above z=4. It was noted in \citet{Lagos2025} that the actual mass threshold for selecting massive quiescent galaxies in \tng was not mass dependent as all were very massive. Whereas \shark is better able to match the high-z massive galaxy number densities, but \citep[as in ][]{Lagos2025} we caution that this is because of the number of massive galaxies with $10^{10}\Msun<M_\star<10^{10.5}\Msun$, whereas all our spectroscopically observed galaxies with z>4 have $M_\star\gtrsim10^{10.5}\Msun$. If we take the \shark number densities for $M_\star\gtrsim10^{10.6}\Msun$ we are likely to see significant offsets to the observed number densities in this work and other studies.

FLAMINGO under predicts at all redshifts with the largest deviation at z>3, but it doesn't decline beyond z=3.5 as rapidly as \tng.

Overall, this motivates the needs for larger observational surveys with spectroscopically confirmed high-z quiescent galaxies, in addition to cosmological simulations with both stronger and faster acting feedback prescriptions.

\newcommand{\sfhwidth}{0.95\columnwidth}

\begin{figure*}
    \centering  \includegraphics[width=\sfhwidth]{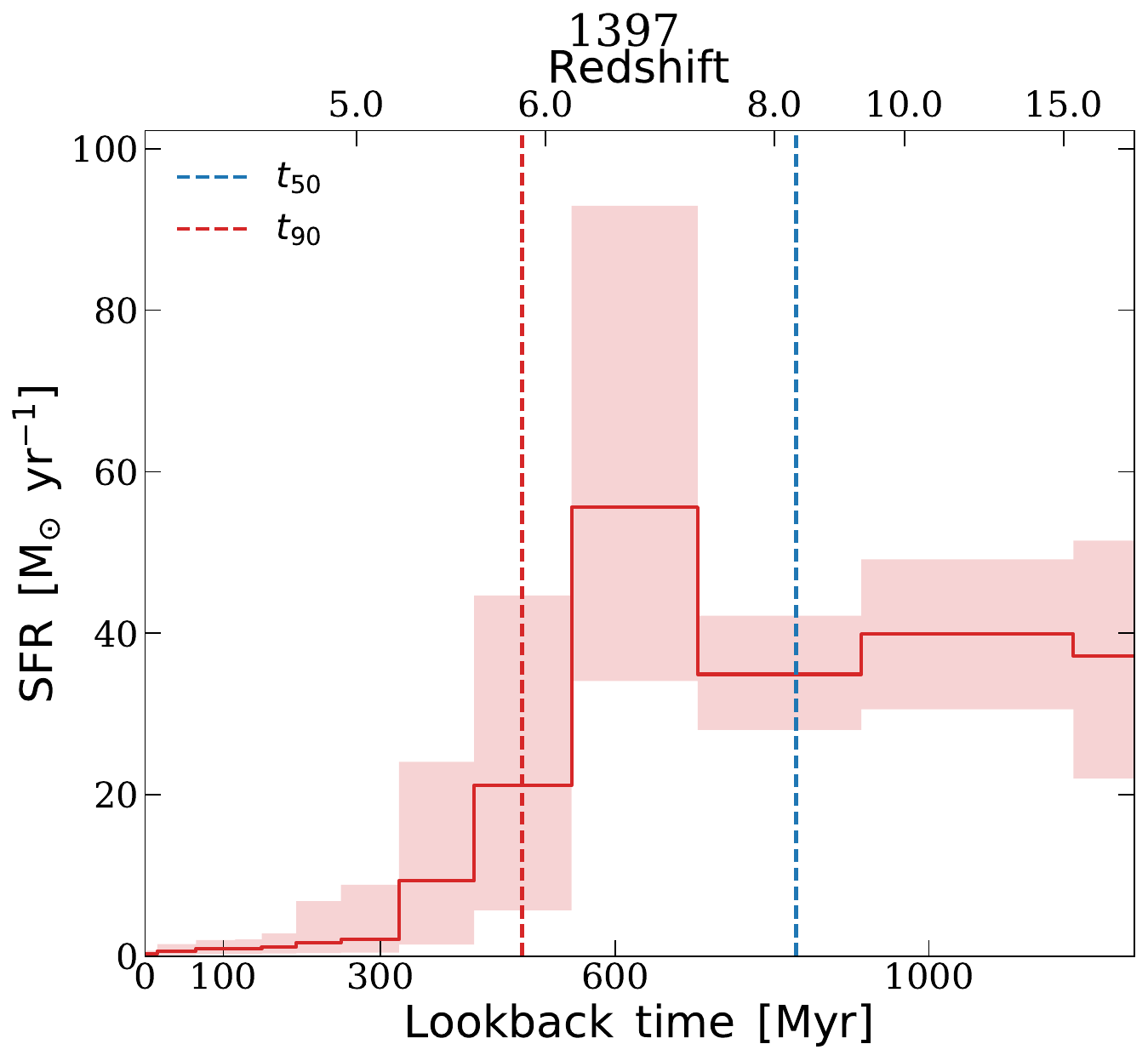}\includegraphics[width=\sfhwidth]{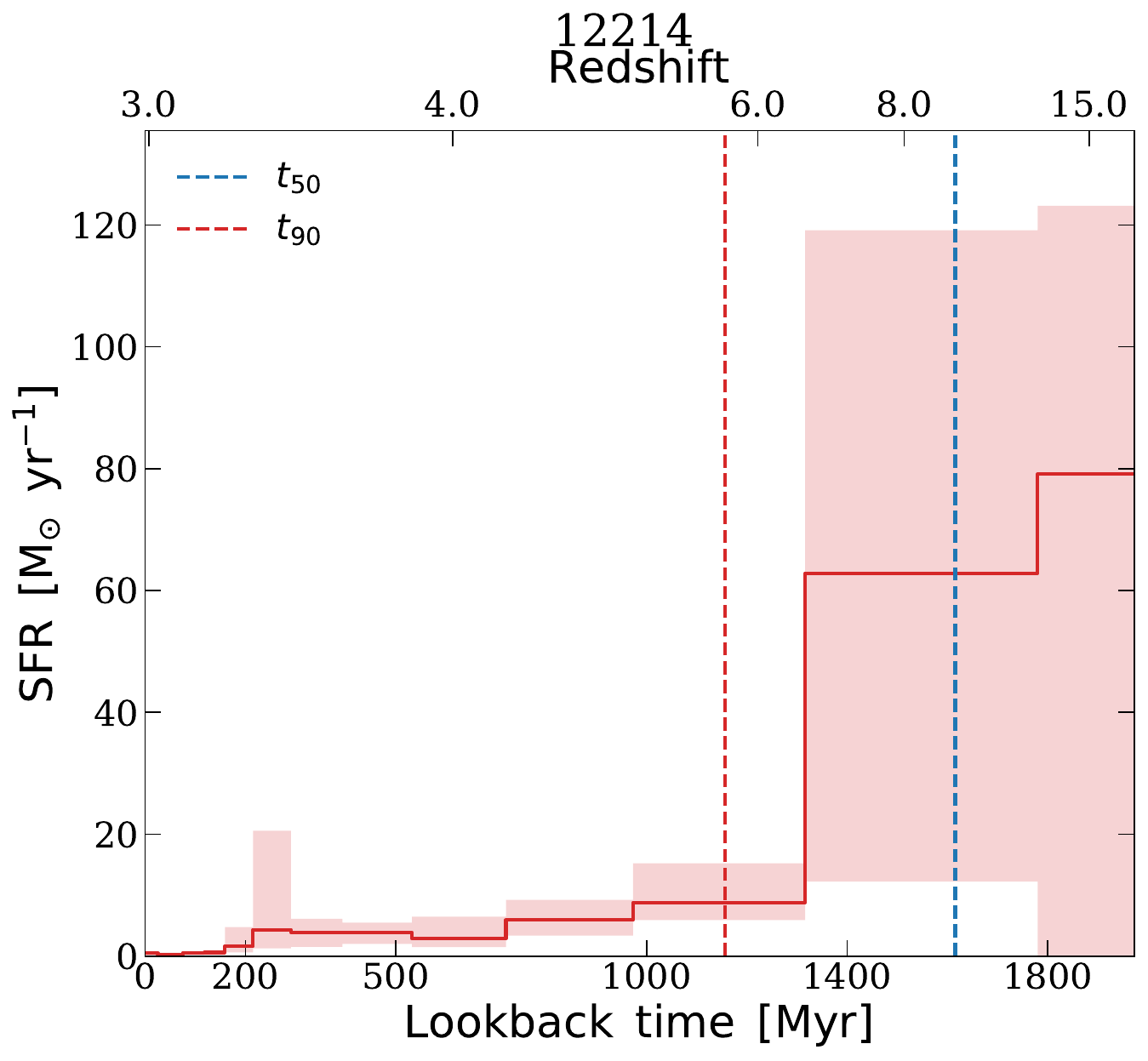}
    \includegraphics[width=\sfhwidth]{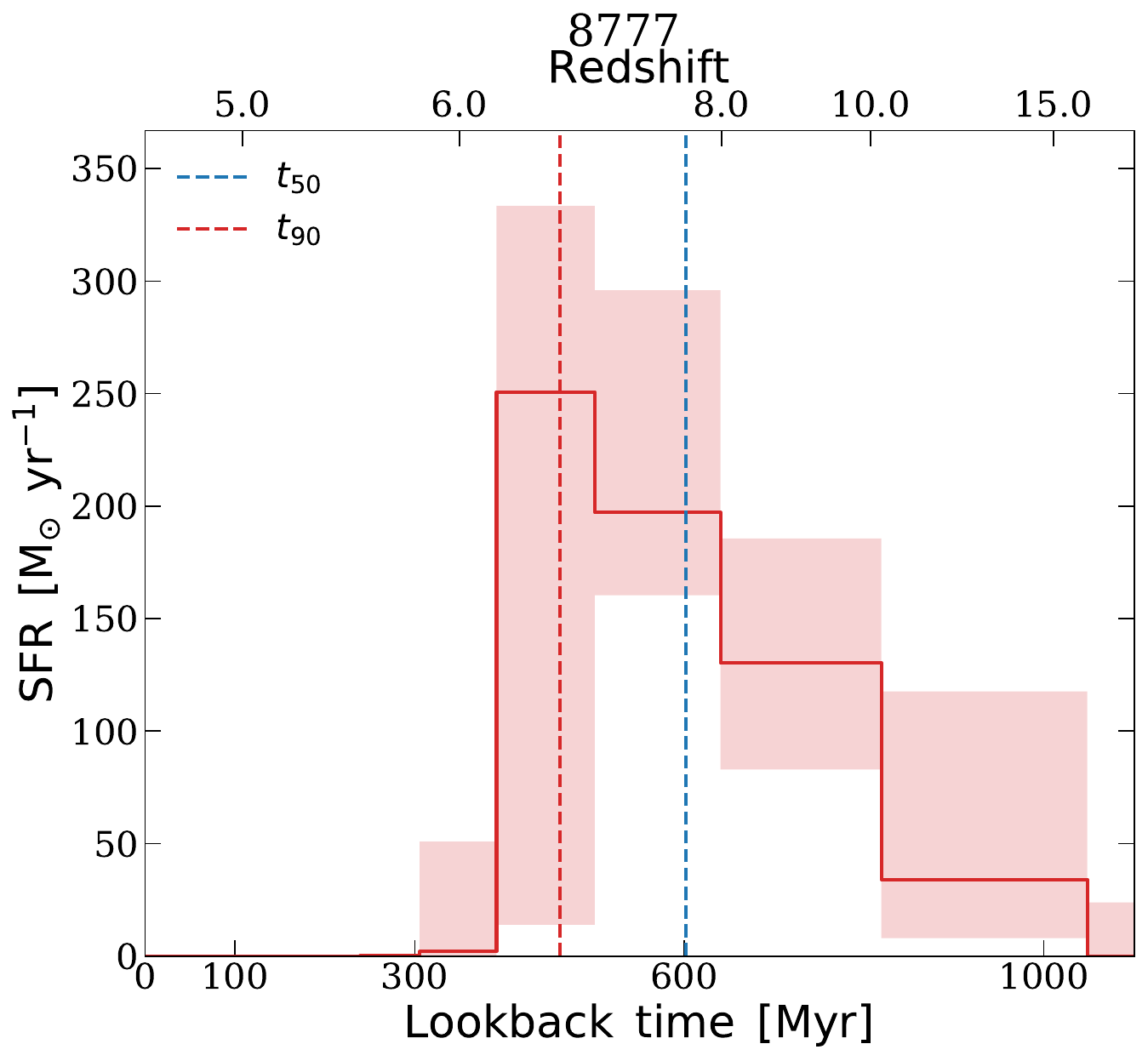}
    \includegraphics[width=\sfhwidth]{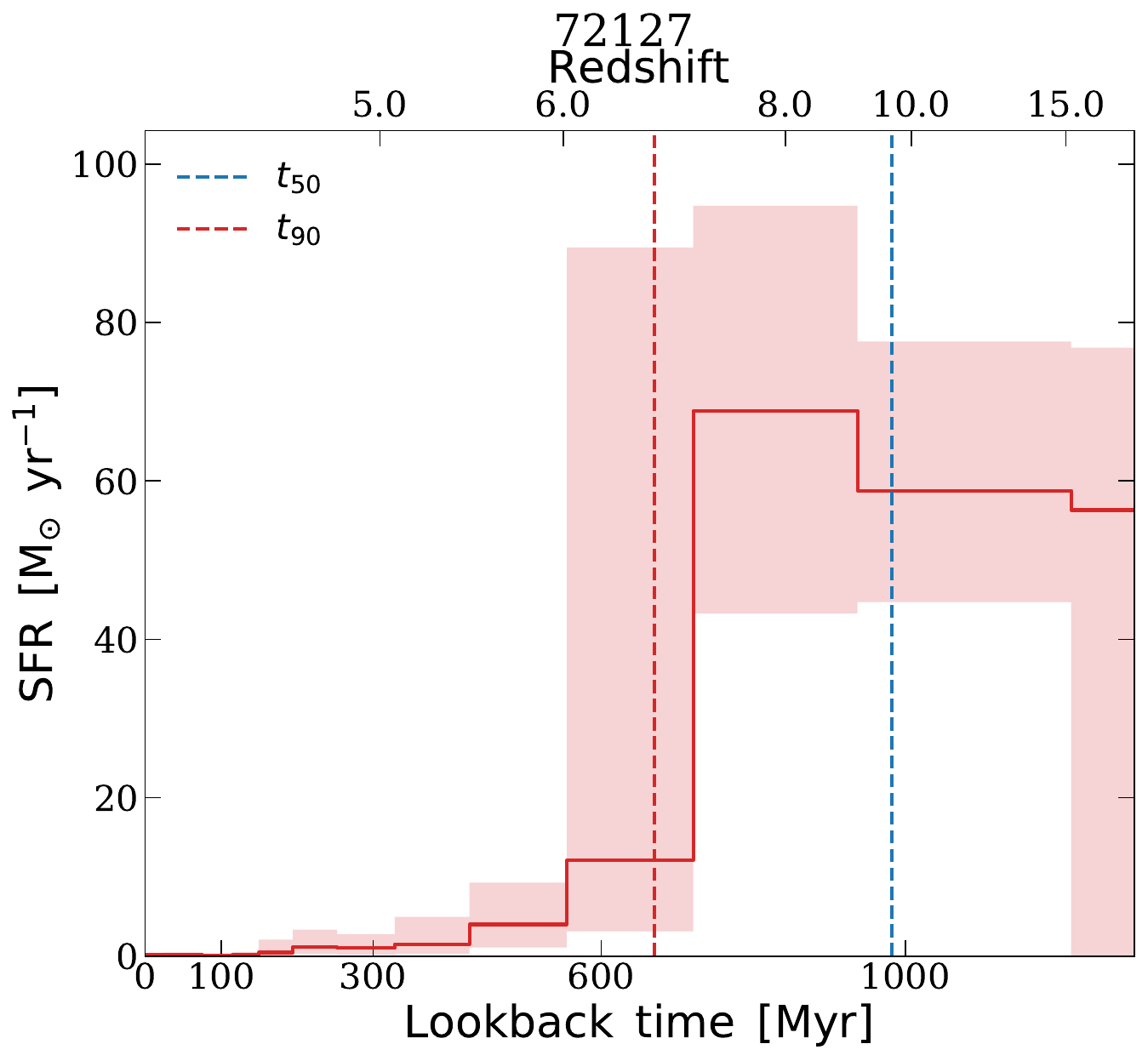}
    \includegraphics[width=\sfhwidth]{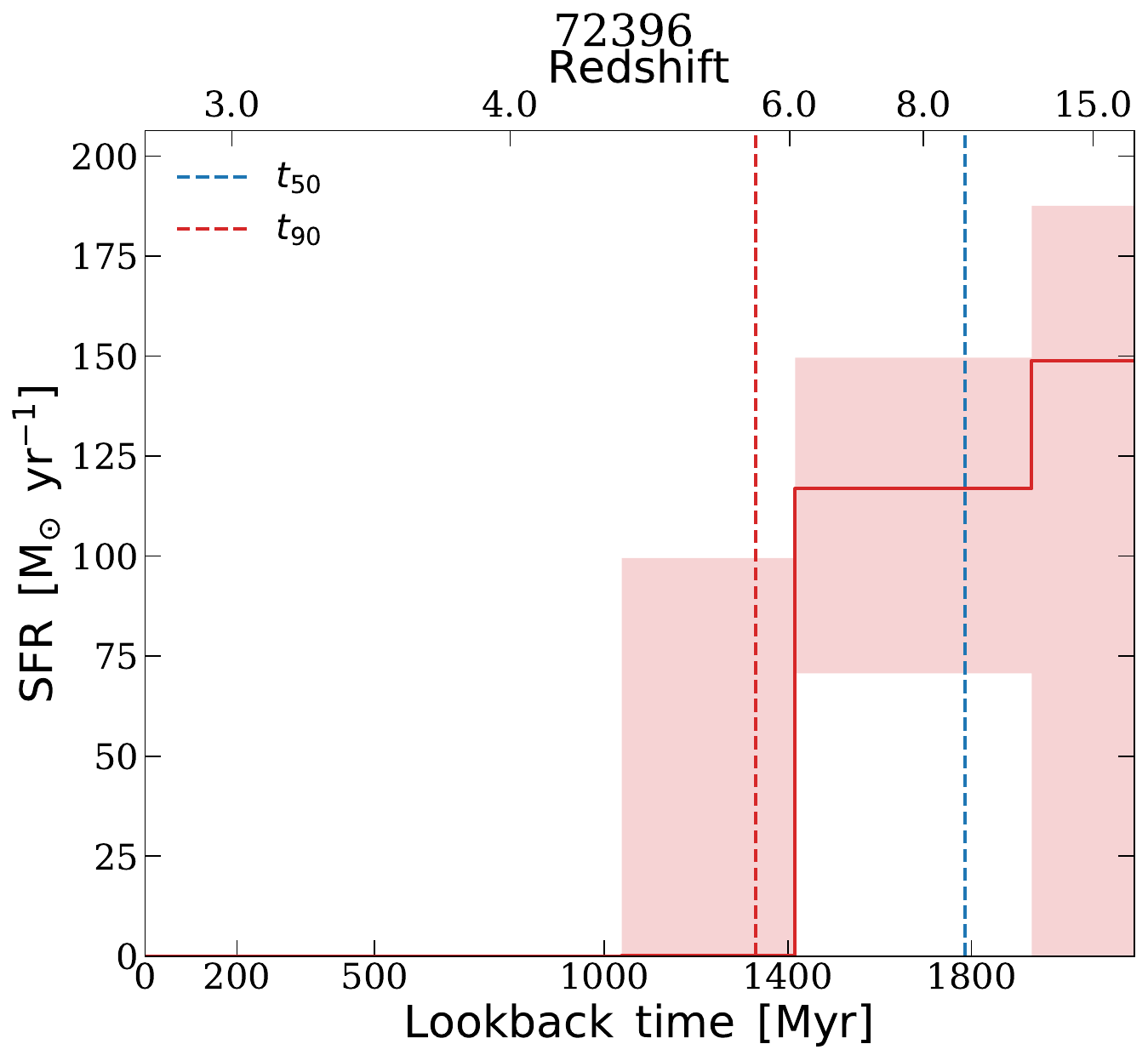}  \includegraphics[width=\sfhwidth]{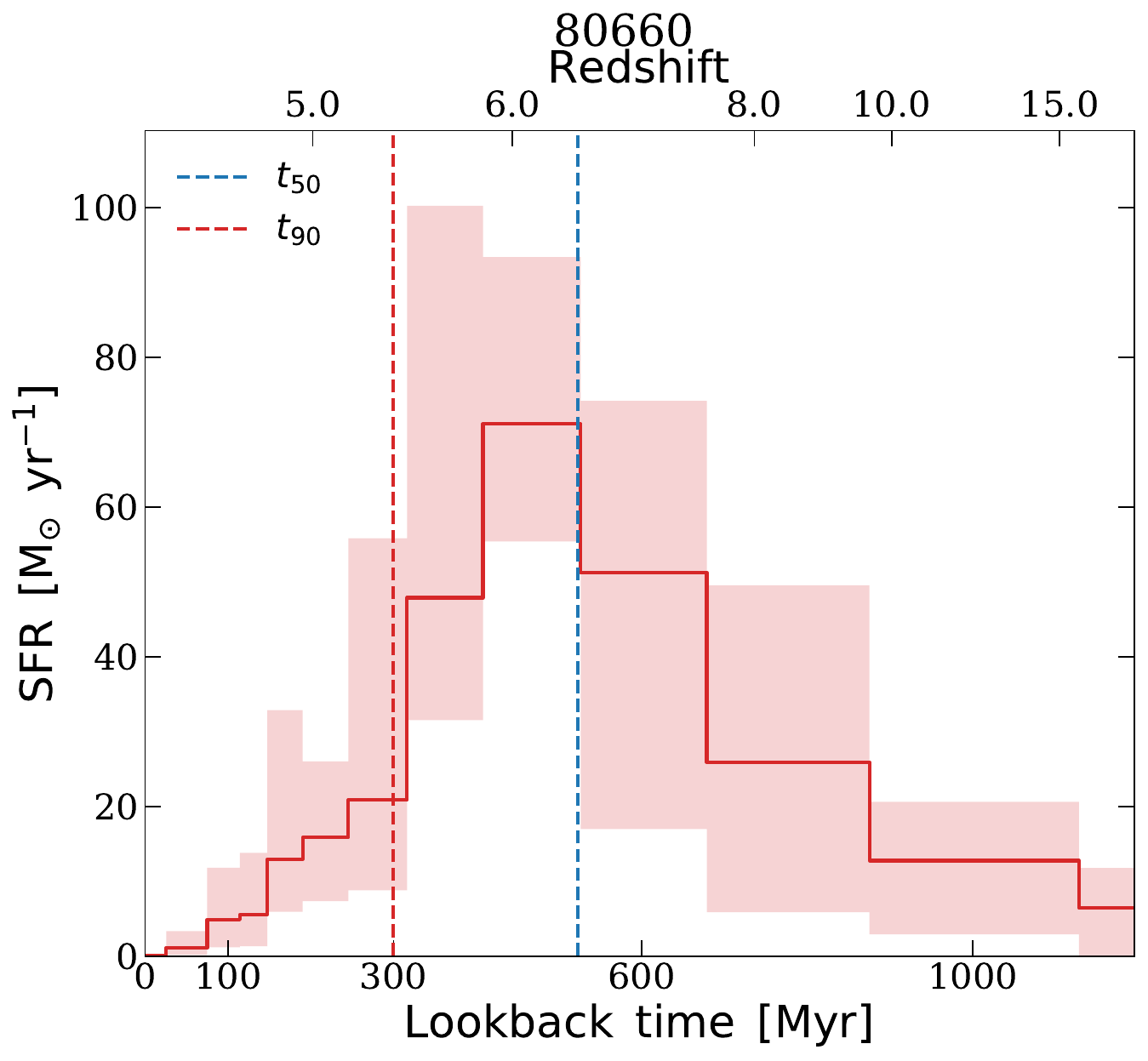}  
    
    \caption{Star-formation rate vs lookback time from observation (redshift) for 6 quiescent galaxies. The red line corresponds to the median of the bootstrapped distribution. The shaded region corresponds to the 16th and 84th percentiles. The blue-dashed line corresponds to $t_{50}$ (the formation time), and the red dashed line corresponds to $t_{90}$ (the quenching time). }
    \label{fig:sfhs_main}
\end{figure*}

\begin{figure*}
    \centering
    \includegraphics[width=2.\columnwidth]{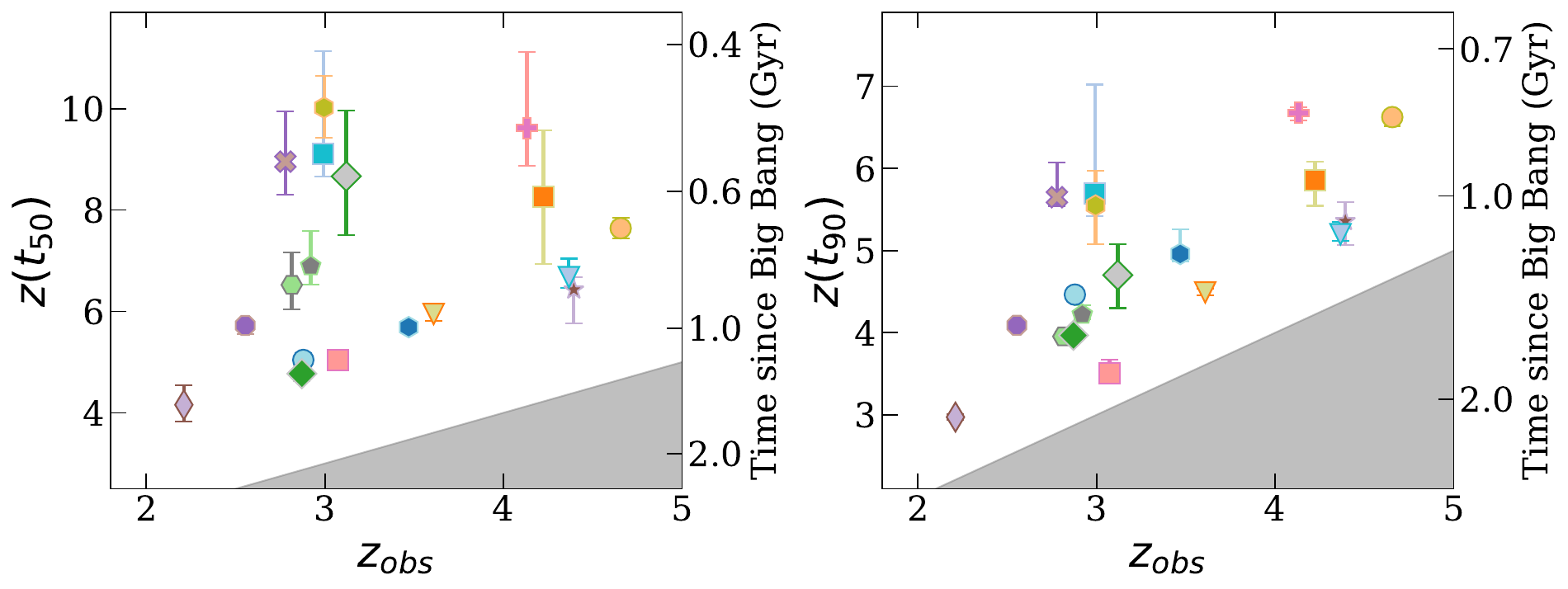}  
    \caption{Formation (z($t_{50})$, left panel) and quenching (z($t_{90}$, right panel) redshifts versus observed redshift for the 18 massive quiescent galaxies. The grey shaded region is the forbidden region based on a 1 to 1 line (i.e. a galaxy cannot form or quench after the redshift at which it has been observed). We find a range of formation and quenching times with several galaxies having formation times above redshift 8 and the majority having formation times above redshift 6. We find the majority of the galaxies appear to have quenched above redshift 4 with a few having quenched above redshift 6.}
    \label{fig:formation_and_quenching}
\end{figure*}

\subsection{SFHs}

Fig.~\ref{fig:sfhs_main} shows a selection of the individual star-formation histories of this sample of massive quiescent galaxies as inferred from our spectra. The remaining SFHs are shown in the Appendix, Figs. \ref{fig:sfhs2} and \ref{fig:sfhs3}.
Technically, these are a composite of the possible SFHs inferred from \texttt{Prospector}, where the red line corresponds to the 50th percentile of the SFR distribution in each time bin, while the shaded region refers to the 16th and 84th percentiles.\footnote{It is understood that the 16th and 84th percentiles do not trace plausible SFHs; for instance, the lowest shading, corresponding to the 16th percentile, is not an actual SFH possibility, rather it is the minimum SFR at a particular time from a combination of different SFHs. This is clear from the fact that a SFH that always followed the 16th percentile in each time bin would form significantly fewer stars than the median, ending up with a total stellar mass significantly below the 16th percentile of the posterior probability of $M_\star$.} The formation and quenching times ($t_{50}$ and $t_{90}$) are indicated by the dashed blue and red lines, respectively. These times are defined by when the galaxy formed 50 and 90~percent of its total formed stellar mass (i.e. not accounting for mass loss due to stellar evolution). 
None of the SFHs show signs of recent activity, but this is imposed by the specific star-formation rate (sSFR) criteria. Some galaxies appear to have formed their stars in a single burst (e.g. 72396) whilst others appear to have a more extended SFH (e.g. 1397). The galaxies with the strong 4000-$\AA$ breaks show the oldest SFHs such as 11252, 12214, and 72396.

More broadly we can see that our SFHs split into two main categories - those that are more extended, with residual star formation over a longer period of time, and those that appear to be consistent with a single burst. This can be clearly seen in the differences between the formation time (dashed blue line) and the quenching time (dashed red line). Galaxies with more extended SFHs have a much greater difference between formation and quenching times, showing that they continued to build up their mass over a long epoch. Whereas for galaxies with a single burst of star formation, they have very similar formation and quenching times; all star-formation activity appears to have occurred within that short period of time  (i.e. $\leq$100Myrs), with the galaxy forming hundreds of solar masses of stars per year. 

None of the galaxies appear to show statistically significant evidence of rejuvenation events, although we remove by selection any substantial rejuvenation event in the last 100~Myr. Galaxies undergoing recent rejuvenation can easily wash out the Balmer break and therefore appear non-quiescent \citep{Witten2024}.  
In spite of this, we do see some galaxies with signs of double peaks in their SFHs, possibly suggesting short periods of reduced activity before the star formation reignites again. However, this is not statistically significant.

\subsection{Formation and quenching times}\label{s.results.ss.downsizing}

\begin{figure*}
    \centering
    \includegraphics[width=1.\columnwidth]{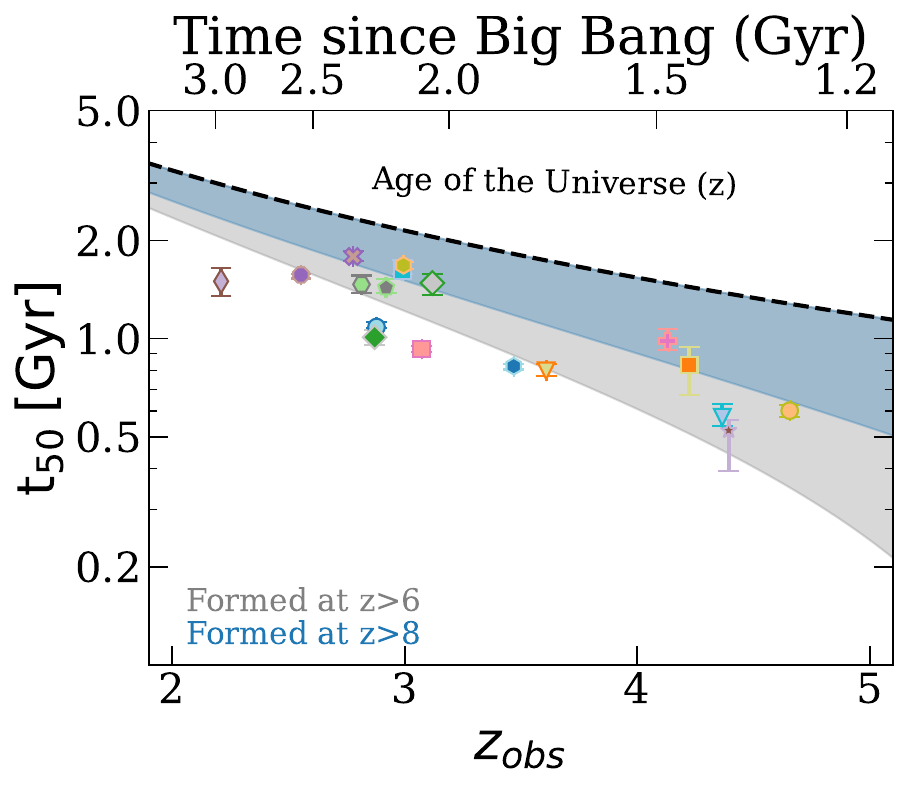}  \includegraphics[width=1.\columnwidth]{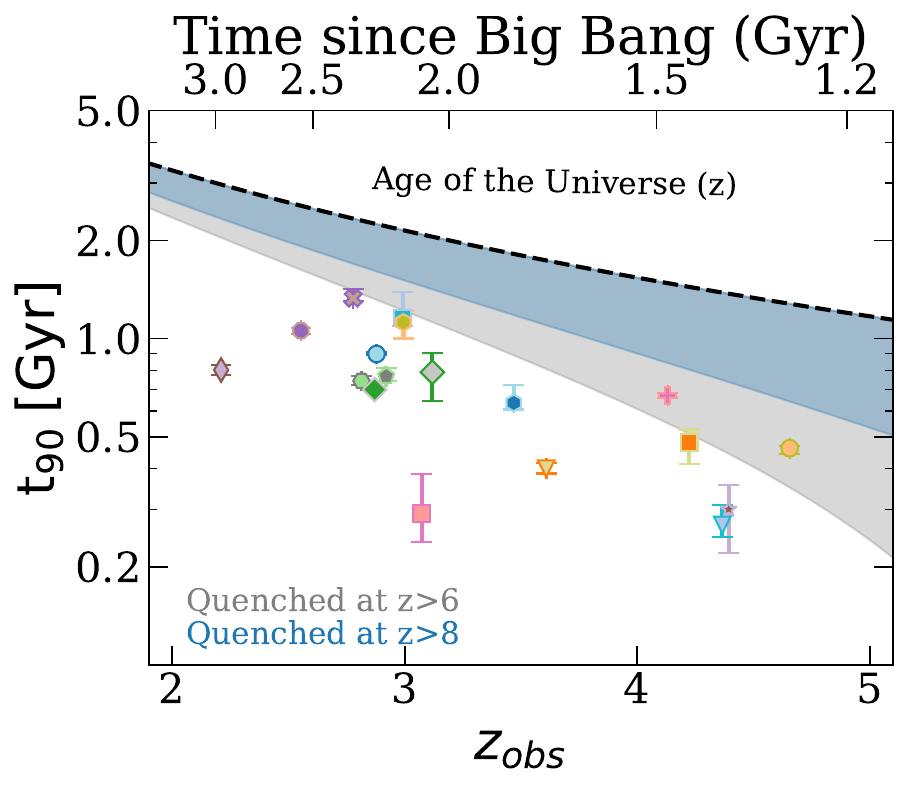}
    \caption{Formation ($t_{50}$, left panel) and quenching times ($t_{90}$, right panel) versus observed redshift for the 18 massive quiescent galaxies. The black dashed line is the age of the universe and the grey and blue shaded regions correspond to galaxies forming or quenching at $z\geq6$ and $z\geq8$, respectively. We find that at least four galaxies in our sample appear to have formation times above redshift 8 which then likely quenched at redshifts 5-7.}
    \label{fig:formation_and_quenching_2}
\end{figure*}

Fig.~\ref{fig:formation_and_quenching} shows the formation redshift ($z(t_{50})$) and the quenching redshift ($z(t_{90})$) vs the observed redshift. The dashed line corresponds to the one-to-one redshift line. 
We see from the figure that we have a range of formation and quenching redshifts, with a significant fraction of the quiescent galaxies forming earlier than redshift 6. This means that these galaxies formed
during some of the earliest stages of the Universe's history, i.e. within the first billion years. 
We also find several galaxies with quenching times above redshifts of 6 suggesting that they have been quiescent for a substantial period of time. This also indicates that they must have built up their mass early on, as they are observed as massive quiescent galaxies. This is consistent with their spectral features showing evidence of 4000-$\AA$ breaks (Section~\ref{sec:4000break}).
This is consistent with previous findings of early formation and quenching times for individual or small samples of galaxies at high-z \citep[e.g.][]{Carnall2024, Glazebrook2024, deGraaff2024} as well as the high-z quiescent galaxy found in \citet{Weibel2024b}.

However, since our samples are spread across a wide range of $z_{\rm obs}$, galaxies with similar quenching redshifts could have been quiescent for various amount of times. Therefore, it can be more appropriate to directly investigate $t_{90}$ or $t_{50}$, rather than the corresponding redshifts, to discuss their ages. This is shown in Fig.~\ref{fig:formation_and_quenching_2}. It nicely demonstrates that a majority of those with earlier quenching redshift have been quiescent for a longer period, thus older on average. It also enables us to better visualise the limits on formation and quenching times set by the age of the Universe (as indicated by the black dashed line).

\subsection{High-z 4000-\texorpdfstring{\AA}{A} Breaks}\label{sec:4000break}

\begin{figure*}
    \includegraphics[width=1\columnwidth]{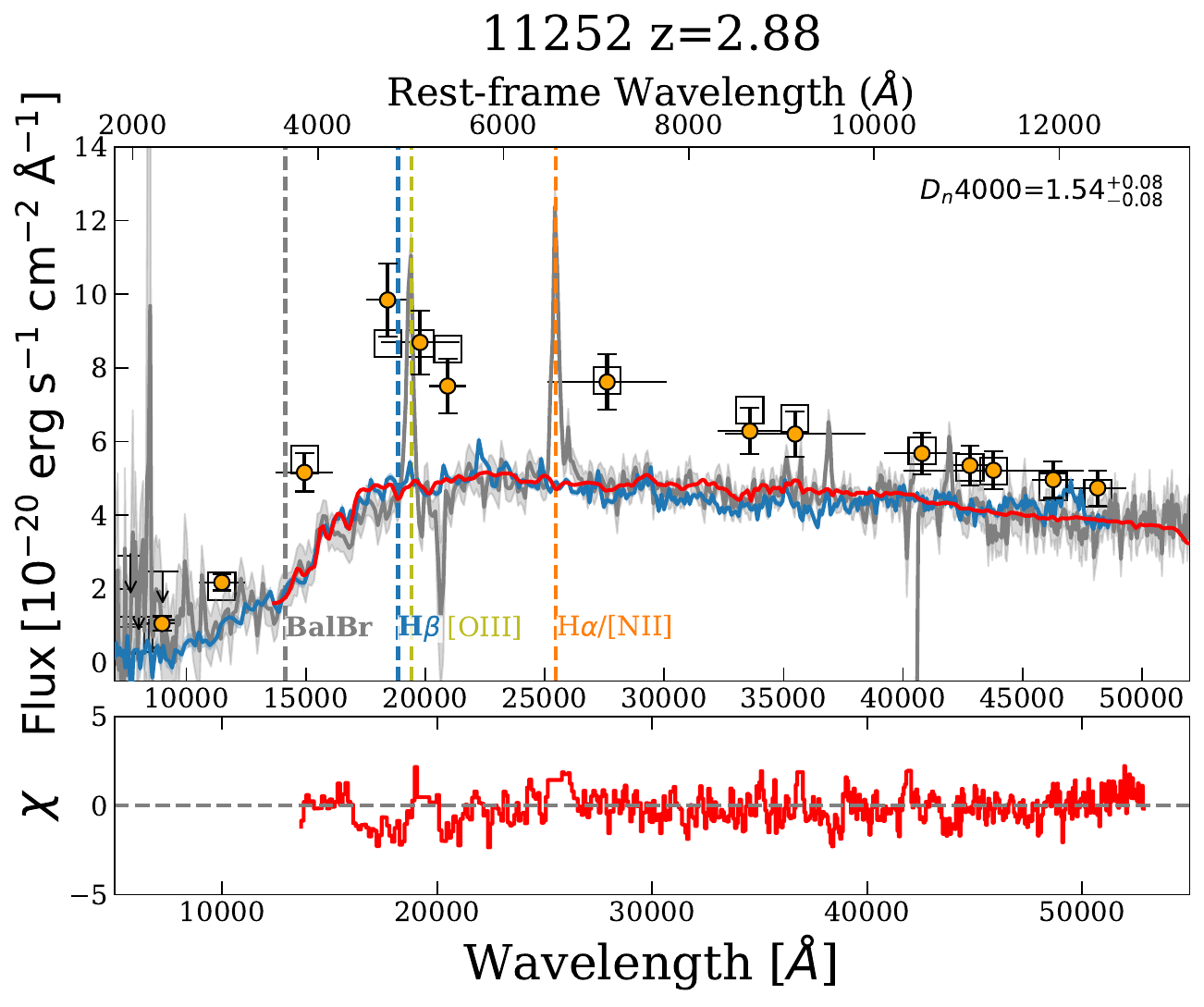}
    \includegraphics[width=1\columnwidth]{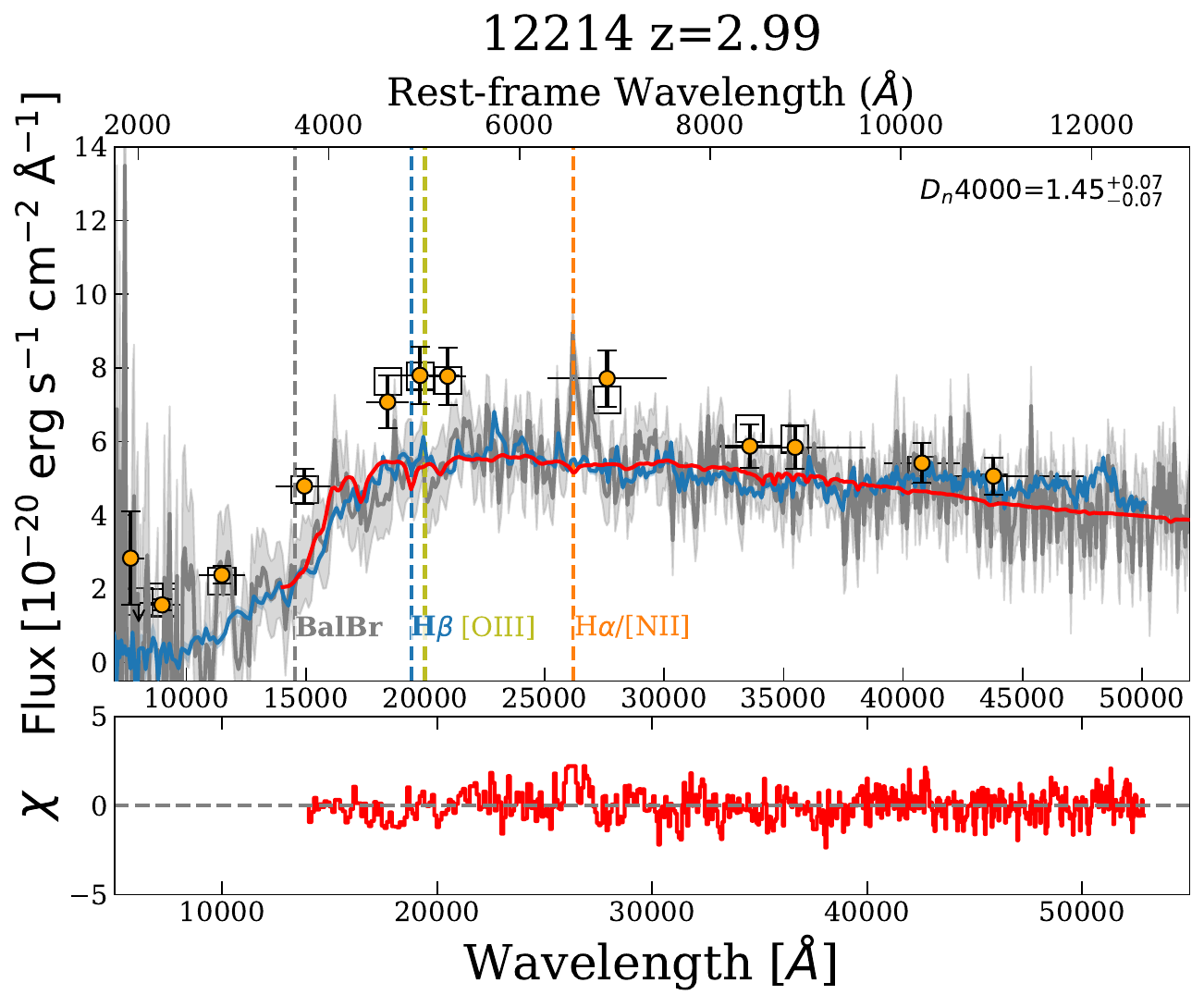}
    \includegraphics[width=1\columnwidth]{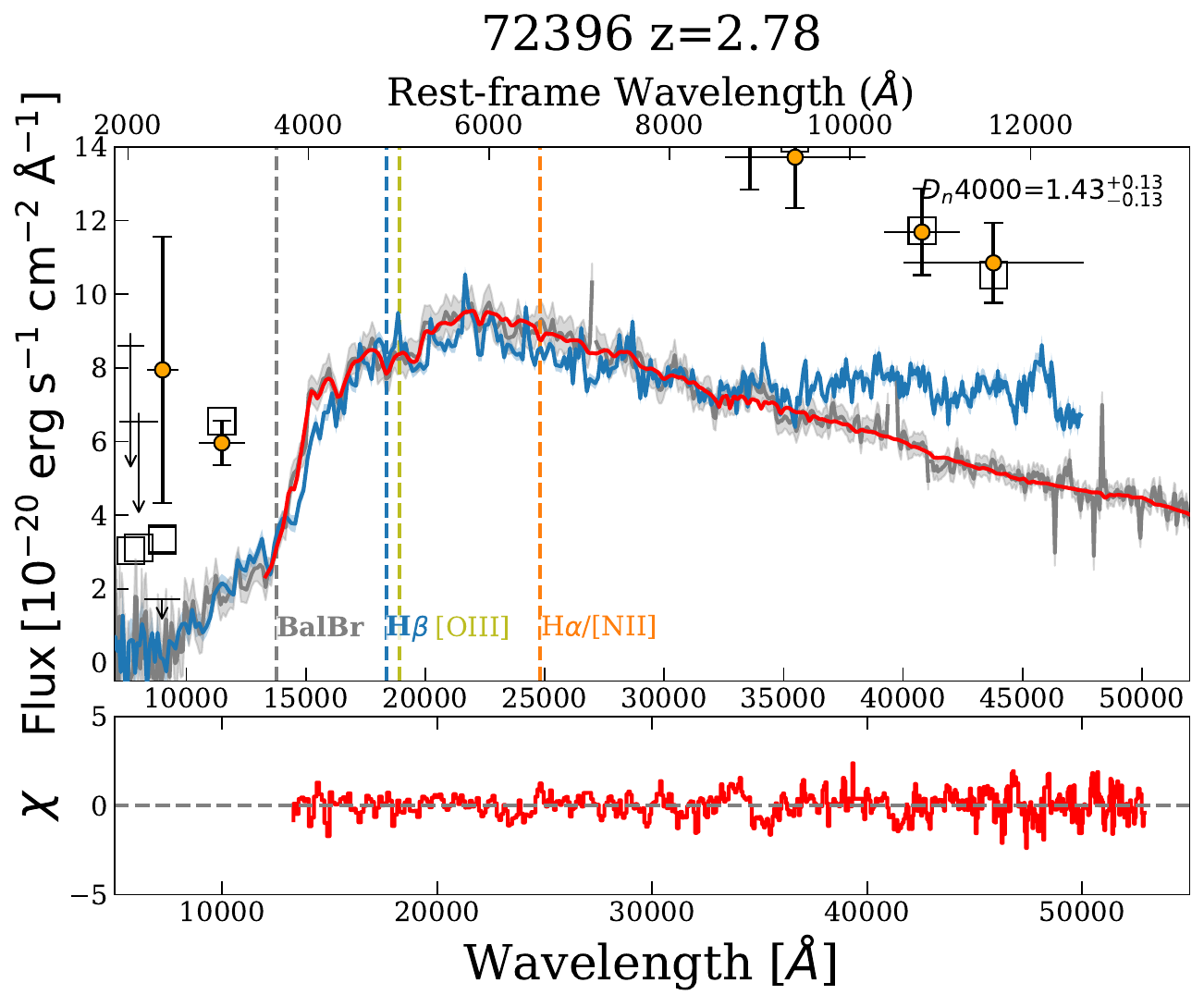}
    \includegraphics[width=1\columnwidth]{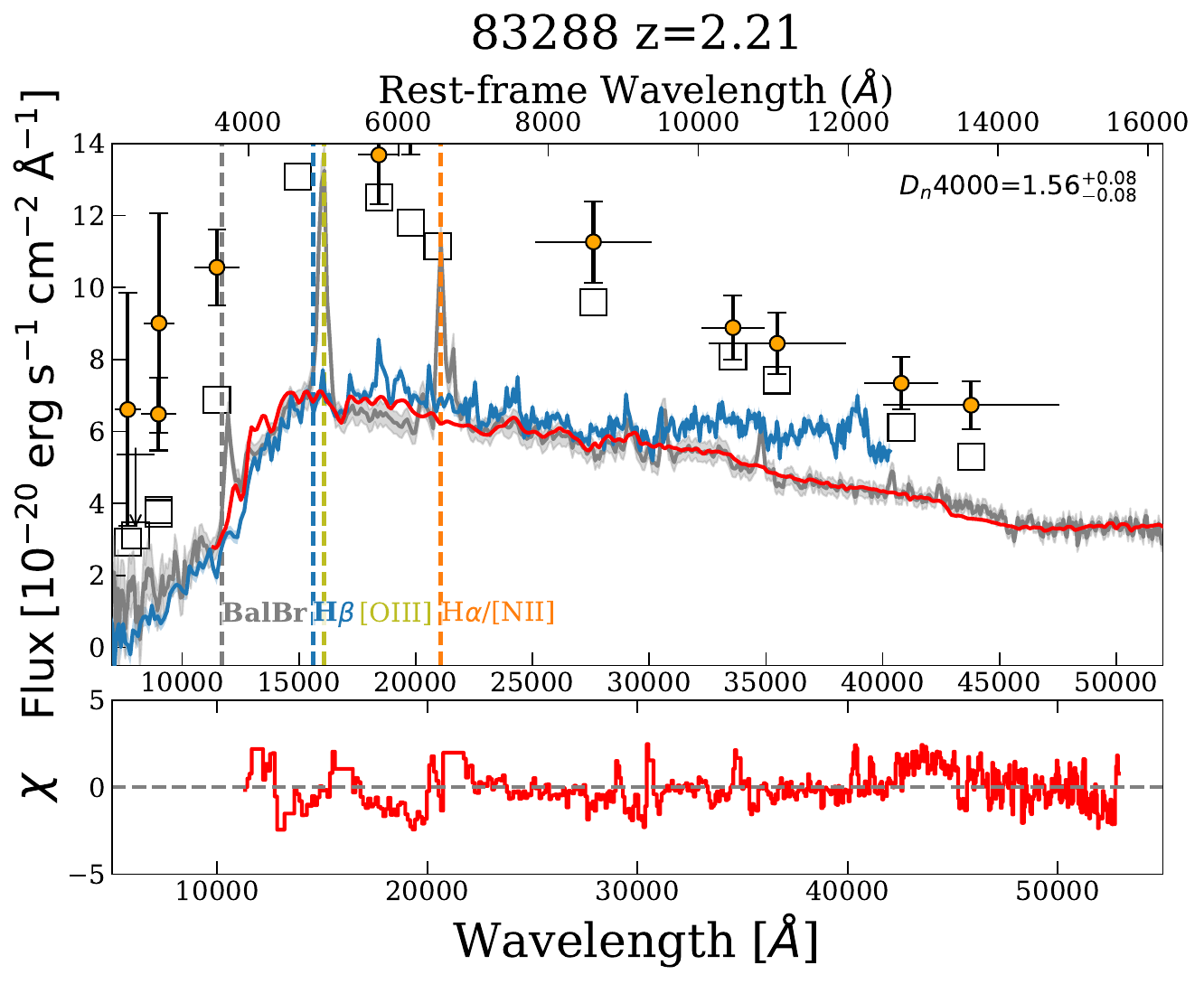}
    \caption{NIRSpec prism spectra and photometry for our four quiescent galaxies showing signs of a strong 4000-\AA break. Overplotted for reference is the galaxy from \citet{Glazebrook2024}, the first 4000-\AA break galaxy discovered at $z>3$. For reference the \citet{Glazebrook2024} galaxy has a 4000\AA break of strength $D_n4000$=$1.56\pm0.18$ which is consistent with the four galaxies shown here.}
    \label{fig:D4000spectra}
\end{figure*}

In this section, we explore the strong 4000-\AA break observed in four galaxies in our sample, as shown in Fig~\ref{fig:D4000spectra}, and why this fits with our early formation timescales for our quiescent galaxies.

The 4000-\AA break is a significant spectral feature observed in galaxies with stellar populations older than $\approx 1\text{--}2$~Gyr. This feature, also sometimes referred to as the D4000 break from the name of an empirical spectral index \citep{BruzualA.1983}, 
is characterised by a discontinuity in the spectrum around 4000~\AA, where there is a marked decrease in flux density on the blue side compared to the red side. However, due to the nature of the 4000-\AA break, it is not a sharp discontinuity.
Another key signifier of a 4000-\AA break is the smaller flux drop that is typically seen around 4300~\AA.

The 4000-\AA break originates primarily from the cumulative effect of numerous absorption lines from neutral hydrogen and ionised metals in the atmospheres of cooler stars. The break is quantitatively defined by the ratio of the average flux density in two narrow wavelength bands on either side of 4000 \AA. Specifically, it is often measured as the ratio between the average flux density in the range 3850 to 3950 \AA and the range 4000 to 4100~\AA \citep[D$_n$4000 index;][]{Balogh1999}.

The strength of the 4000-\AA break (i.e. D$_n$4000) serves as a robust diagnostic tool for determining the age and metallicity of stellar populations in galaxies. Older stellar populations exhibit a stronger D$_n$4000 break due to the higher prevalence of cool stars and increased metal blanketing. Conversely, younger stellar populations, which contain a higher fraction of hot blue stars, exhibit a weaker D$_n$4000 break. A D$_n$4000 above a value of 1.4 is also related to the metallicity of stellar populations, as well as to their age \citep{Kauffmann2003a}.

We report D$_n$4000 indices for the quiescent galaxy sample in Appendix Table \ref{tab:values_errors}. We find a range of values from 1.23 to 1.62.

Fig. \ref{fig:D4000spectra} shows the spectra for the four galaxies in our sample that show features consistent with a strong 4000\AA break. Overplotted in blue is the galaxy from \citet{Glazebrook2024} which has been shown to have a strong 4000\AA break {/bf ($D_n 4000=1.56\pm0.18$)} consistent with an old stellar population and high formation and quenching time. We can clearly see that the breaks match between the four galaxies within our sample and that of the \citet{Glazebrook2024} galaxy. The $D_n 4000$ values measured for our 4 galaxies are consistent with those of \citet{Glazebrook2024}. This further supports the early formation and quenching times shown in Figs. \ref{fig:formation_and_quenching} and \ref{fig:formation_and_quenching_2}.

\subsection{\textit{In-situ vs ex-situ} star formation in FLAMINGO}\label{s.results.ss.insitu}

\begin{figure*}
    \centering
    \includegraphics[width=2\columnwidth]{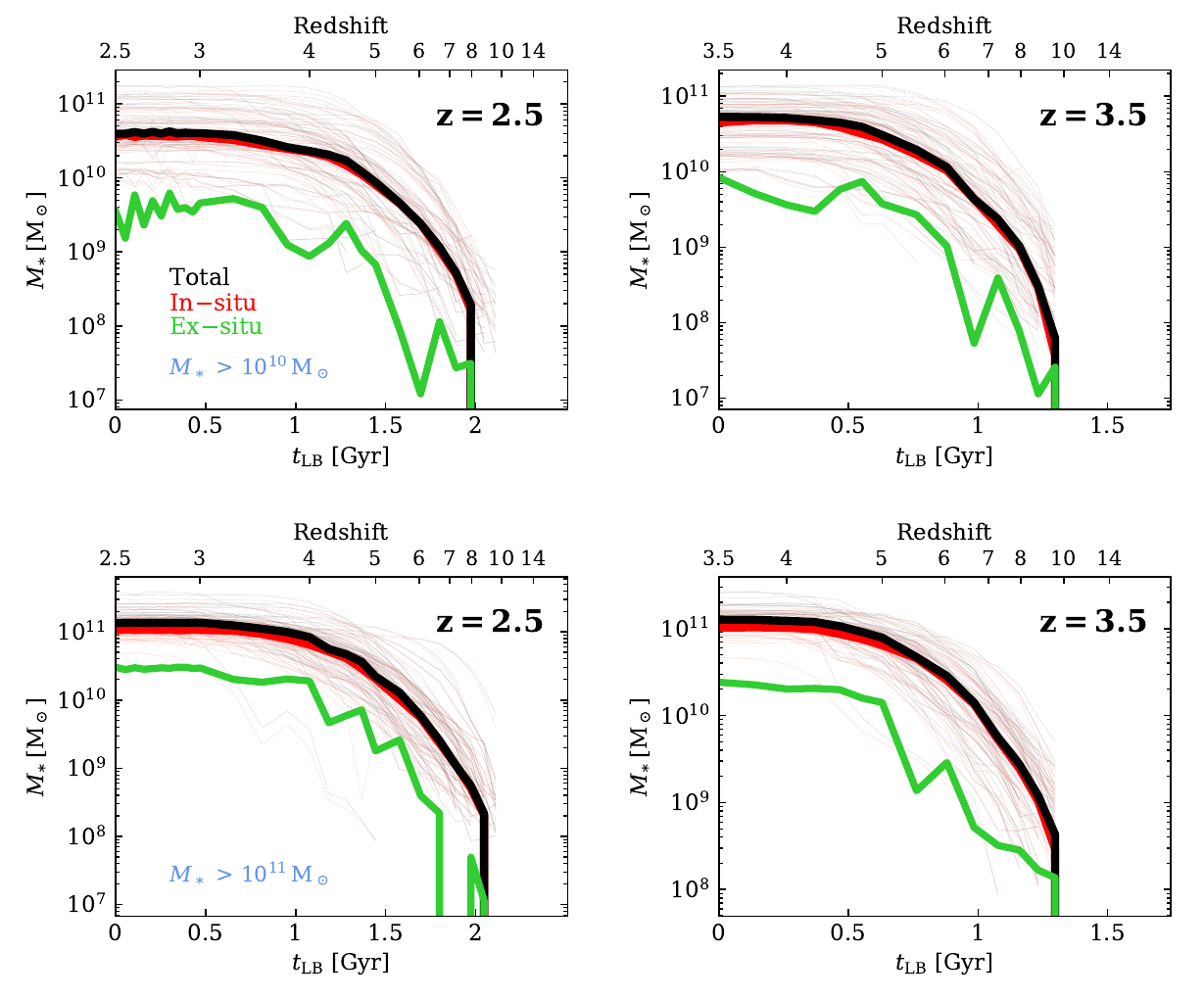}
    \caption{Stellar mass vs lookback time from observation (redshift) for massive quiescent galaxies from the FLAMINGO simulation in two redshift bins (left, z=2-3, right z=3-4.5) and two mass bins (upper $\rm M_\star\geq10^{10}M_\odot$, lower $\rm M_\star\geq10^{11}M_\odot$). The same criteria of Eq. \ref{eq:quenching} were used to select the quiescent galaxies. The black line is the median total stellar mass, and the red line is the median in-situ mass, while the green line shows the ex-situ mass obtained by subtracting the latter from the former. The thin curves show the individual mass evolution for 50 randomly chosen mock galaxies that satisfy the criteria. We can clearly see that the black lines and red lines are almost identical, meaning that almost all the mass of these galaxies was formed in-situ, showing that these galaxies have on average not built up their mass via major mergers.}
    \label{fig:in-situ}
\end{figure*}

A key limitation of a reconstructed SFH is the inability to trace where stars formed, i.e. whether they were born from gas already in the galaxy (\text{in situ}), or if they were accreted during a merger event (\textit{ex situ}).
However, numerical simulations can offer insight into the likely formation pathways. In this section, we explore the \textit{in-situ} vs \textit{ex-situ} star formation for massive quiescent galaxies in FLAMINGO.
Fig.~\ref{fig:in-situ} shows the mass build-up for the massive ($M_\star\,{\geq}\,10^{10}\,{\rm M}_\odot$) and very massive ($M_\star\,{\geq}\,10^{11}\,{\rm M}_\odot$) galaxies from the simulation at redshifts of 2.5 and 3.5. The black curve shows the total stellar mass as a function of redshift, whilst the red curve shows the amount built up \textit{in-situ}. Both the medians and individual curves for 50 randomly selected samples that satisfy the same quenching criteria as in Section~\ref{ssec_FLAMINGO_sample} are shown.

The \textit{in-situ} mass from the simulation is calculated as follows. First, for each simulated galaxy, we identify its main progenitors across cosmic time using the simulation merger tree. Next, for each star particle associated with the main progenitor at each given snapshot, we identify the redshift of its birth, which is available from the simulation product. Then we go to the first snapshot after the particle's birth and see if at that time the particle belongs to the main progenitor or not. If it belongs to the main progenitor, then we identify the star particle as \textit{in-situ}. Due to the finite time intervals of ${\simeq}\,100\,$Myr between the simulation snapshots, we cannot rule out the cases that the star particle was born \textit{ex-situ} but merged into the galaxy between the birth and the first snapshot after the birth. In principle, this may therefore lead to an overestimation of the \textit{in situ} fraction. Given the low \textit{ex-situ} or merger fraction overall as indicated by the results, it is unlikely to affect the results significantly. 

The key takeaway from Fig.~\ref{fig:in-situ} is that, on average, the \textit{in-situ} mass build-up explains the vast majority of the total mass formed. In contrast, the \textit{ex-situ} mass build-up via mergers appears to be contributing always less than 40\% to the galaxies' growth, even though the \textit{ex-situ} fraction slowly increases with galaxy mass and time. This can also be seen by the smoothness of the black median curve. In the case of frequent and dominant major dry-mergers, we would expect to see an almost stepwise increase in the mass by factors of up to two, due to the almost instantaneous increase of the mass of the resulting galaxy. We do not see this from the simulation results. The simulations predict, instead, that only a minority of the galaxies in our observed sample will have undergone a major gas-poor merger event over each ${\simeq}\,$100\,Myr. In fact, the mock galaxies have undergone about one and a half (one) major dry-mergers in total by $z\,{=}\,2.5$ (3.5) on average, where major dry-mergers are defined as mergers with the stellar mass ratio greater than 10:1. This corresponds to 0.6 such mergers per Gyr. In addition, not even all these mergers contribute significantly to the stellar mass budget at $z_{\rm obs}$, because the impact from the mergers with a long lookback time will be dominated quickly by the \textit{in-situ} growth since then. Given the low rate of 0.6 per Gyr, only a negligible fraction of galaxies at a given $z_{\rm obs}$ are expected to have had major dry-mergers recently enough to possess a significant \textit{ex-situ} mass budget. This is why the black and red median curves are so close to each other. 

On the other hand, they are typically found to undergo minor mergers, which are much more frequent and thus contribute more significantly to the stellar mass. However, the resolution of FLAMINGO does not allow us to probe the impact of mergers with a mass ratio lower than 10:1 at $z\,{\gtrsim}\,7$ for objects like our samples. In addition, assessing the impact of minor mergers depends on the definition of such mergers, as well as science goals. 

The case of wet major mergers is slightly more complicated. This would result in a stepwise increase in the total baryon budget, but the star-formation resulting from any gas inflow would then be counted as \textit{in-situ} star-formation. This could very likely "wash-out" any stepwise increase seen in the figure, assuming that the newly acquired gas would be consumed gradually over a long period, not instantaneously. Therefore, we caution that we cannot rule out the impact of wet major mergers in the same way as we can rule out dry major mergers (although at early epochs it is difficult to distinguish between wet mergers and gas accretion). 

This is important to determine likely quenching mechanisms. There has previously been a debate about whether major mergers can be involved in quenching massive galaxies \citep[e.g.,][]{Ilbert2013}. In this scenario, a major merger would funnel gas into the centre of the galaxy, triggering a starburst, which would rapidly use up the gas and simultaneously cause strong AGN feedback, thereby quenching the galaxy. However, this scenario has been disfavoured \citep[e.g.,][]{Grogin2005, McAlpine2017,RodriguezMontero2019}. Overall, while major mergers are likely to cause significant starbursts, they are no longer thought to play a significant role in actually quenching massive galaxies. This is supported by the findings of \cite{D'Eugenio2023a} for an early quenched galaxy that could be studied in detail: the quenching process left the dynamics and morphology of the galaxy unaffected (disc-like), implying that it could not have happened via a catastrophic merging event.

According to the simulation predictions, most
massive quiescent galaxies at $z>2\text{--}5$ did not undergo a major merger event recently. This is in qualitative agreement with the compact sizes of observed galaxies, which suggest no major contribution of mergers. Therefore, the quenching of these massive galaxies is most likely related to an internal process. For galaxies of this mass, that is almost definitely going to be related to some form of AGN feedback \citep[e.g.,][]{Man2018}.

\subsection{AGN incidence}

The duration of active phases in galaxies is debated \citep{Alexander2012, Harrison2018, Harrison2024}, but timescales are generally thought to be much shorter than the quenching timescale of galaxies, of order 1--10~Myr. In fact, both theory and observations suggest that AGN-driven quenching is related to the time-integrated accretion history, as traced by the SMBH mass \citep[e.g.,][]{Davies2020,Bluck2022,Bluck2023, Piotrowska2022, Brownson2022,Baker2024}.
Nevertheless, by observing quiescent galaxies at earlier and earlier epochs, we necessarily `squeeze' the timescales of quenching and AGN activity. Specifically, there is less time available over which to spread the time-integrated accretion history, meaning we have a high chance to detect AGN activity.
To ensure the most complete census possible, we use a combination of four diagnostics to probe AGN. The first is X-ray detections based on Chandra Legacy data of GOODS-S and GOODS-N fields. 
%An X-ray detection at z$\geq$2 means the presence of an AGN. 
For GOODS-N, we use the catalogue and classifications from \cite{Xue2016}, whilst for GOODS-S we use the catalogue and classifications from \cite{Luo2017}. We cross-match the R.A. and Dec. to within one arcsec. 
This gives us three matches, all of which are classified as AGN. These are galaxies 199773, 197911 \citep[as also identified in][]{Circosta2019,D'Eugenio2023a, Scholtz2024}, and 83288. 
However, non-detection in X-rays does not rule out the presence of an AGN, due to the possibility of substantial obscuration, or even intrinsically weak X-rays emission \citep[e.g.,][]{Circosta2019,Mazzolari2024, Mazzolari2024b,Maiolino2024a, Juodzbalis2024a, Lyu2024}.  This has been ascribed to many different physical processes, but these are beyond the scope of this paper; we refer the reader to \citet{Maiolino2024a} for an in-depth summary.

\begin{figure}
  \includegraphics[width=\columnwidth]{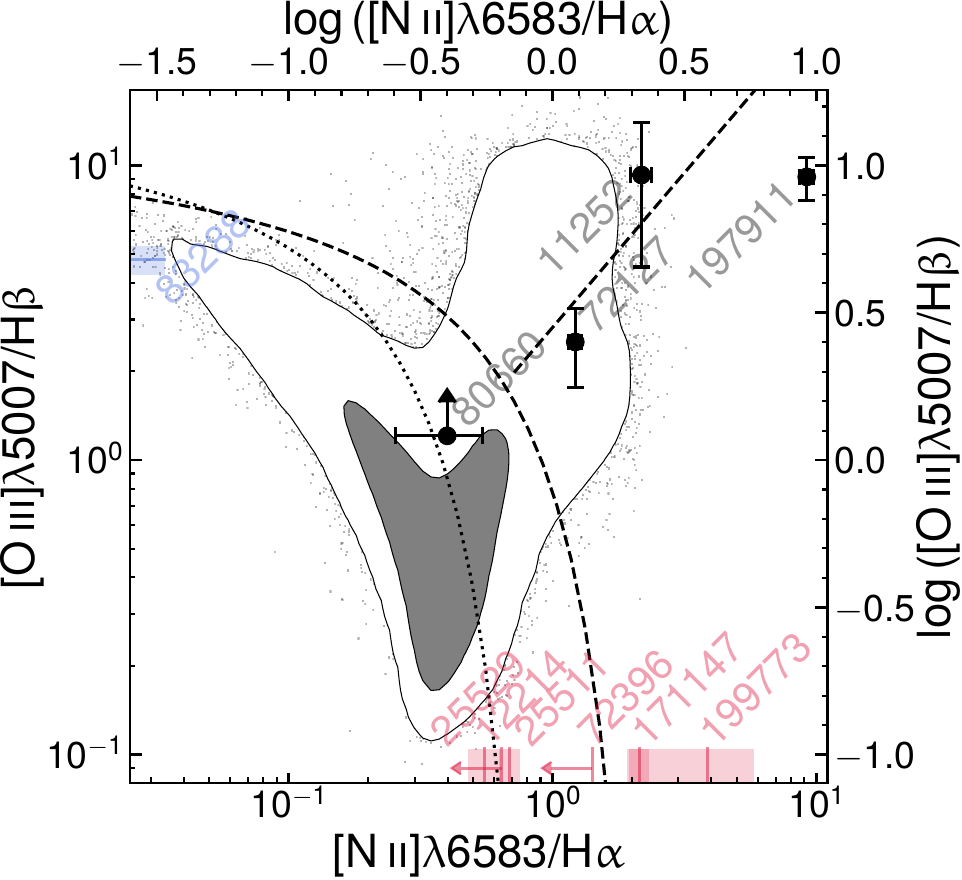}
  \caption{BPT line-ratio diagram, showing the incidence of AGN in our sample. The contours are data from SDSS, the demarcation lines are from \citet{Kewley2001} (dotted), \citet{Kauffmann2003} (dashed) and \citet{Schawinski2007} (dashed extension). The measurements to the bottom have no available [OIII]$\lambda$5007/H$\beta$; the measurement to the left has no available [NII]$\lambda$6583/H$\alpha$.}\label{fig:bpt}
\end{figure}

In addition to X-ray selection, we use the classical optical BPT diagram \citep{Baldwin1981} with diagnostic dividing lines from \cite{Kauffmann2003}.
We used the flux from the R1000 gratings to separate out H$\alpha$ and [NII]. The results are shown in Fig. \ref{fig:bpt}. Note that several spectroscopically confirmed AGN have been reported to occupy the star-forming region of the BPT diagram at redshifts $z\gtrsim5$ \citep[e.g.,][]{Ubler2023,Maiolino2023}. However, only one of our BPT measurements occupies a region consistent with high-redshift AGN (83288), and this galaxy is at $z=2.2$, below the redshift where the BPT becomes unreliable \citep{Scholtz2023}.
We find several AGN candidates which are 011252, 072127, 1171147, 197911, and 199733. Three of these are classified on the basis of the four emission lines, while another two are classified as LIER/AGN based on their high [NII]/H$\alpha$ ratios only.

Two of our sample show broad H$\alpha$ lines corresponding to a broad-line region. These are 8777 \citep[previously identified in][]{Carnall2023Nature} and 11252.

We also use the Near-Infrared (NIR) to help identify AGN. We use MIRI data where available. This identifies 11252, 197911, and 199773 as NIR-selected AGN (all of which are selected as AGN by other diagnostics).

Table \ref{tab:table_agn} provides a summary of our AGN selection criteria. We find we observe AGN activity in 8/18 massive quiescent galaxies in our sample.

\begin{table}
    \centering
    \caption{Table showing AGN incidence and detection method. Empty values (---) mean that no data is available; for the BPT columns, this means that no spectra with sufficient spectral resolution were available. The question marks indicate that the data are present but insufficient to classify the object.}
    \label{tab:table_agn}
    \setlength{\tabcolsep}{2pt}
    \resizebox{\columnwidth}{!}{%
    \begin{tabular}{llllll}
\hline
NIRSpec ID & X-ray & BPT & BLR & NIR &  AGN  \\
\hline                    
011252     &   0   & 1   & 1   &  1  &   1   \\
012214     &   0   & ?   & 0   &  ?  &   0   \\
012619     &   0   & --- & 1   &  0  &   1   \\
025511     &   0   & ?   & 0   & --- &   0   \\
025529     &   0   & ?   & 0   & --- &   0   \\
027461     &   0   & ?   & 0   & --- &   0   \\
072127     &   0   & 1   & 0   & --- &   1   \\
072396     &   0   & ?   & 0   & --- &   0   \\
080660     &   0   & ?   & 0   & --- &   0   \\
083288     &   1   & --- & --- & --- &   1   \\
171147     &   0   & 1   & 0   & --- &   1   \\
197911     &   1   & 1   & 0   &  1  &   1   \\
199733     &   1   & 1   & 0   &  1  &   1   \\
200733     &   0   & --- & --- & --- &   0   \\
\hline                    
8777       &   0   & ---  & 1   &  0  &   1   \\
1397       &   0   & ---   & 0   & --- &   0   \\
8290       &   0 & ---   & 0   & --- &   0   \\
6620       &   0   & ---   & 0   & --- &   0   \\ 
\hline
\end{tabular}
}
\end{table}

\subsection{\texorpdfstring{Ly$\alpha$}{Lya} Emission in a redshift 4.1 quiescent AGN host galaxy}

072127 is a particularly interesting target \citep{Kokorev2024,Jones2024}. It is extremely compact, with a half-light radius of just 0.3 kpc, comparable to 8777 \citep{Carnall2023abundances}. It shows clear signatures of being an AGN host \citep[like 8777, as further shown in][]{Ji2024Hotdust}, but it displays prominent narrow-line emission, making it a type-2 AGN (unlike 8777, which shows only weak narrow-line emission). \citet{Kokorev2024} recently reported broad-H$\alpha$ emission, indicating a type-1 AGN, which would further confirm our case for an AGN.
However, we note that our analysis does not confirm a strong detection \citetext{Juod\v{z}balis I., in~prep.}. 
In addition to narrow-line optical emission, 72127 also shows extremely strong Ly$\alpha$ emission in the prism spectrum \citetext{Fig.~\ref{fig:spectra} and \citealp{Jones2024}}, the second such instance in a high-redshift quiescent galaxy after RUBIES-EGS-QG-1 \citep{UrbanoStawinski2024}.

\subsection{Morphology and sizes}\label{sec:morphology}

Another important aspect to explore is the morphology of high-z quiescent galaxies \citep[e.g. see][]{Kartaltepe2023, Ferreira2023, Baker2024b}. 
In the more local to redshift 2 Universe, galaxy sizes and \sersic indices are important tests of both theory and cosmological models \citep[for a review see][]{Conselice2014}. This is important to link concepts such as merger rates, growth, and size evolution. As described in Section~\ref{s.analysis.ss.morphology}, we use \textsc{PySersic} to fit the galaxy images in the F444W filter. We chose this filter because it is the least contaminated by emission lines from the AGN. The downside is that we are most affected by the PSF at this wavelength. To test these effects we also fit the galaxies in the F200W filter and find that the results for the sizes remains consistent.

\begin{figure}
\centering
    \includegraphics[width=1\columnwidth]{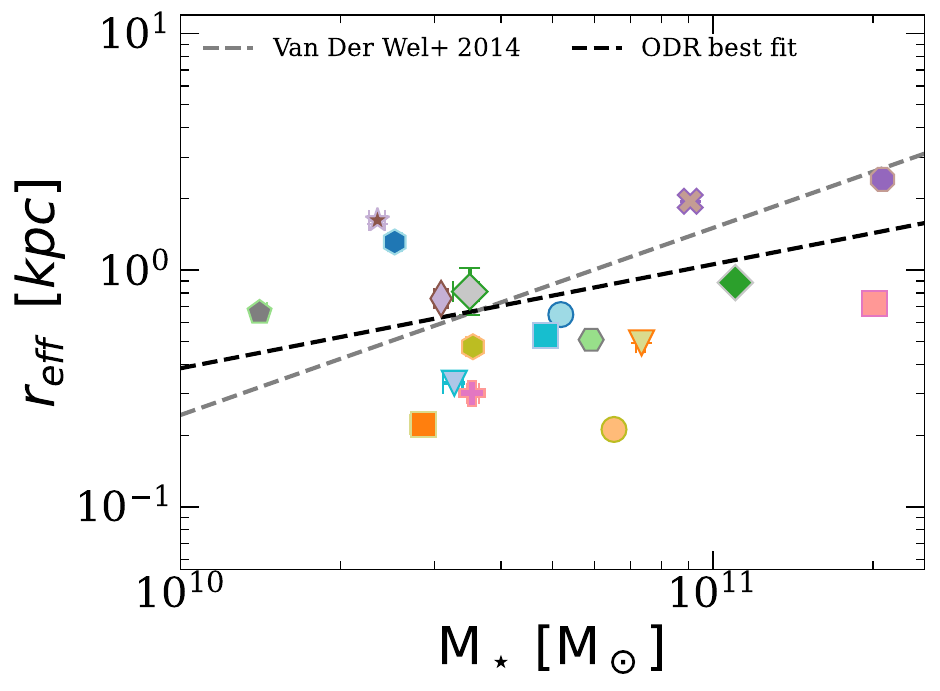}
 
    \caption{Half-light radius vs stellar mass for the 18 massive quiescent galaxies.
    The grey dashed line is the best-fit relation from \citet{vanderWel2014}
    We see we obtain a large scatter which is unsurprising as we have a wide redshift range.}
    \label{fig:sizemass}
\end{figure}

Fig. \ref{fig:sizemass} shows the effective radius in kpc vs. the stellar mass for the 18 massive quiescent galaxies. For comparison, the size--mass relation from \citet[][for early-type galaxies at redshift 2.75]{vanderWel2014} is overplotted. We find that the relation almost completely bisects the galaxies in our sample. Interestingly, we do not appear to see two distinct size--mass relations based on stellar mass \citep[as was seen in a larger photometric sample in ][]{Ji2024}, but this is likely due to the smaller size of our spectroscopically confirmed sample.

We find a large amount of scatter in our size measurements. Despite this, we use orthogonal distance regression (ODR) to provide a best fit to our sample. We find a shallower best-fit relation than that of \cite{vanderWel2014}, but warn that our sample contains higher-redshift galaxies and we expect the size-mass relation to have a redshift dependence. Therefore, we caution that the exact shape of our size-mass relation is likely not statistically significant.
The relation obtained is $\rm r_{eff}/kpc= 10^{-0.11\pm0.08}\times (M_\star/5\times10^{10}M_\odot)^{0.44\pm0.21}$ with a scatter of 0.4 dex.
We include the effective radii for the quiescent galaxies in Appendix, table \ref{tab:values_errors}.

In addition to galaxy sizes, we can also explore the distribution of \sersic indices for the sample. Quiescent galaxies traditionally show more concentrated light distributions corresponding to \sersic indices greater than 2. Often these take the form of the traditional de-Vaucouleurs elliptical-style galaxies. We again draw our \sersic indices from fits to the quiescent galaxies in the F444W filter. We find a range of \sersic indices from two to eight (eight being the upper bound for the fits). The majority of the galaxies show high \sersic indices, consistent with them being ellipticals. The \sersic indices are reported in Appendix table \ref{tab:values_errors}.

\section{Discussion}

We selected a sample of 18 massive quiescent galaxies from the JADES survey and from PID~2198 \citep{Barrufet2021} that have both NIRCam photometry and NIRSpec prism spectroscopy, complemented by medium-resolution grating spectroscopy from JADES and PID~1207 for 2 galaxies. We demonstrated that these galaxies are both quiescent and massive. Our primary goal is to study internal quenching mechanisms; for this reason, we consider only central galaxies, based on the absence of massive neighbours. We reject two lower-mass and one higher-mass quiescent satellite galaxies on the basis of their proximity to higher-mass galaxies at the same redshift (Section~\ref{s.sample.ss.jades}, and for more details on the satellites, see Appendix~\ref{app:satellites}).

\subsection{Number densities}\label{s.disc.ss.numberdens}

The number density of quiescent central galaxies is one of the most basic tests of galaxy evolution and cosmology. The existence of massive, quiescent galaxies at $z>3$ is no longer considered a problem for cosmology \citep[e.g.,][]{Glazebrook2017}. However, their existence and abundance provide compelling constraints on our models of star formation efficiency \citep{Glazebrook2024,deGraaff2024,Carnall2024} and SMBH-driven quenching \citep{Xie2024}. Although a few objects may still be considered outliers \citep[e.g.,][]{Carnall2023Nature,Glazebrook2024,deGraaff2024}, samples of quiescent galaxies require a revision of our models \citep{Carnall2023abundances,Valentino2023,Barrufet2021,Carnall2024,Park2024,Nanayakkara2024}.
In this study, we used the superior precision of spectroscopy to study quiescent galaxies down to a stellar mass limit of $\Mstar \gtrsim 10^{10}~\Msun$. In Section~\ref{s.results.ss.numberdens}, we compared the number densities of quiescent galaxies in our spectroscopically corrected photometric sample with that found in FLAMINGO, one of the largest-box hydrodynamical cosmological simulations to date \citep{Schaye2023,Kugel2023}.
This is a crucial check -- many works have previously explored the number densities of high-z quiescent galaxy candidates \citep[as identified photometrically;][]{Alberts2023, Carnall2023abundances, Ji2024, Valentino2023}, but we are the first to combine a large, spectroscopically corrected photometric sample with a large-volume hydrodynamic simulation. Our approach removes both possible interlopers and contaminants and enables us to rule out a major role of cosmic variance \citep[e.g.,][]{Carnall2023abundances}. We are also able to account for quiescent galaxies missed photometrically with UVJ colour selection (see Section \ref{s.sample} and Fig. \ref{fig:uvj}).

In addition, the large simulation box allows us to safely probe cosmic variance up to 3,$\sigma$, as demonstrated in Fig.~\ref{fig:Number densities}. 
Our observational analysis reveals a significant excess of massive quiescent galaxies compared to FLAMINGO, both at $z=2\text{--}3$ (2$\sigma$), and even more so at $z=3\text{--}4.5$ (3$\sigma$). This cements the findings of earlier studies \citep[which used photometry and a range of other numerical simulations; e.g.,][]{Valentino2023, Lagos2025} and gives us important information on the feedback prescriptions of the simulations. 
This comparison can be seen in more detail in Sec. \ref{s.results.ss.comp_num_den}, where we compare our spectroscopically corrected number densities to other observational studies, cosmological simulations and SAMs. We again find that our observed number densities are generally consistent with other observations at high-z, showing an overabundance compared to the cosmological simulations. 
In particular, our findings suggest that theoretical models require 1) stronger and/or more efficient quenching of massive galaxies, 2) acting as early as redshifts of 6 and earlier. Energy arguments suggest that this earlier quenching is most likely due to AGN feedback, given that other mechanisms struggle to counteract cosmic gas accretion \citep[e.g.,][]{Silk1998}.
We also note the usual caveat that increased resolution would be important to fully probe the parameter space of the typical quenching criteria.

\subsection{AGN feedback}\label{s.disc.ss.agn}

That AGN feedback may be stronger and/or more efficient at earlier epochs is suggested not only by the unexpectedly high number density of quiescent galaxies but also by other lines of evidence. In our sample, we find a large fraction of AGN, suggesting continued fuelling of the central SMBH even after hundreds of Myr of quiescence. This is confirmed by several other findings of nebular emission in quiescent galaxies, both at $z>4$ \citep[e.g.,][]{Carnall2023abundances,deGraaff2024}, in samples near Cosmic Noon \citep{Park2024,Bugiani2024}, and even at later epochs \citep[][though these authors argue that low-level star formation could explain their findings]{Maseda2021}. Our work is the first sample study at $z=2\text{--}5$, finding an AGN fraction of 50~percent. A comparison to the AGN incidence in a control sample of star-forming galaxies of the same mass is urgently needed \citep[initial estimates hover around 20\%][]{Scholtz2023}.

We stress that finding AGN activity is not a smoking gun for AGN-driven quenching; simple logic posits that ongoing AGN cannot have caused quenching hundreds of Myr earlier, especially given the short duration of AGN episodes, which is thought to be a few Myr at most \citep[at least at lower redshifts; e.g.,][]{Harrison2018}. Evidence from statistical studies at $0<z<2$ finds that quiescence is most closely related to the central stellar velocity dispersion \citep{Bluck2022,Bluck2023, Baker2024}, which is a proxy for SMBH mass \citep[e.g.,][]{Magorrian1999, Ferrarese2000,Kormendy2013}, itself a measure of the time-integrated SMBH activity. These studies suggest that it is not any particular AGN episode that quenches a galaxy, but the cumulative effect of all episodes, which heat the galaxy halo and prevent further gas accretion (preventive feedback) \citep{Baker2024}. This finding and its interpretation is supported both by studies of small samples of local galaxies with directly measured SMBH masses, as well as by theoretical predictions \citep{Davies2020,Bluck2022,Piotrowska2022}. Recently, evidence in this direction has been reported even by observations of molecular gas (or lack thereof) in a quiescent galaxy at $z=3$ \citep{Scholtz2024}. Nevertheless, even in this scenario preventive-feedback scenario, we would still expect a higher incidence of AGN at high redshift, simply because at earlier cosmic times there is less time available to distribute the time-integrated accretion. In addition, maintenance-mode feedback may be required to maintain galaxies quiescent over long periods of time -- particularly at early cosmic epochs, when gas density and mass accretion rates were higher.
However, we cannot exclude the possibility that AGN-driven quenching was different in earlier epochs \citep[e.g.,][]{Xie2024, Maiolino2023}. Although existing studies based on JWST imaging confirm the trends found at lower redshifts \citep{Bluck2023}, these early studies are still dominated by quiescent galaxies at Cosmic Noon, while quiescent galaxies at earlier epochs may still be too few in number to detect signatures of different trends, as supported e.g. by numerical simulations \citep{Remus2023,Valenzuela2024}.

Another independent line of evidence supporting stronger and/or more efficient AGN feedback at high redshift is the discovery of widespread neutral-gas outflows in quiescent galaxies at Cosmic Noon \citep{Davies2024} and even earlier epochs \citep{Belli2024,D'Eugenio2023}. These studies confirm the continued fuelling of SMBHs in quiescent galaxies, and underscore the role of ejective feedback -- at least in maintaining galaxies as quiescent.

A final possibility is that SMBH growth is more rapid than expected at high redshift. This is supported by the discovery of SMBHs that are `overmassive' relative to the stellar mass of their host galaxy, when compared to local scaling relations \citetext{\citealp{Ubler2023,Juodzbalis2024,Maiolino2023,Mezcua2024, Kokorev2023, Kocevski2023}; but see e.g. \citealp{Sun2024} or \citealp{Stone2024} for a different view}. These overmassive black holes are a strong sign that AGN feedback in the early Universe could have been more rapid than what we initially expected, as it outpaces the growth rate of the host galaxy. Alternatively, we could interpret this finding as further evidence supporting stronger/more efficient SMBHs feedback, which may have inhibited star formation in the host galaxy.

Studies of the cold gas reservoir of quiescent galaxies will be crucial to pin down how exactly quenching works \citep{Williams2021,Whitaker2021} -- particularly at $z>2\text{--}3$ \citep{Scholtz2024}.

\subsection{Star-formation histories}\label{s.disc.ss.sfh}

Exploring the SFHs of our sample we see a clear divide between galaxies that appear to have formed all their mass within a single short burst and those that appear to have a much more extended star-formation period. The `burst' ones, similar to those observed by \citet{Carnall2024}, appear to have managed to form hundreds of solar masses worth of stars per year for a short period of time. Something then appears to have rapidly shut off this star formation, and these galaxies have remained dormant ever since. Meanwhile, the more extended SFH galaxies appear to have had a much more gradual decline in star formation, corresponding to a likely much slower-acting feedback process.

This kind of divide between fast-quenching and slower-quenching mechanisms has previously been seen for populations of lower-redshift quiescent galaxies \citep[e.g.][]{Belli2019, Wu2018}.

To some extent this can help inform us as to possible quenching mechanisms. In the case of a single AGN outflow being the dominant driver we would expect a very short quenching timescale and much more of a `single-burst' SFH. Whereas it is possible to envisage the cumulative effects of AGN feedback \citep[e.g. ][]{Bluck2022, Piotrowska2022, Baker2024} working to quench a galaxy over an extended period of time. In addition, according to their SFHs, many of our galaxies have remained quenched for extended periods of time, in some cases since z$\sim$6. %This requires a mechanism to prevent further gas accretion and continue the galaxies' quiescent phase, which may well be provided by preventative forms of AGN feedback. 

This requires a mechanism to prevent further star formation and continue the galaxies' quiescent phase. Such mechanism cannot be star formation feedback, because even relatively small star-formation episodes would be easily visible due to their low M/L ratios. In contrast, preventative AGN feedback may be able to stop gas accretion and thus explain the protracted quiescent phase.
Other quenching mechanisms are possible, e.g. stellar feedback, cosmic rays, environment and more, but these are generally considered less likely quenching causes in these most massive central high-z quiescent galaxies, but have been shown to possibly be important in lower mass high-z galaxies \citep{Gelli2025}. 

Our results match nicely with the results of \citet{Park2024} who explored the quenching of massive galaxies at Cosmic Noon with the Bluejay survey. They found evidence for three different types of SFHs: old galaxies quenched at early epochs; galaxies rapidly and recently quenched after a flat/bursty formation history; and galaxies that were quickly and recently quenched after a short and intense burst of star formation (starburst). We appear to find all three scenarios within our sample. We find the early-formed, early-quenched galaxies, which in our case are dominated by the 4000-\AA break galaxies (e.g. 11252, 72396), plus galaxies quickly quenched after a longer SFH (corresponding to our extended SFH galaxies, e.g. 199773, 27461) and also the short starbursts \citep[e.g. 80660, 25511, or 8777 as found also by][]{Carnall2023Nature}. However, in addition to the three types found by \citet{Park2024}, we also find evidence for galaxies that appear to have quenched more slowly, such as 83288 and 200733. 

The formation and quenching epochs within our high-z galaxy sample are generally at much higher redshifts consistent with results found for individual galaxies and much smaller samples \citep[e.g. ][]{Glazebrook2024, Carnall2023Nature, deGraaff2024, Weibel2024b}. Part of the reason is due to selection; our sSFR cut removes more recently quenched galaxies (i.e. those having quenched within the 100~Myr prior to observation). However, despite this `adverse' selection, we still find many more quiescent galaxies at these redshifts than would have been expected in the pre-JWST era (Section~\ref{s.disc.ss.numberdens}), in agreement with earlier studies \citep[][]{Carnall2023abundances, Valentino2023}.

As part of our analysis we did explore whether we saw any correlation between formation and quenching time and other galaxy properties such as stellar mass for our 18 quiescent galaxies. This can probe important galaxy effects such as 'downsizing' \citep{Cowie1996,Thomas2005,Ilbert2013}. We see a tentative trend towards more massive galaxies having an older formation and quenching time, but due to the small nature of our sample, we find that it is not statistically significant. Larger samples of spectroscopically confirmed high-z massive quiescent galaxies will be required to properly test these trends.

We find no evidence for rejuvenating galaxies \citep{Witten2024}, or galaxies that appear "mini" or "rapidly" quenched \citep{Looser2023, Looser2024, Dome2024, Baker2025}, although again we caution that this is due to our selection criteria. 
In particular, while some models predict that quiescent galaxies at high redshift are likely to rejuvenate \citep{Remus2023}, our data may not be sensitive to bimodal SFHs if both peaks are older than 100~Myr (as required by our selection criteria). Investigating recently rejuvenated galaxies is easier when including currently star-forming galaxies, which increases the baseline in logarithmic look-back time over which the two peaks of SFR can be separated.
As for mini or rapidly quenched galaxies, medium-band photometric studies find that these systems may be abundant in GOODS-S \citep{Trussler2024}. However, they are by definition rapidly and recently quenched galaxies, with quenching times of a few tens of Myr \citep[e.g.,][]{Looser2024, Baker2025} -- which we specifically select against.
In addition, mini-quenched galaxies are expected to have a much lower mass than those we are probing and to be more common at higher redshifts than what we consider here \citep{Dome2024,Ceverino2018,Ceverino2021,Lovell2023}.
For these reasons, we find no tension between our lack of mini-quenched systems and previous studies \citep[e.g.,][]{Trussler2024}.

Interestingly, four galaxies within our sample exhibit 4000-\AA breaks, an unequivocal indicator of an aged stellar population. These breaks are distinct from the Balmer break, which typically describes stellar populations that are a few hundreds of Myr old. In contrast, the 4000-\AA break typically appears in populations around 1~Gyr old and older, depending on metallicity.

These 4000-\AA break galaxies have been seen before in high-z studies, most notably in the case of ZF-UDS-7329 \citep{Glazebrook2024}. This galaxy was observed at a redshift of 3.205, when the Universe was around 2 Gyr old. The key conundrum with these breaks is that they imply a much earlier formation time, as they are only seen in old stellar populations, which provides strong constraints to galaxy formation. Indeed, \citet{Glazebrook2024} show that there is a lack of massive enough dark-matter haloes to host their galaxy at the redshift at which it should have built up its measured stellar mass. 

The galaxies in our sample that show 4000-\AA breaks appear to be less extreme than the \citet{Glazebrook2024} example. They are all at lower redshifts and less massive. However, the fact that we see four of them suggests that they may be much more prevalent than previously thought. Their formation epochs of $z>8$, coupled with early quenching around redshift 5--8, require rapid quenching. They also require no recent star formation, which would otherwise erase the 4000-\AA break \citep[similar to what is described for Balmer breaks;][]{Witten2024}.
Fascinatingly, even in the 4000-\AA break subset, two of them turn out to be AGN hosts (in agreement with the overall sample fraction of 50~percent; Section~\ref{s.disc.ss.agn}). This suggests that there is still some form of gas reservoir/fuel remaining, not only in recently quenched galaxies, but even in these maximally old galaxies.

\subsection{\textit{In-situ vs ex-situ} star formation}\label{s.disc.ss.insitu}

Using the FLAMINGO simulation, we were able to explore the mass build-up of massive quiescent galaxies, finding that primarily the mass is formed \textit{in situ} (Section~\ref{s.results.ss.insitu}), with only a small fraction of \textit{ex-situ} stars. The simulations predict that these galaxies typically experience one or two major dry-mergers, or only 0.6 per Gyr. They can still undergo minor mergers that may contribute more than major mergers to the mass budget. However, the numerical resolution of FLAMINGO does not allow us to probe down to minor mergers. This result, in combination with the high AGN incidence (Section~\ref{s.disc.ss.agn}), presents an interesting scenario. 50 percent of our massive quiescent galaxies host an AGN (and  the true number is likely to be higher). This requires the presence of gas to fuel the AGN, but this gas cannot have come from the classic gas-rich, major-merger scenario, which would efficiently channel gas into the galaxy's centre. This leaves minor mergers or halo cooling as sources of gas. 

However, as previously mentioned, there remains the possibility that we cannot accurately pick up wet mergers due to star-formation resulting from the gas of the wet merger. This would then be counted as \textit{in-situ} star-formation, possibly washing out any sudden increase in the stellar mass budget. 
Alternatively, AGN feedback is ineffective in expelling all of the gas from the galaxies. This would leave a significant gas reservoir that, after spending some time at large galactocentric distances, could return to continue to fuel the AGN.

\subsection{Limitations of the simulations}\label{s.disc.ss.sim}

Although we exploited the large box size of FLAMINGO simulations as a crucial advantage to explore the cosmic variance, it comes at several expenses as well, including the resolution. Increasing a numerical resolution normally leads to a lower quenched fraction, because of more star forming gas being resolved, and assignment of non-zero SFR to galaxies whose SFR fall below a poorer resolution. Therefore, given the direction of the current discrepancy, namely fewer quenched objects predicted by the simulations than observations, our simulation results for the number densities can be considered rather as a conservative limit to the tension in principle. 
For example, the formation of a single star particle in 100 Myrs corresponds to a sSFR = 1.3$\times10^8$ $\rm M_\odot$ / $100\times10^8 \rm yr$ / $\rm 2.5\times10^{10}\;M_\odot$= $5.2\times10^{-11}\;\rm yr^{-1}$ at the observed lower mass limit. Our sSFR quenching criterion corresponds to a sSFR $\rm \leq 0.2 / 3\times10^9\;yr^{-1}$ = $\rm 6.7\times10^{-11}\;yr^{-1}$ at $z=2$, which is clearly not well resolved at the lower mass limit.

However, there are a few specifics of FLAMINGO that complicate such interpretation and the impact of resolution. First, the simulations take so-called `weak convergence' scheme where the values of the model parameters are adjusted when changing the numerical resolution. This means that the simulations were tuned to best match the observational constraints at \textit{each} resolution independently. Although this is generally good, because one does not need to care about an absolute numerical convergence, it complicates assessing the significance and even the direction of resolution effect. Namely, it is practically impossible to know whether more or less quenched galaxies will be found when the resolution increases. Second, the simulations fit to the stellar mass function (SMF) at $z\,{\simeq}\,0$ only up to $\rm M_\ast\,{=}\,10^{11.5}\,{\rm M}_\odot$, to calibrate the model parameters. While this is due to the great uncertainties of the SMF at the massive end, this may impact the reliability of the predictions assuming a large fraction of our samples are possibly progenitors of those massive galaxies not used by the simulations as constraints. 

There is another limitation of using simulations, in general, for studying rare, high-$z$ objects. Many of the state-of-the-art hydrodynamical cosmological simulations use $z\,{\simeq}\,0$ observations of normal populations of galaxies as primary constraints for the calibration of their models, as is also the case for FLAMINGO. This complicates assessing the accuracy and reliability of their predictions on rare, extreme galaxies at high redshifts, since there could be a wide range of evolutionary paths that can lead to very similar low-redshift scaling relations, hence degeneracy. We refer the reader to \citet{Lim2021} for a more extensive discussion of the limitations involved in comparisons of high-$z$ extreme objects between observations and models in general.

\section{Conclusions}

We explored spectroscopically confirmed massive quiescent galaxies from the JADES survey to investigate their star-formation histories, quenching timescales, and AGN incidence. We also explored a representative sample from the FLAMINGO simulations to compare number densities and mass build-up.
Our key results are the following.
\begin{itemize}
    \item We find 18 massive spectroscopically confirmed quiescent galaxies, 5 of which are above redshift 4, some of which are clearly quiescent but do not fulfil the traditional UVJ cut (but are still selected via a sSFR criterion).
    \item We observe more spectroscopically confirmed massive quiescent galaxies than predicted from the FLAMINGO simulations, even more than allowed by the 3\,$\sigma$ cosmic variance (especially at $z=3\text{--}4.5$ and $z=4\text{--}5$). This has important implications for theoretical modelling, in particular for the feedback prescriptions used in the simulations.
    \item We find both short and long star-formation histories suggesting the prevalence of both short and long timescale quenching mechanisms.
    \item We find four galaxies with clear evidence of 4000 \AA breaks, a signifier of an old evolved stellar population. These galaxies have high formation and quenching redshifts and are particularly interesting objects. 
    \item We find that most of our massive quiescent galaxies formed above redshift 5, with some of these likely to have formed above redshift 6. The majority of our sample quenched at redshifts 4 to 6 with a few galaxies having quenched even earlier.
    \item Using the FLAMINGO simulation, we explore the stellar mass build-up of massive quiescent galaxies similar to our samples. We find that the mass of these galaxies is almost entirely built up \textit{in-situ} with \textit{ex-situ} mass (via mergers) contributing little to the total mass budget. This is explained by these galaxies on average appearing to have not undergone a major merger event recently. 
    \item Using a combination of diagnostics, AGN activity is found within 8/18 galaxies in our observed sample. This means that these galaxies have had a source of fuel for the AGN so are likely to contain some non-negligible gas reservoirs.

    \item The sizes and morphologies of the observed quiescent galaxies are explored. Every quiescent galaxy in our sample appears to be some form of spheroid with a \sersic index above 2. Many appear to have high-\sersic indices above 5, showing a compact central light distribution. 
    In many cases, it is likely to trace the AGN component. 
    
\end{itemize}

\section*{Acknowledgements}

WMB, FDE, RM, T.J.L, and JS acknowledge support by the Science and Technology Facilities Council (STFC), by the ERC through Advanced Grant 695671 ``QUENCH'', and by the UKRI Frontier Research grant RISEandFALL.
This work was supported by a research grant (VIL54489) from VILLUM FONDEN.
RM also acknowledges funding from a research professorship from the Royal Society.
SA acknowledges grant PID2021-127718NB-I00 funded by the Spanish Ministry of Science and Innovation/State Agency of Research (MICIN/AEI/ 10.13039/501100011033)
AJB acknowledges funding from the ``FirstGalaxies' Advanced Grant from the European Research Council (ERC) under the European Union’s Horizon 2020 research and innovation program (Grant agreement No. 789056).
S.C acknowledges support by European Union’s HE ERC Starting Grant No. 101040227 - WINGS.
XJ acknowledges support by JWST/NIRCam contract to the University of Arizona, NAS5-02015
BER acknowledges support from the NIRCam Science Team contract to the University of Arizona, NAS5-02015, and JWST Program 3215.
The authors acknowledge use of the lux supercomputer at UC Santa Cruz, funded by NSF MRI grant AST 1828315.
H{\"U} gratefully acknowledges support by the Isaac Newton Trust and by the Kavli Foundation through a Newton-Kavli Junior Fellowship.
The research of CCW is supported by NOIRLab, which is managed by the Association of Universities for Research in Astronomy (AURA) under a cooperative agreement with the National Science Foundation.
CNAW acknowledges support from JWST/NIRCam contract to the University of Arizona NAS5-02015.
JZ acknowledges support from JWST/NIRCam contract to the University of Arizona NAS5-02015.
This work used the DiRAC@Durham facility managed by the Institute for Computational Cosmology on behalf of the STFC DiRAC HPC Facility (www.dirac.ac.uk). The equipment was funded by BEIS capital funding via STFC capital grants ST/K00042X/1, ST/P002293/1, ST/R002371/1 and ST/S002502/1, Durham University and STFC operations grant ST/R000832/1. DiRAC is part of the National e-Infrastructure.
This work is based [in part] on observations made with the NASA/ESA/CSA James Webb Space Telescope. The data were obtained from the Mikulski Archive for Space Telescopes at the Space Telescope Science Institute, which is operated by the Association of Universities for Research in Astronomy, Inc., under NASA contract NAS 5-03127 for JWST. These observations are associated with programs \#1180, \#1181, \#1210, \#1286 \#1895, \#1963, and \#3215. . The data is from the JADES survey with the associated MAST doi:10.17909/8tdj-8n28.

%%%%%%%%%%%%%%%%%%%%%%%%%%%%%%%%%%%%%%%%%%%%%%%%%%
\section*{Data Availability}

The data underlying this article will be shared on reasonable request
to the corresponding author. The JADES data is publicly available at https://archive.stsci.edu/hlsp/jades/.



\bibliographystyle{mnras}
\bibliography{refs} % if your bibtex file is called example.bib

%\begin{thebibliography}{99}
%\end{thebibliography}

%%%%%%%%%%%%%%%%%%%%%%%%%%%%%%%%%%%%%%%%%%%%%%%%%%

%%%%%%%%%%%%%%%%% APPENDICES %%%%%%%%%%%%%%%%%%%%%

\appendix

\section{How deep can we go? Brief exploration of quiescent satellite galaxies}\label{app:satellites}

\begin{figure}
  \includegraphics[width=\columnwidth]{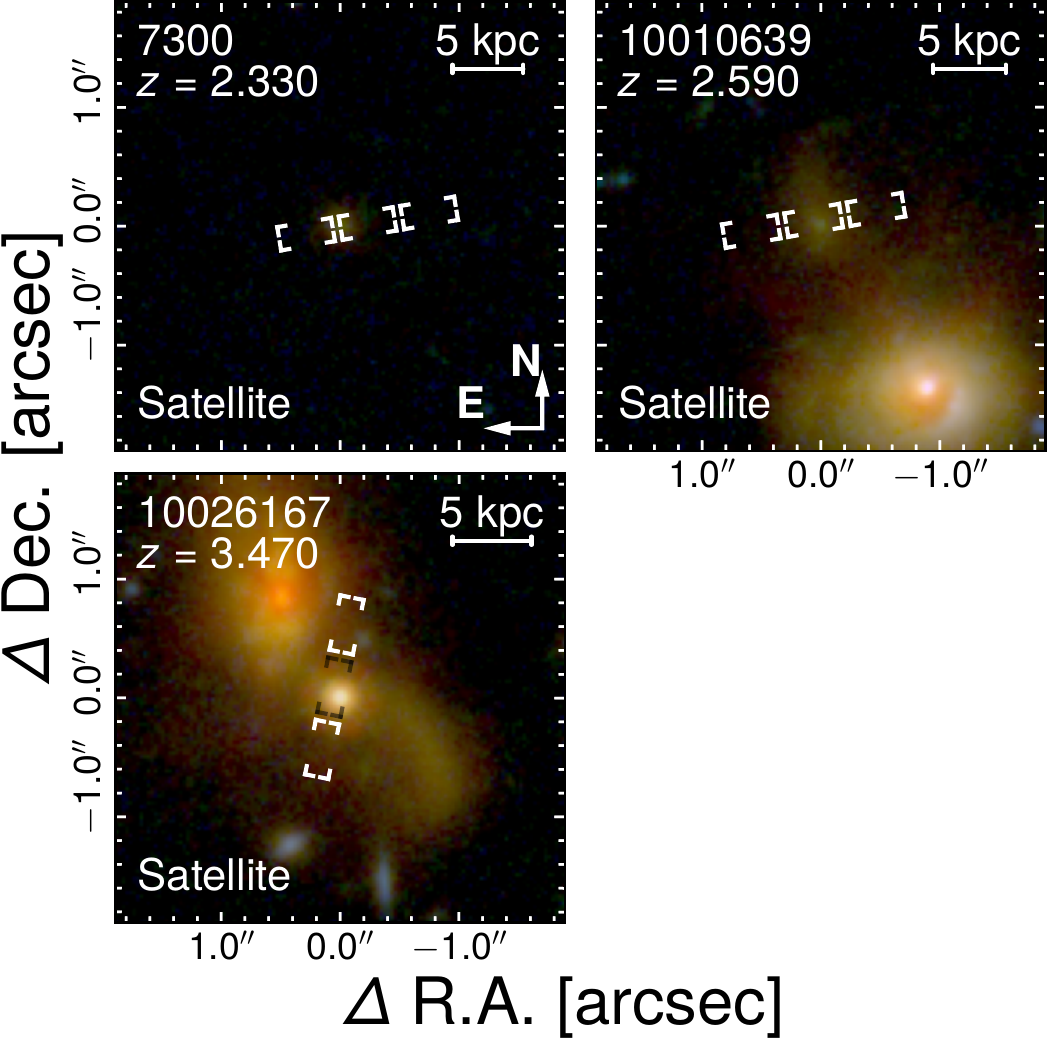}
  \caption{False-colour images of three quiescent galaxies observed in JADES, but not part of our sample due to being spectroscopically confirmed satellites. In addition, IDs 7300 and 10010639 do not meet our mass-cut.
  The deep tiers in JADES are sufficiently sensitive to measure the SFHs of even low-mass satellites. The lack of low-mass quiescent centrals suggests that quiescence in centrals is subject to a threshold in stellar-mass or in a quantity that correlates with stellar mass.
  Red, green and blue colours are from F444W, F200W and F090W NIRCam imaging.
  All symbols are the same as Fig.~\ref{fig:rgb}.
  }\label{fig:rgb.satellites}.
\end{figure}

\begin{figure*}
    \includegraphics[width=\columnwidth]{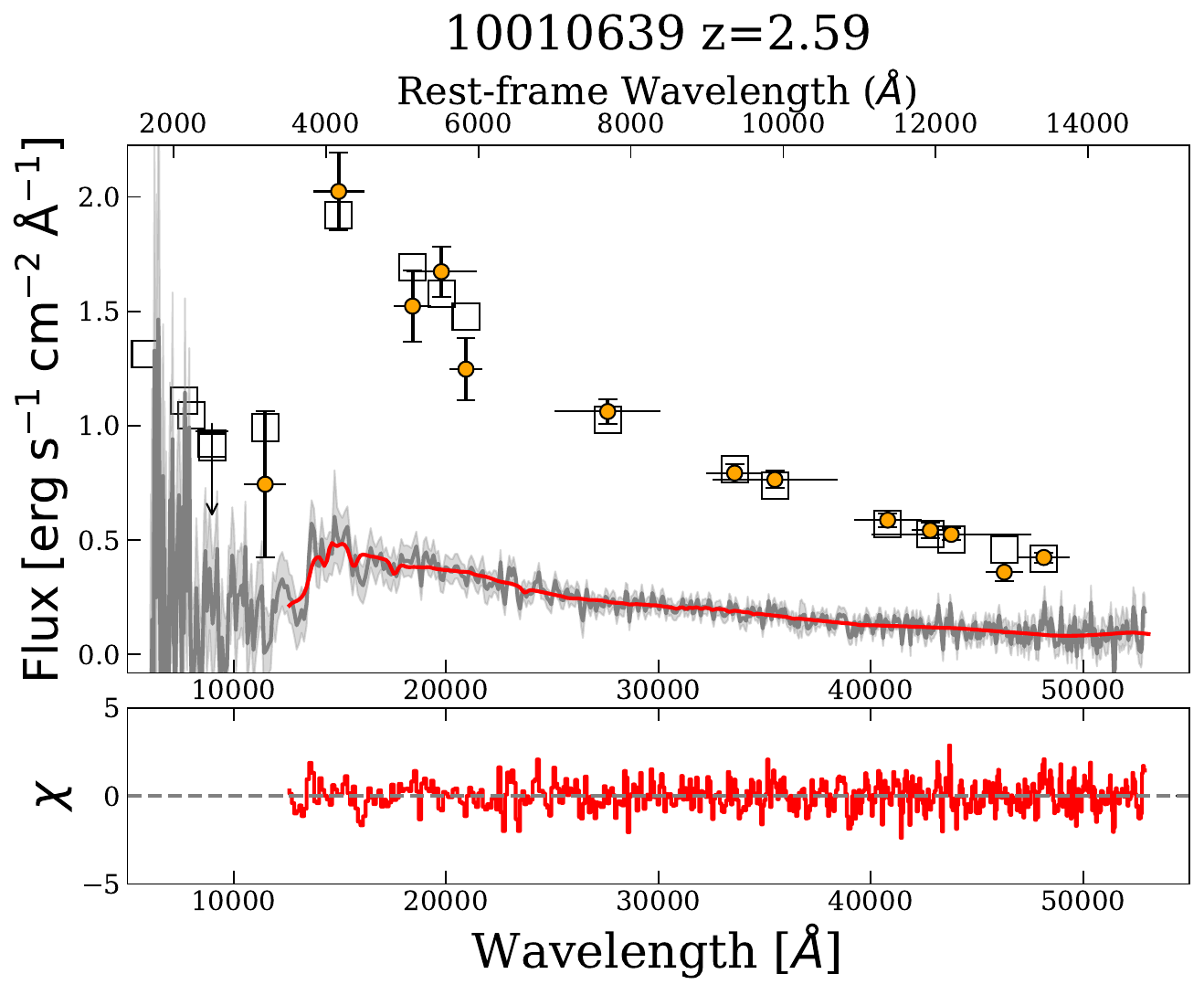}
    \includegraphics[width=\columnwidth]{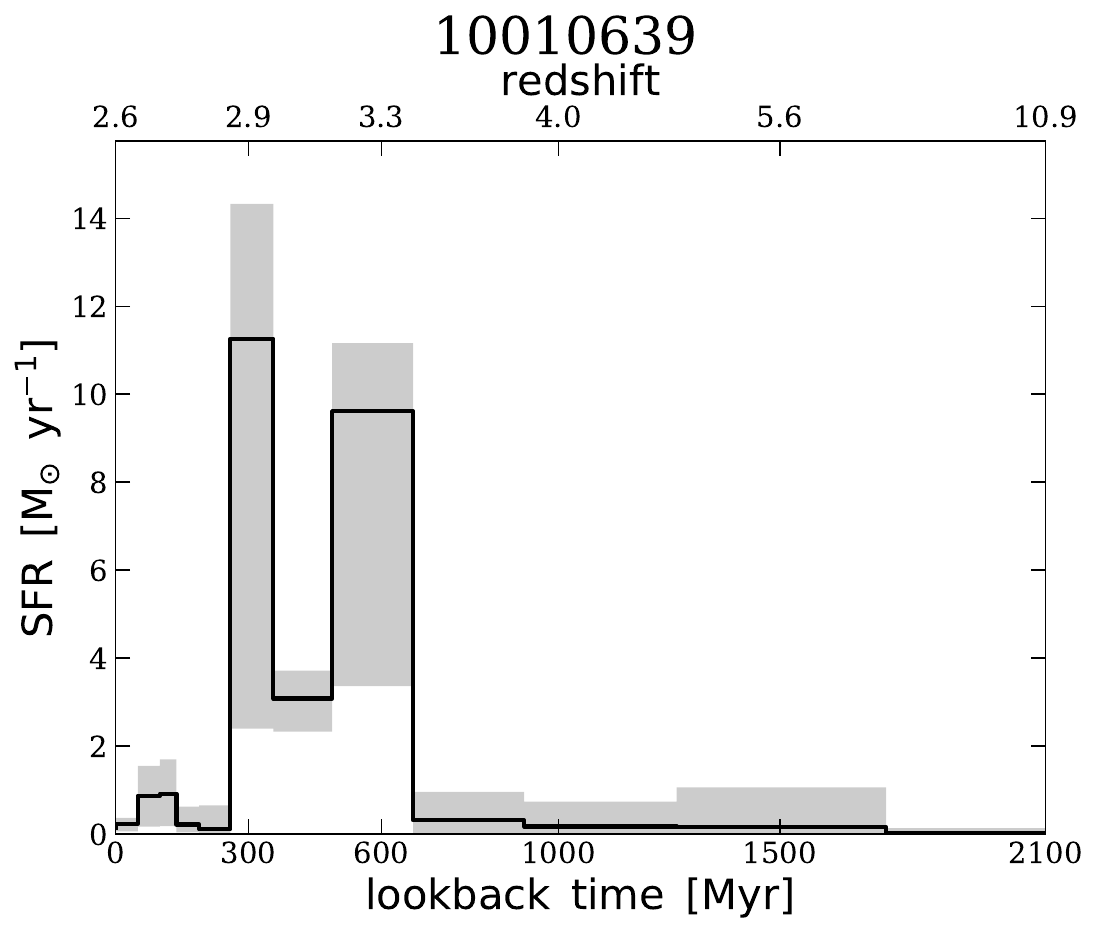}
    \includegraphics[width=\columnwidth]{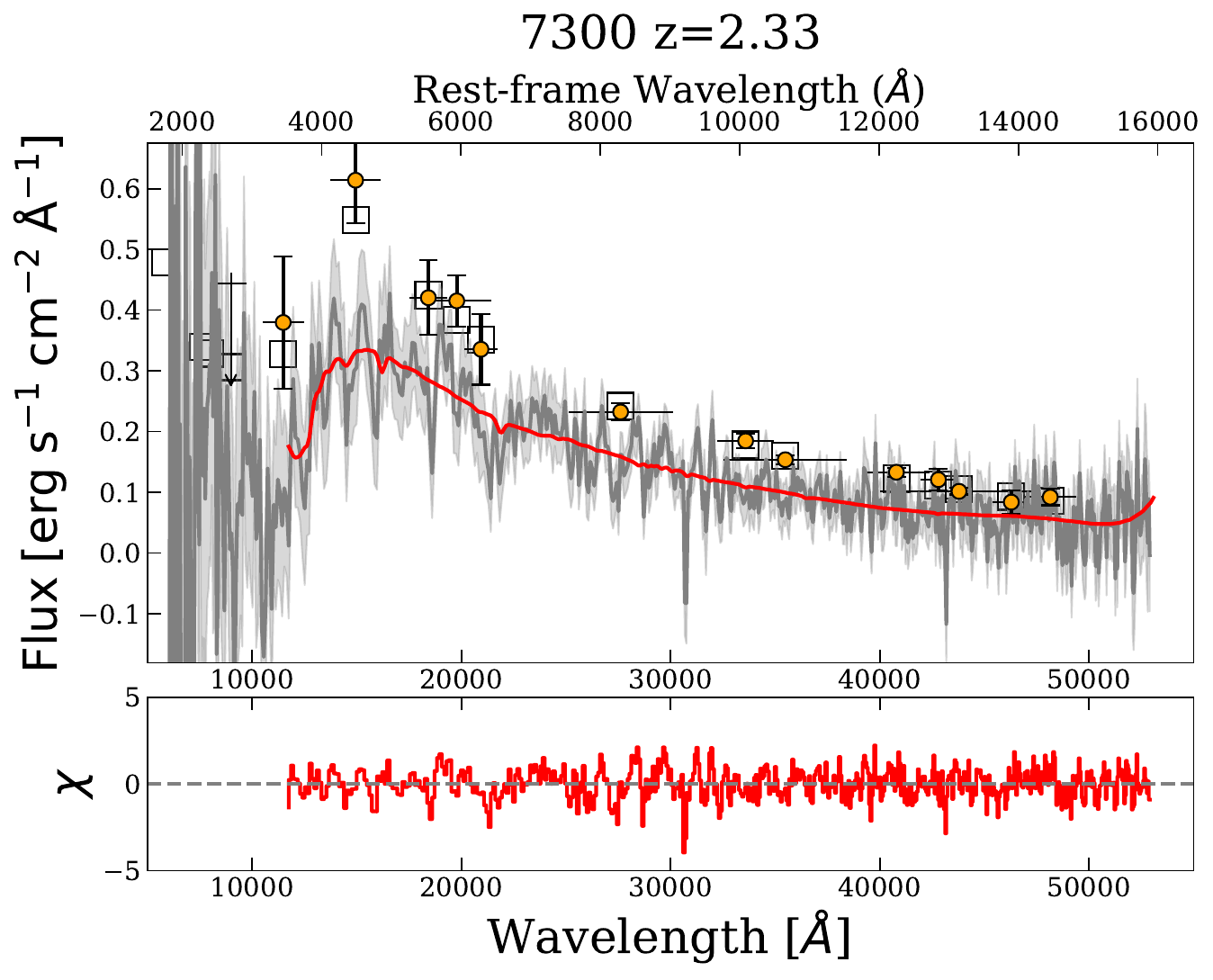}
    \includegraphics[width=\columnwidth]{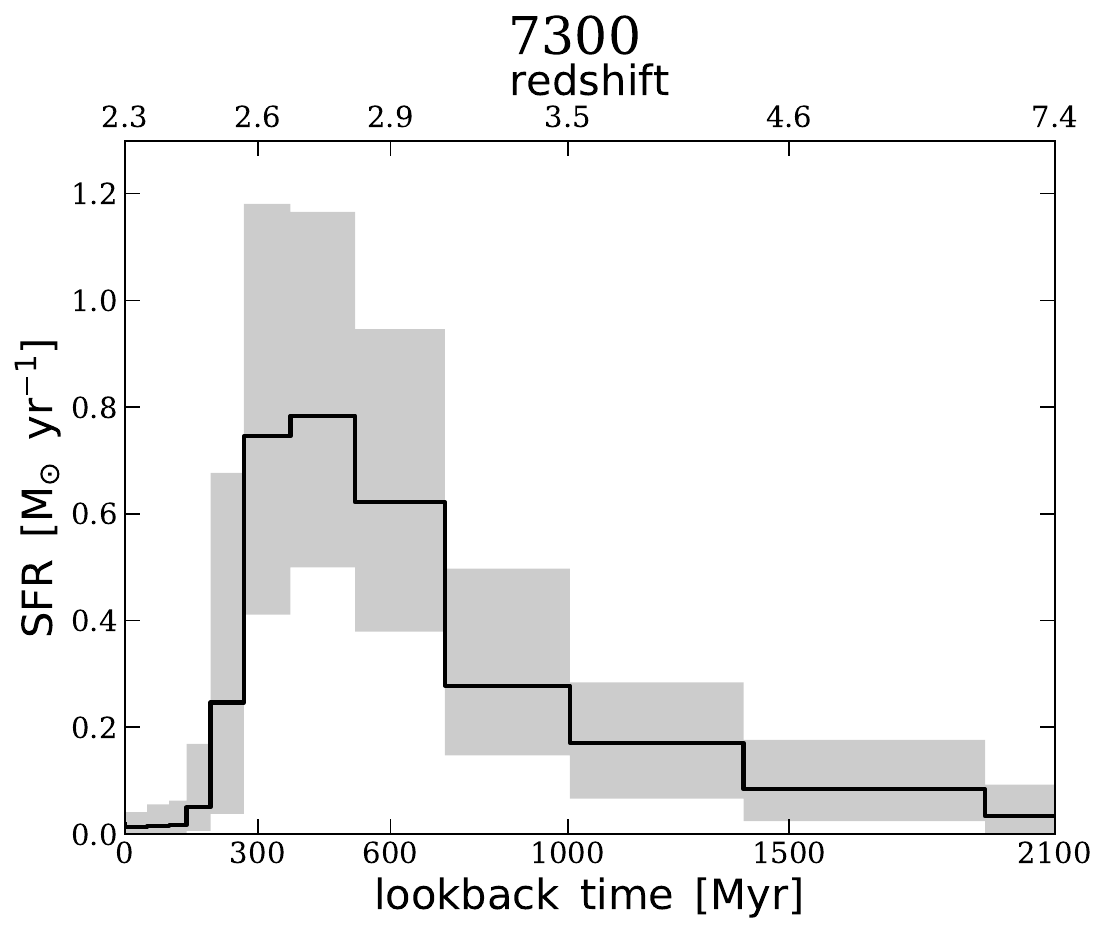}
    \includegraphics[width=\columnwidth]{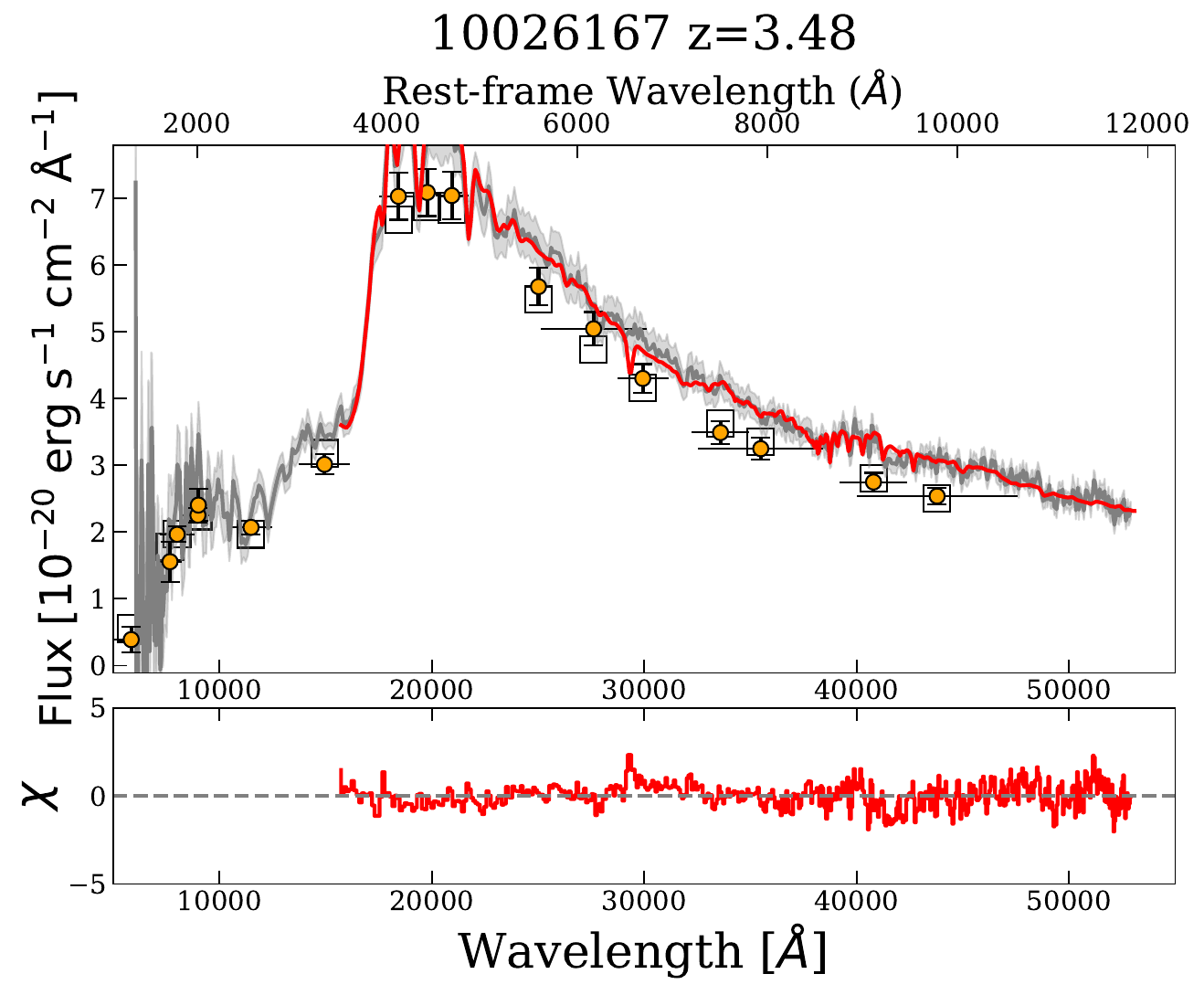}
    \includegraphics[width=\columnwidth]{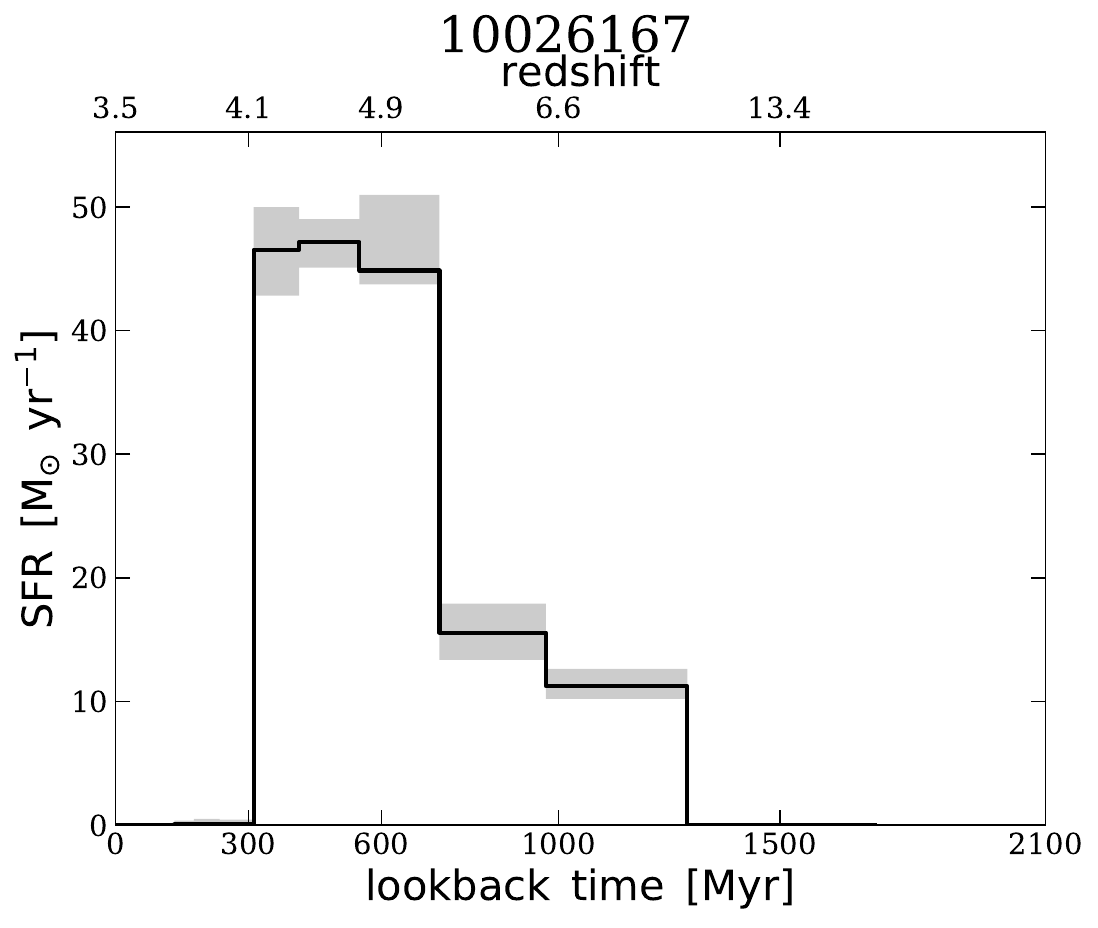}
    
    \caption{Left panels: NIRSpec prism spectra for the three high-redshift quiescent satellite galaxies found in addition to the centrals in our main sample. The grey is the intrinsic spectrum and errors, while the red corresponds to the best-fit spectrum. The yellow points are the photometry, and the black squares the best-fit model photometry. The chi values correspond to the best-fit spectrum. We marginalise over the emission lines because we cannot fit these for AGN - this is not taken into account in the $\chi$ figure. Right panels: SFHs for the three satellite galaxies showing their quiescent nature.}
    \label{fig:sats}
\end{figure*}

Although the goal of this paper is to explore the massive quiescent galaxy population within GOODS-S and GOODS-N, in the process of these explorations, we also uncover two lower-mass (likely satellite) quiescent galaxies, 10010639 and 7300 in addition to a higher-mass satellite galaxy 10026167. 
These satellite galaxies are shown as RGB images in Fig. \ref{fig:rgb.satellites}.
Galaxy 7300 has previously been identified and studied in \citet{Sandles2023}, so it is reassuring to rediscover it. We can see it is compact and faint in the imaging. Galaxy 10010639 is as of yet unpublished and appears to also be low-mass. We can clearly see a large central galaxy to the south-east in the imaging.
However, 10026167 appears to be a higher-mass quiescent satellite galaxy with a stellar mass of $M_\star=10^{10.4}\Msun$. In order to check that this is not a central, we fit the photometry of the massive dusty star-forming neighbour 175041 (upper left of the RGB image) in order to obtain a stellar mass estimate. We find 175041 has a stellar mass of $M_\star=10^{11.3}\Msun$ so is clearly a massive central. It also shows signs of considerable star-formation activity with an $\rm SFR_{10}=22^{+8}_{-6}M_\odot /yr^{-1}$. This means that we can be confident that 10026167 is in fact a quiescent satellite of this central galaxy (that is itself not quiescent).
Exact explorations of these three satellite galaxies' environment is beyond the scope of this project but will be followed up in the future. 

Fig. \ref{fig:sats} shows the spectra + photometry and star-formation histories for the two low-mass and one higher mass quiescent satellite galaxies. We can clearly see from the star-formation histories that all three galaxies are quiescent with no signs of recent star-formation. The three appear to have gone through short bursts of star-formation lasting roughly 300Myrs before quenching. We find a stellar mass of $10^{9.4}\Msun$ for 10010639 and $10^{8.6}\Msun$ for 7300, whilst as previously mentioned 10026167 has a stellar mass of $10^{4}\Msun$. These galaxies provide a fascinating avenue for further follow up research to better understand possible environmental quenching.

\section{Tables}

\include{quiescent_tab_18_12Sep}

Table \ref{tab:values_errors} is a table showing the NIRSpec ID (JADES in the case of the JADES spectra and DJA in the case of the four \cite{Barrufet2021} spectra), right-ascension and declination, and the best-fit properties including redshift, stellar mass, formation redshift ($z_{50}$), half-light radius, \sersic index, Balmer break strength and D$_n$4000 index.

We measure the Balmer break strength in $F_\nu$ following the procedure in \citet{Roberts-Borsani2024} and \cite{Witten2024}. Briefly, this is as follows: we use a window between 3500 --3630\AA ($\rm F_\nu(3565\AA)$) and a window between 4160-- 4290\AA ($F_\nu(4225\AA)$) in order to select regions that do not overlap with strong emission lines. We then take the ratio between the two to give the Balmer break strength such that
\begin{equation}
    \rm Balmer\;break\;=F_\nu(4225\AA)/F_\nu(3565\AA).
\end{equation}
The Dn4000 index is measured as defined in \cite{Balogh1999}.

\section{Extra Spectra}

\begin{figure*}
    \centering
    \includegraphics[width=1.\columnwidth]{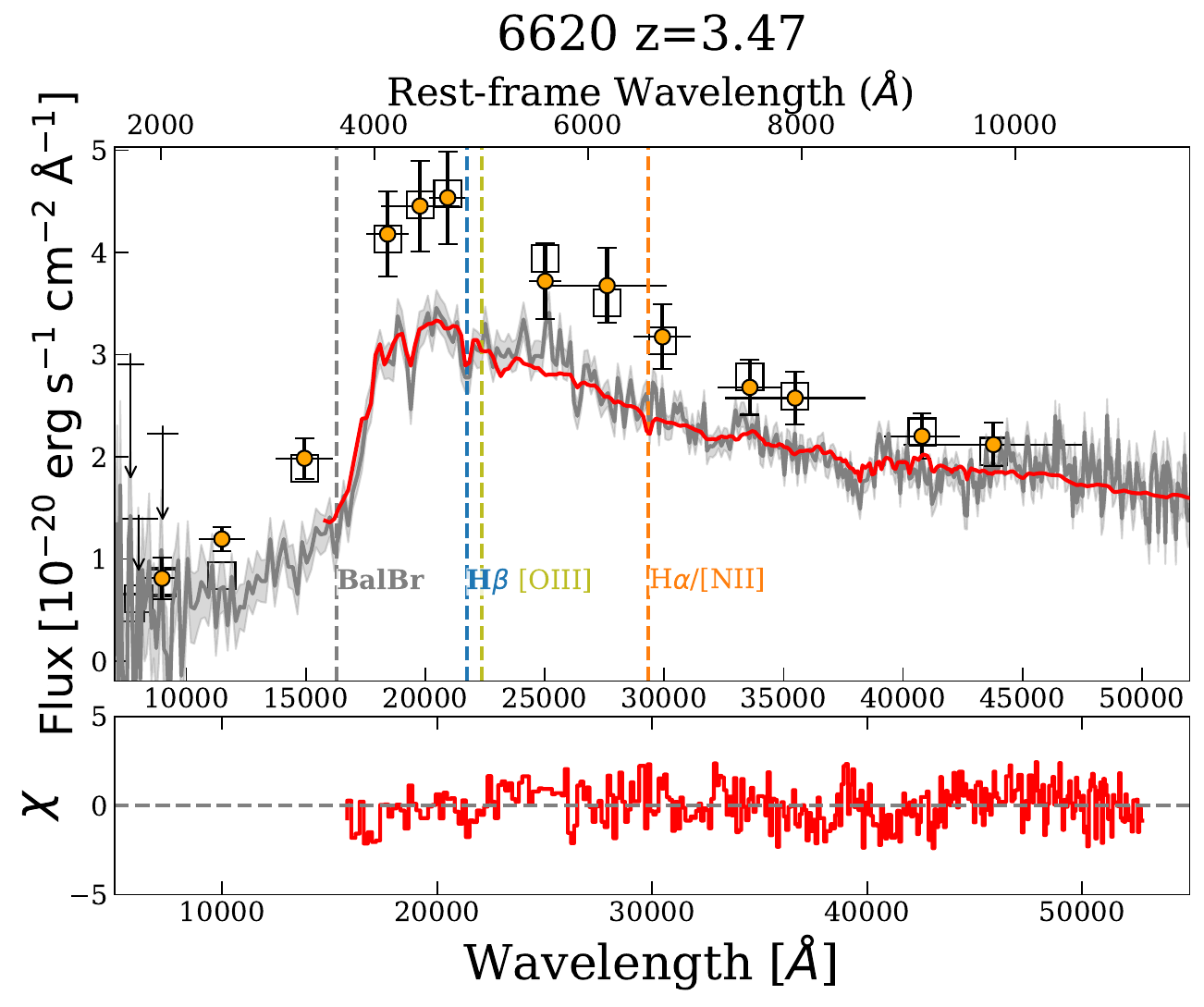}  
    \includegraphics[width=1.\columnwidth]{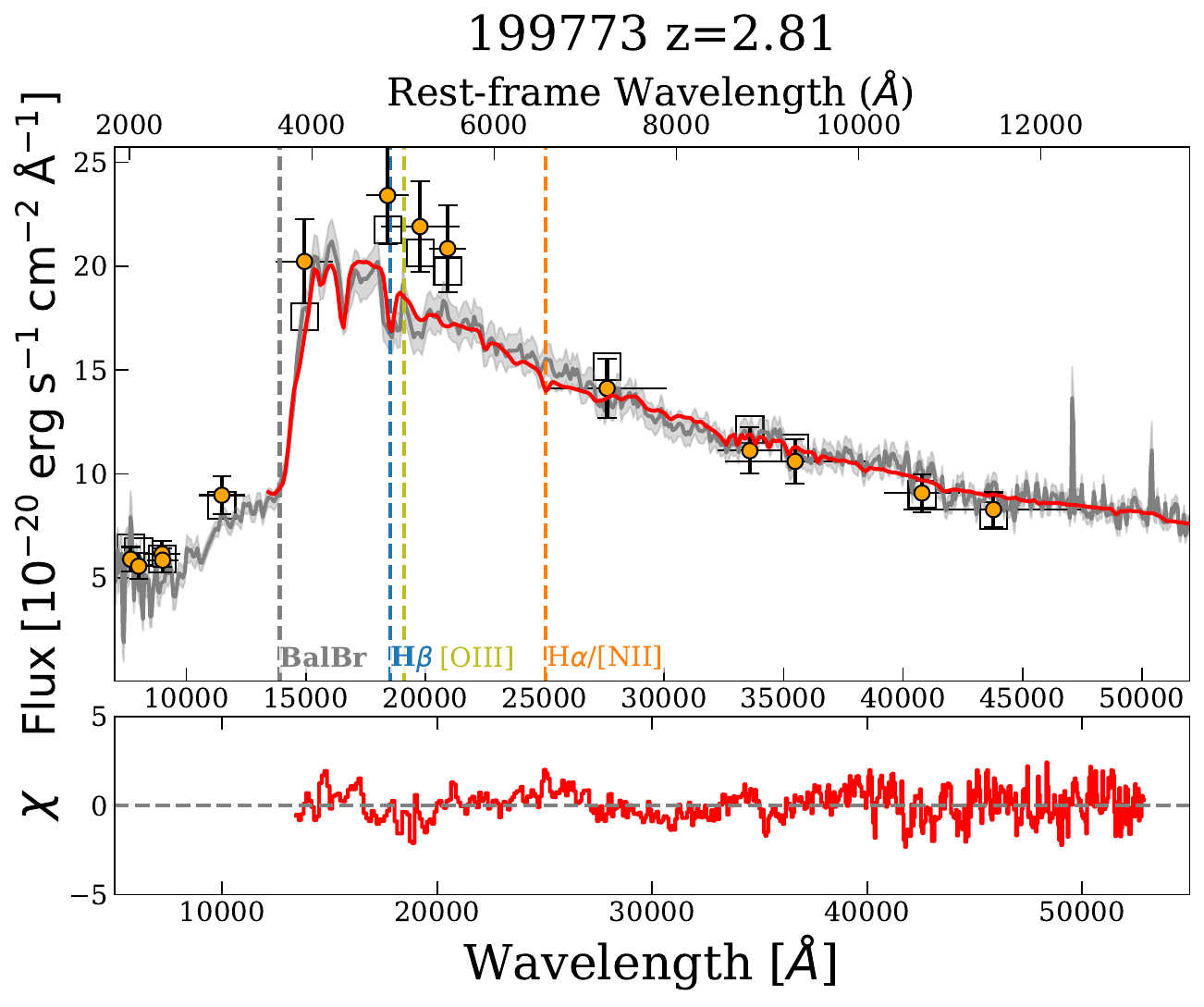}
    \includegraphics[width=1.\columnwidth]{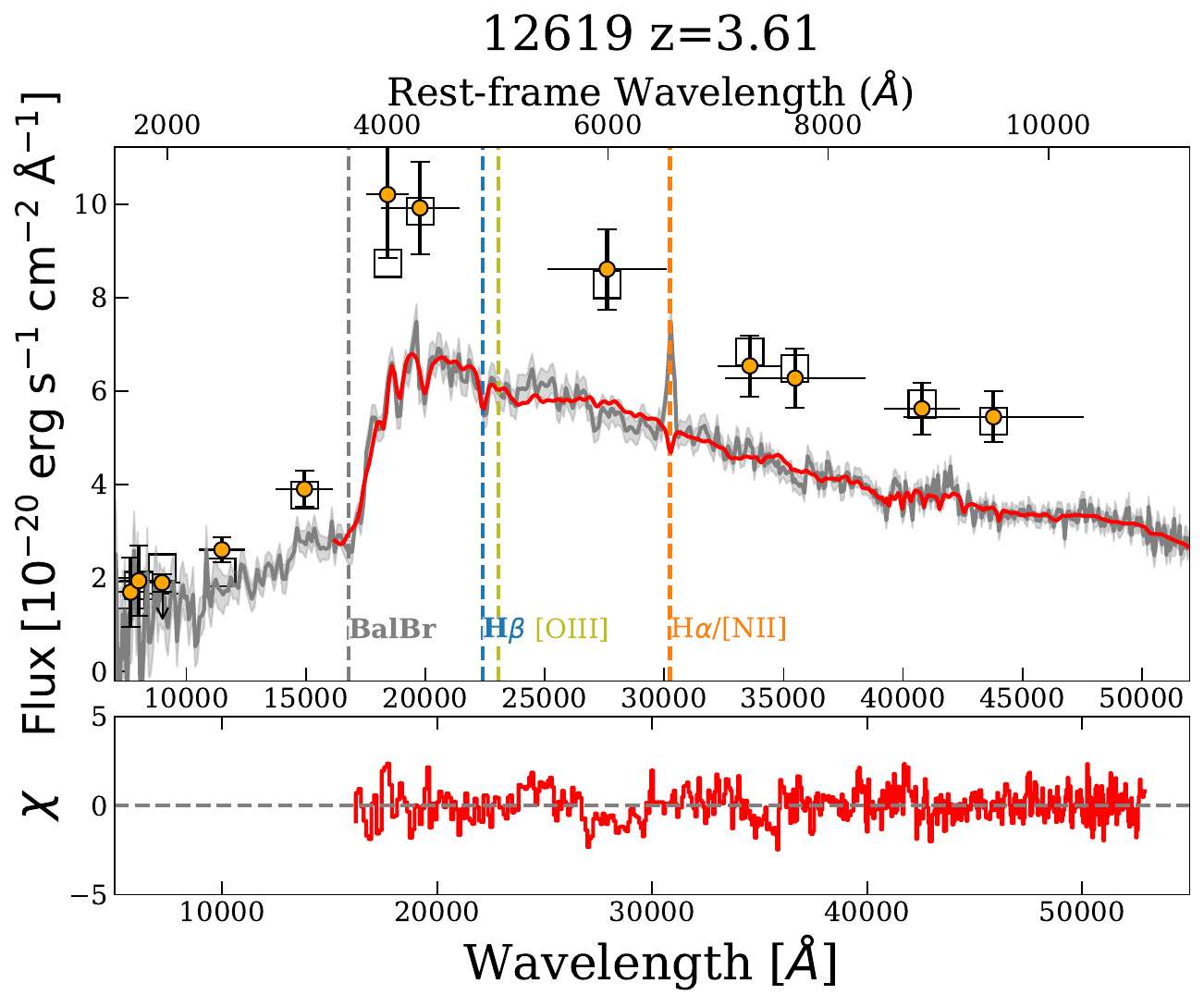}
    \includegraphics[width=1.\columnwidth]{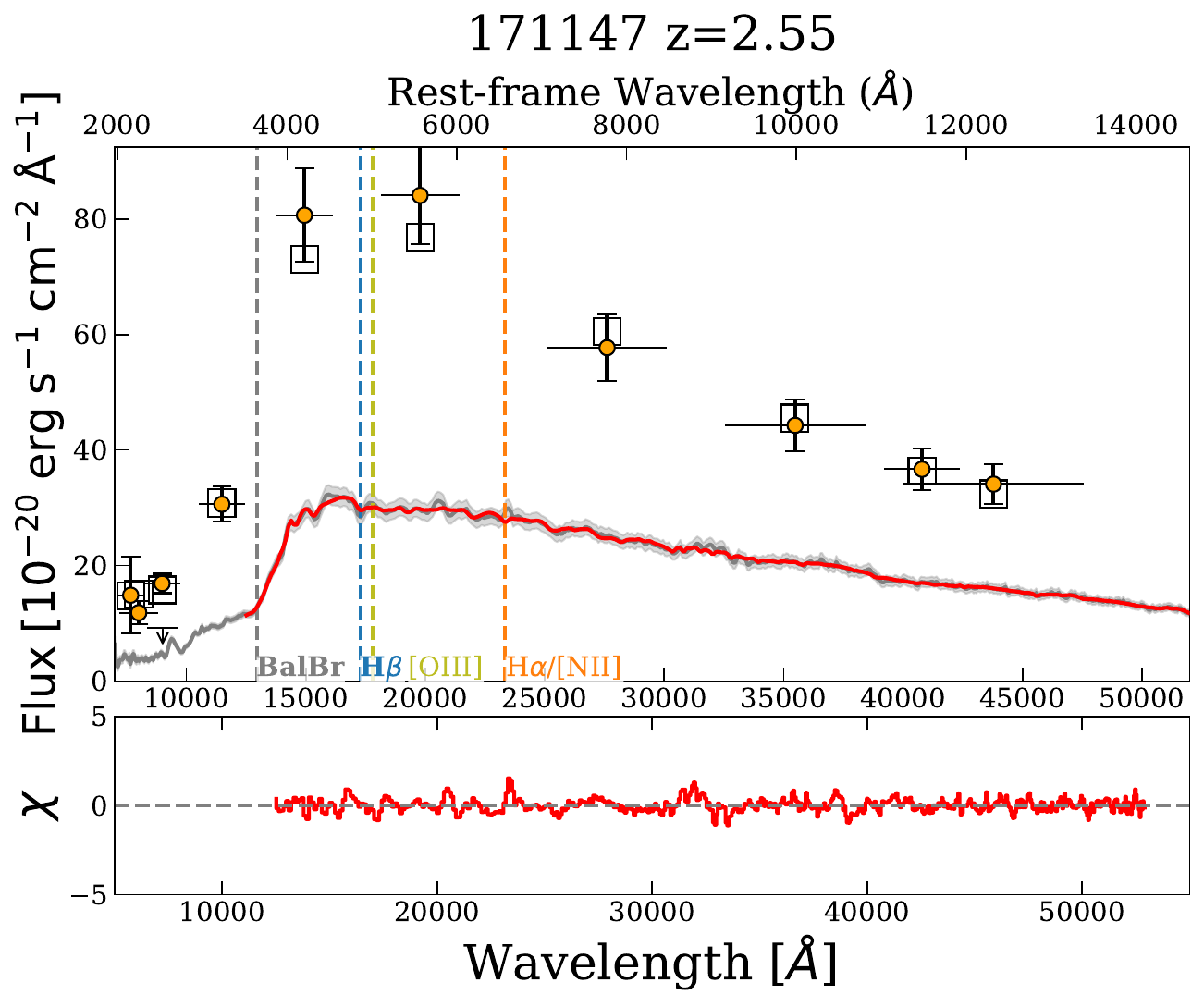}
    \includegraphics[width=1.\columnwidth]{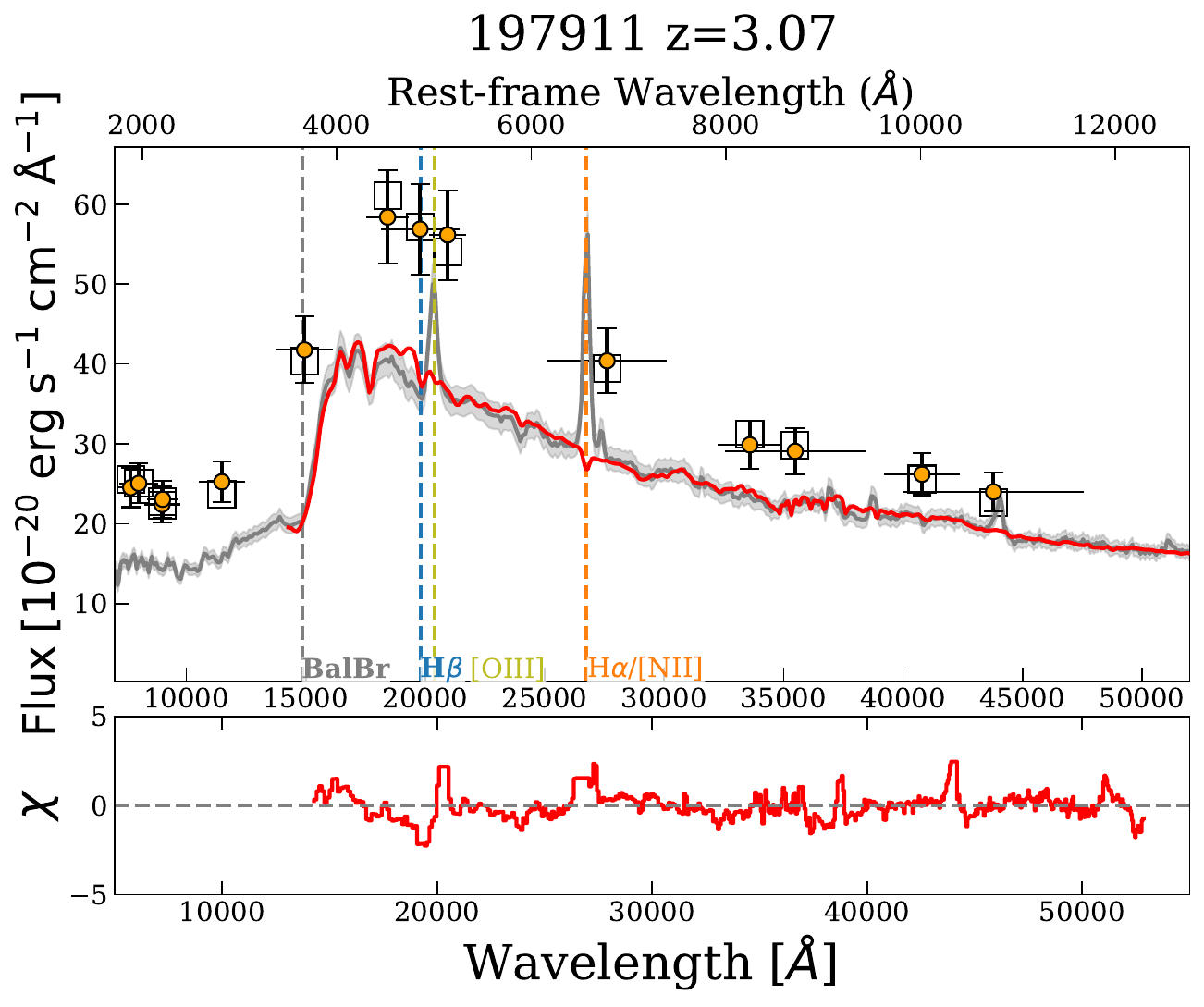}
    \includegraphics[width=1.\columnwidth]{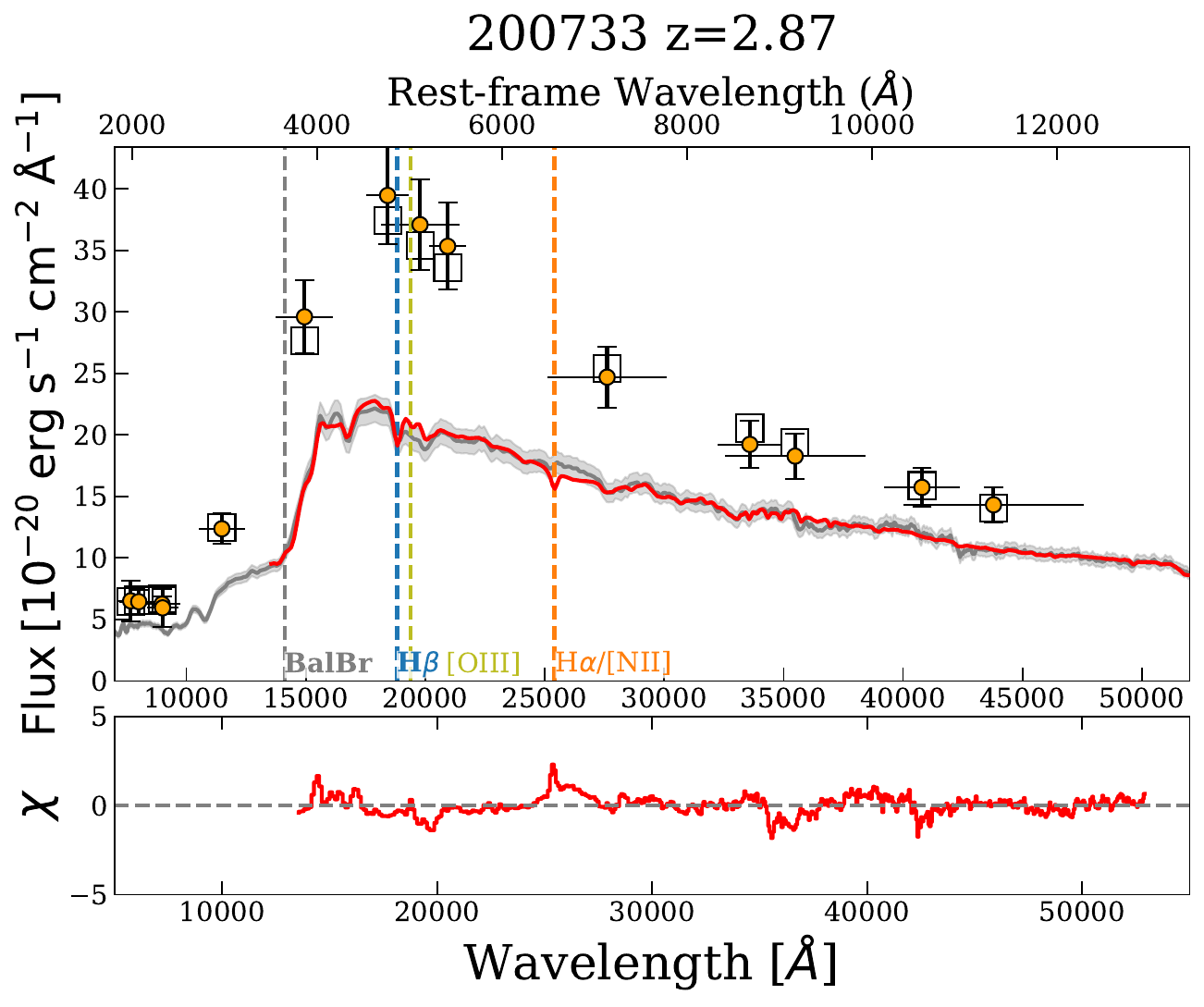}
    
    \caption{NIRSpec prism spectra for six high-redshift quiescent galaxies in the sample. The grey is the intrinsic spectrum and errors, while the red corresponds to the best-fit spectrum. The yellow points are the photometry, and the black squares the best-fit model photometry. The $\chi$ values correspond to the best-fit spectrum. We marginalise over the emission lines because we cannot fit these for AGN - this is not taken into account in the $\chi$ figure. }
    \label{fig:spectra_1}
\end{figure*}

\begin{figure*}
    \centering
    \includegraphics[width=1.\columnwidth]{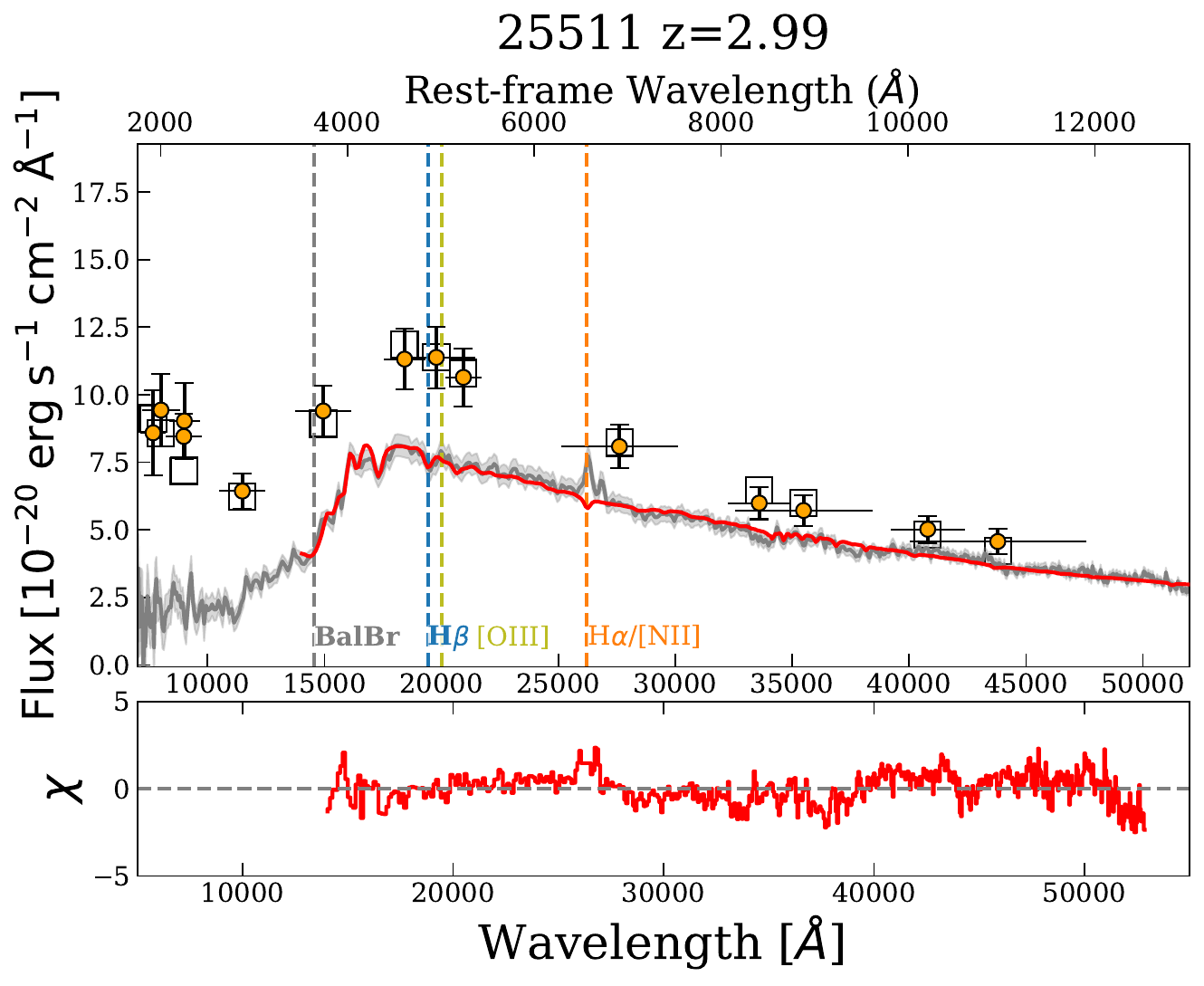}  
    \includegraphics[width=1.\columnwidth]{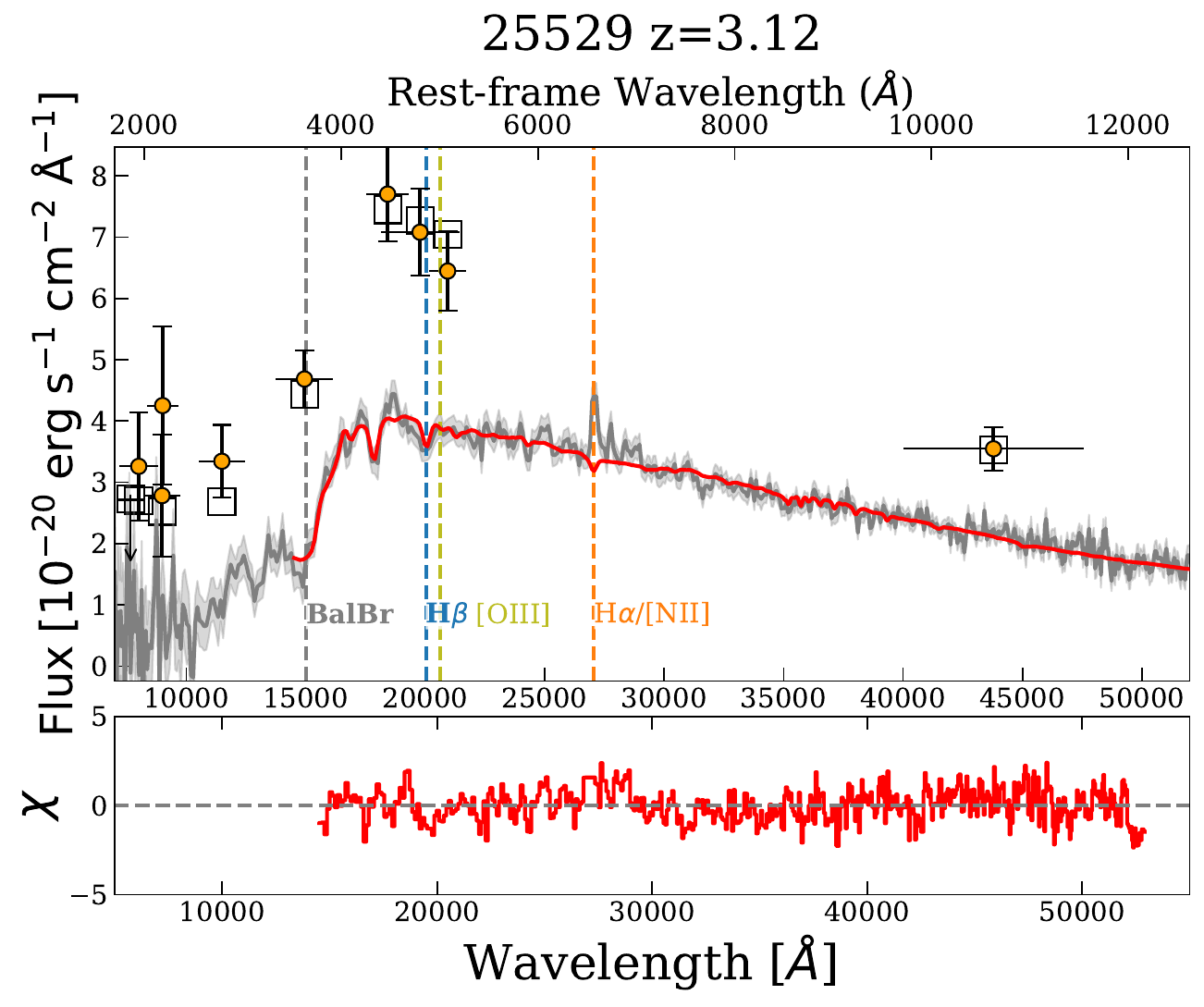}
    \includegraphics[width=1.\columnwidth]{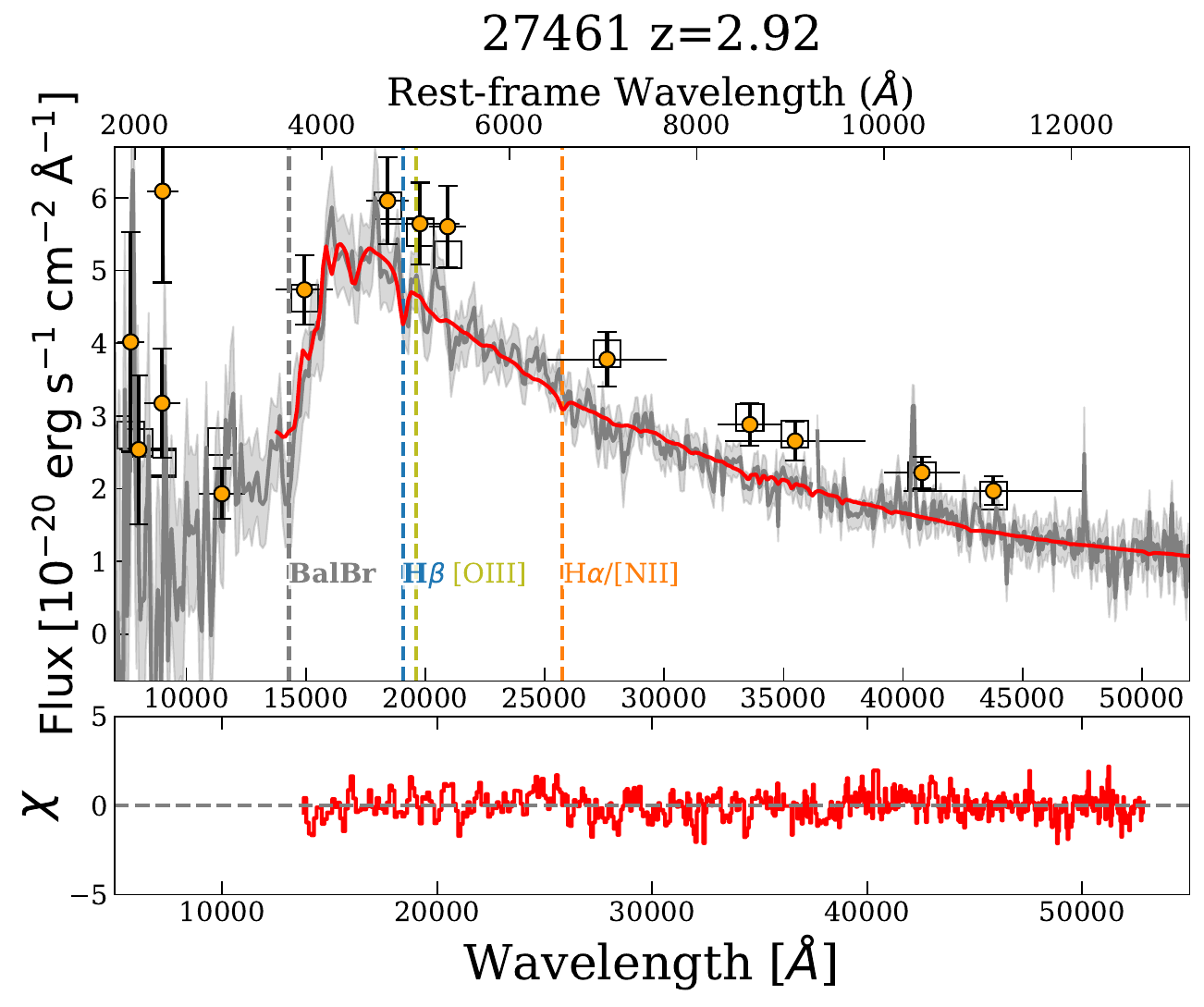}
    \caption{NIRSpec prism spectra for three more high redshift quiescent galaxies in the sample. The grey is the intrinsic spectrum and errors, while the red corresponds to the best-fit spectrum. The yellow points are the photometry, and the black squares the best-fit model photometry. The chi values correspond to the best-fit spectrum. We marginalise over the emission lines because we cannot fit these for AGN - this is not taken into account in the $\chi$ figure. }
    \label{fig:spectra2}
\end{figure*}

Figs. \ref{fig:spectra_1} and \ref{fig:spectra2} contain the observed spectra and photometry and best-fit model spectra and photometry for the remaining nine galaxies in our sample. The other nine galaxies' spectra and photometry appeared in Figs. \ref{fig:spectra} and \ref{fig:D4000spectra}. Once again, the observed spectra and errors are in grey, and the observed photometry is given by the yellow points. The best-fit model spectra is in red, and the best-fit model photometry is given by the black squares. The lower panel shows the $\chi$ values of the fit to the spectra. 

\section{Extra Star-formation Histories}

\begin{figure*}
    
    \centering
    \includegraphics[width=\sfhwidth]{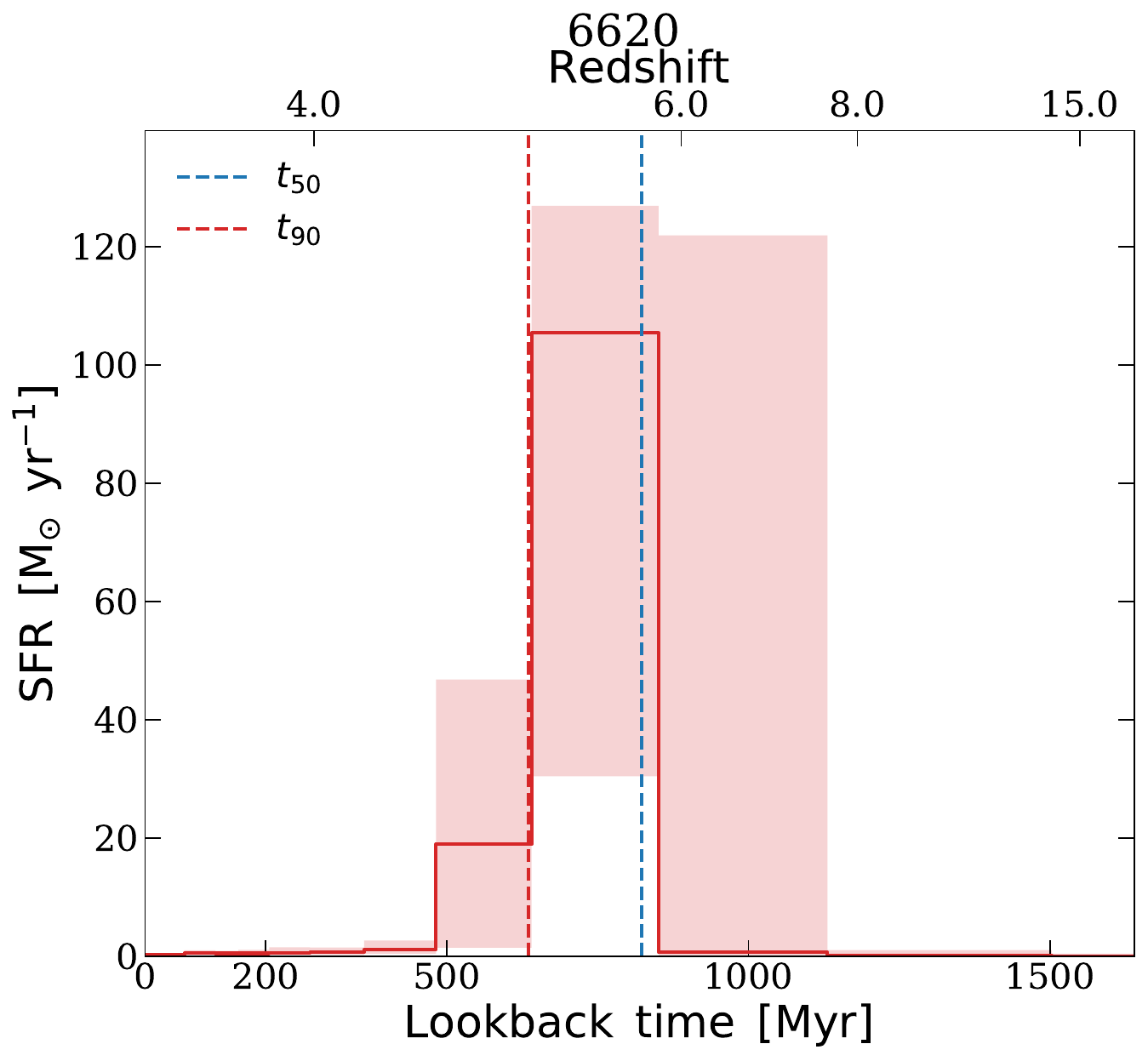}
    \includegraphics[width=\sfhwidth]{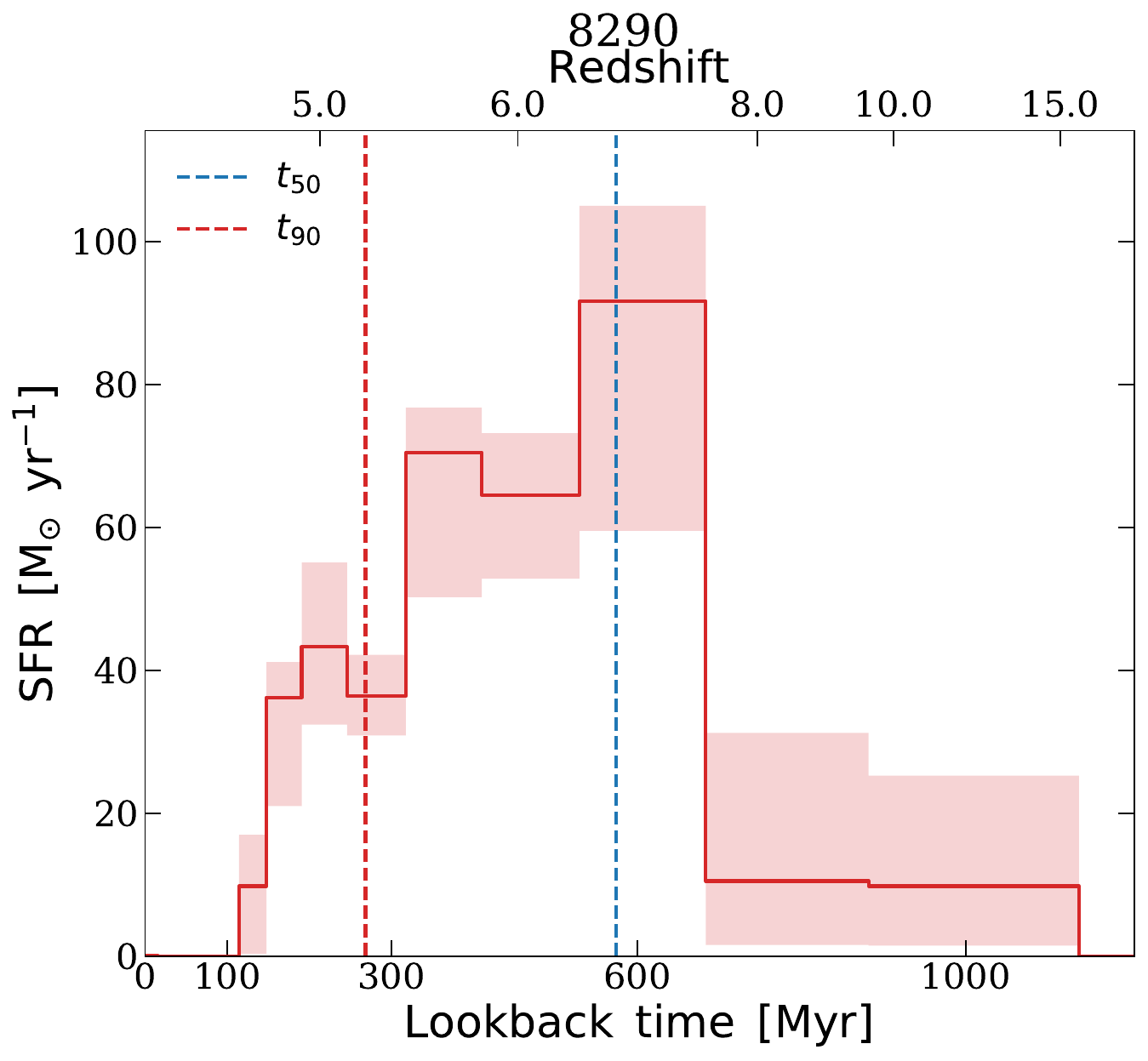}
    \includegraphics[width=\sfhwidth]{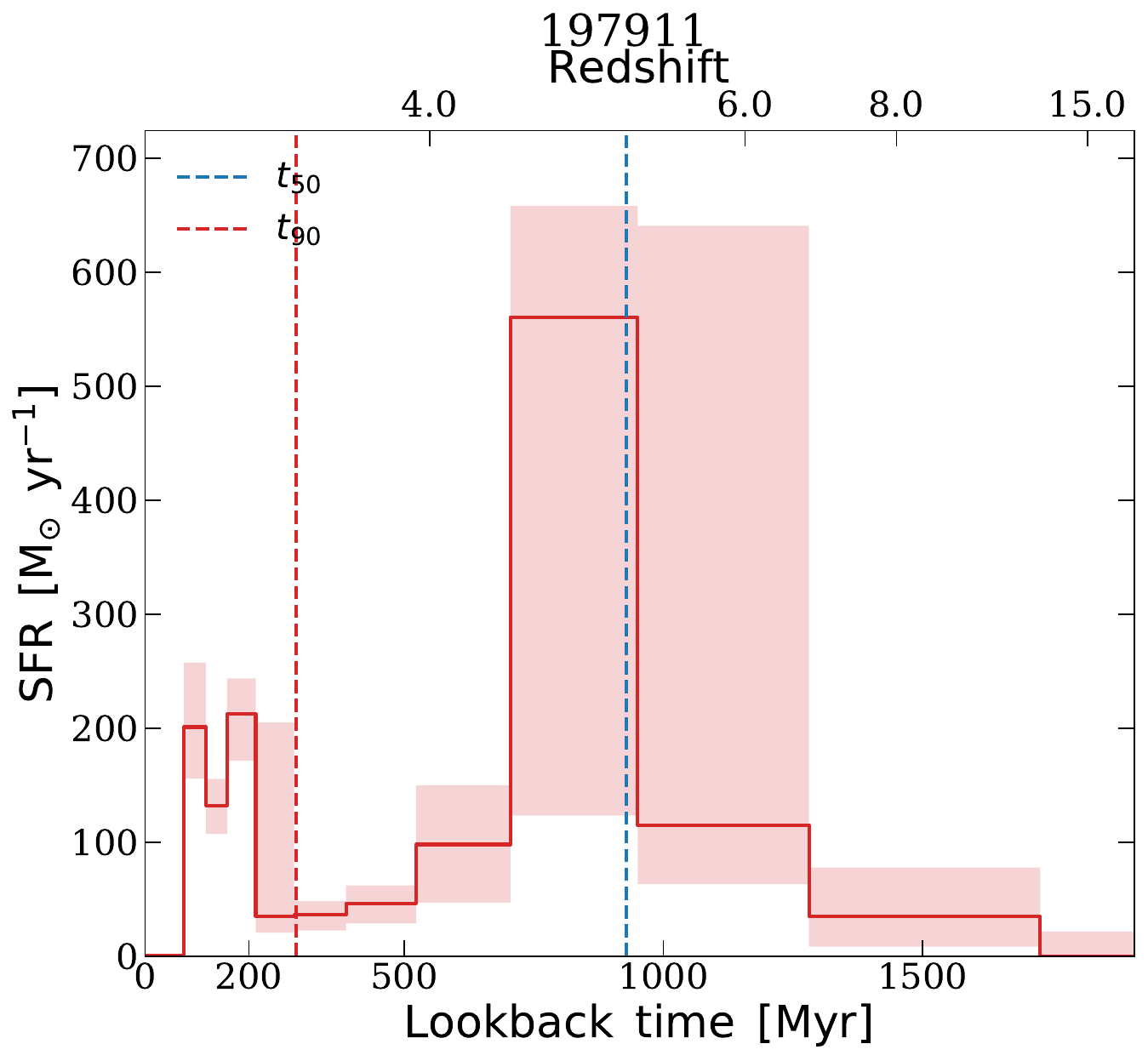}
    \includegraphics[width=\sfhwidth]{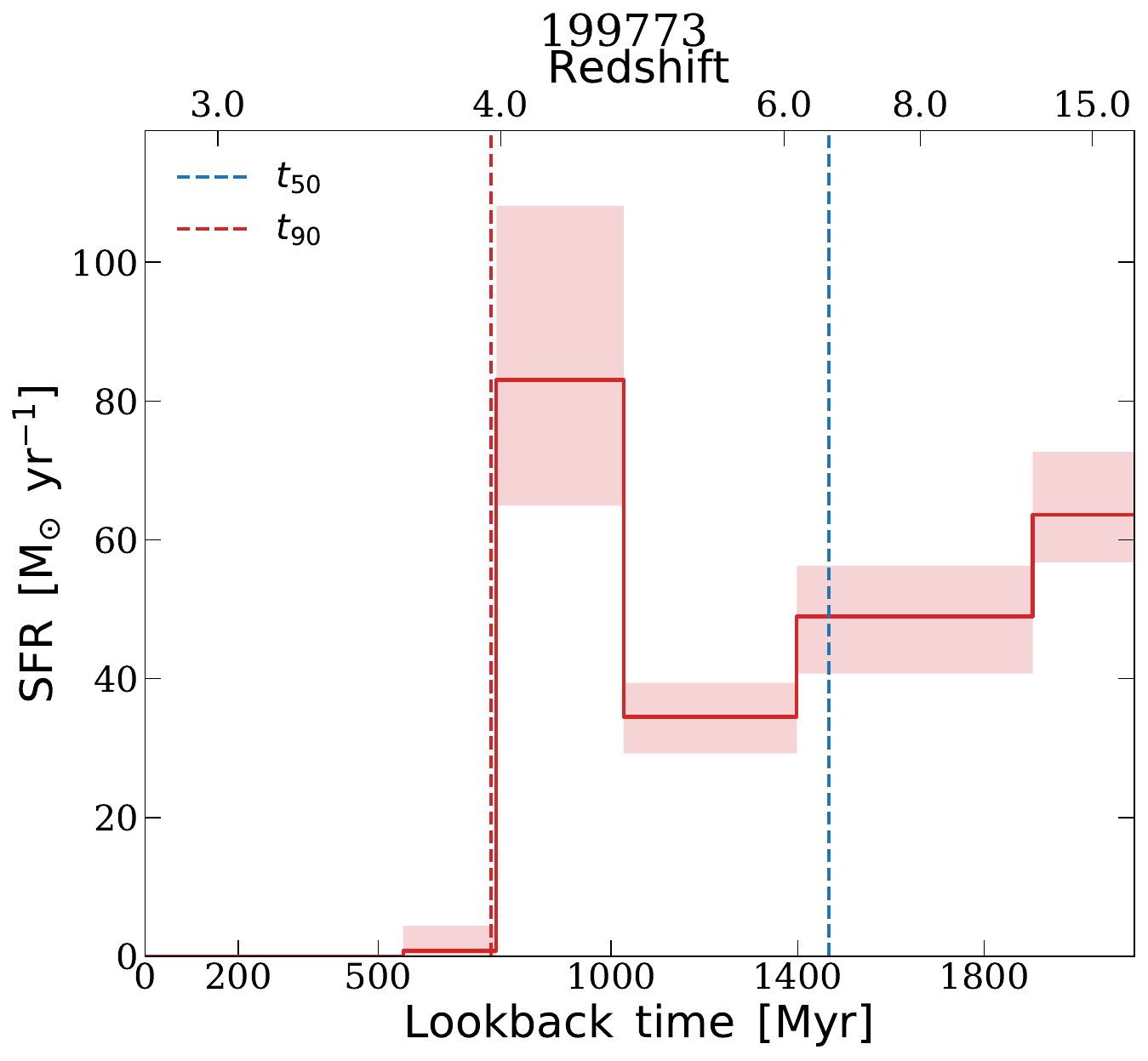}
    \includegraphics[width=\sfhwidth]{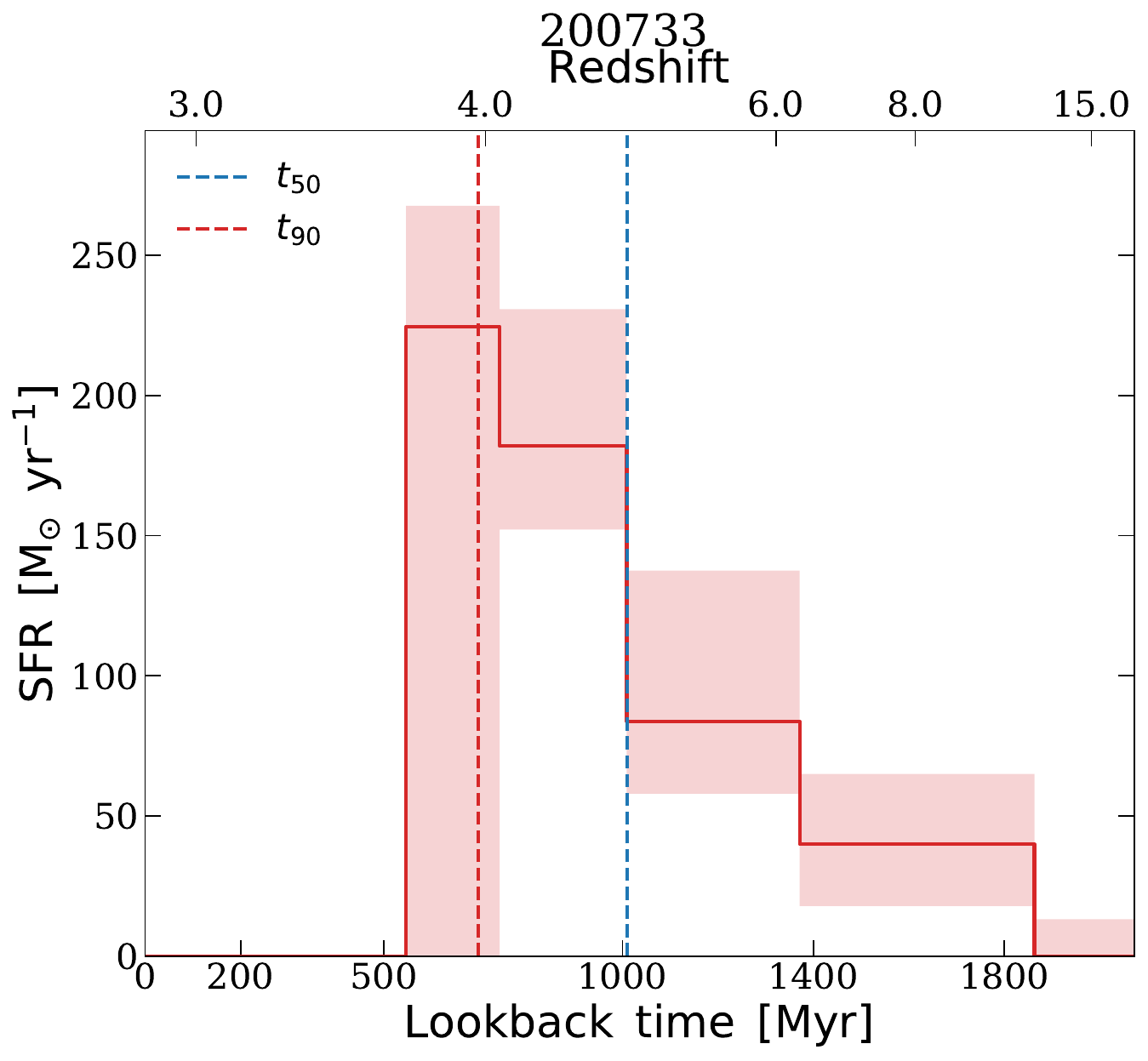}
    \includegraphics[width=\sfhwidth]{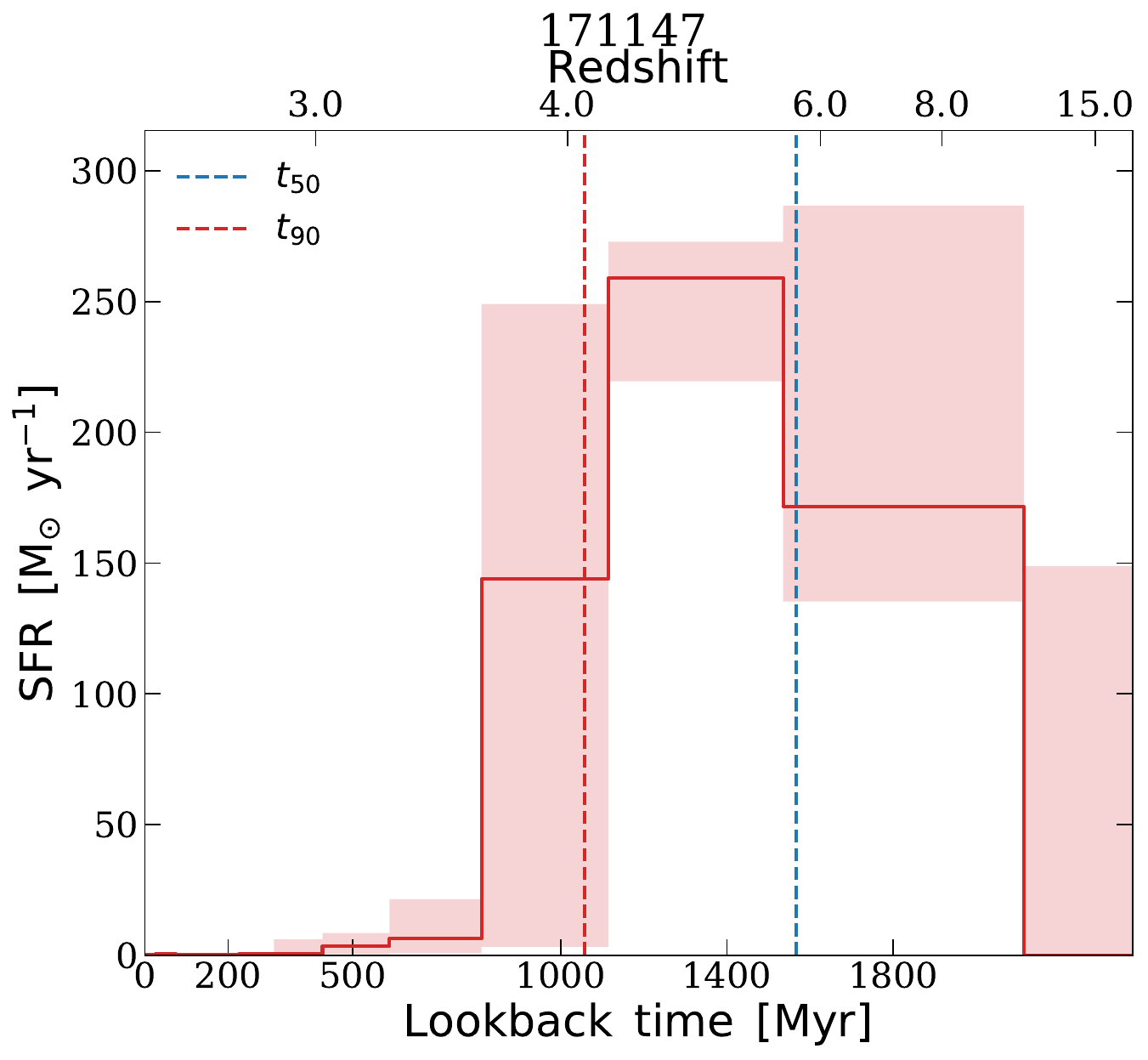}

    \caption{Star-formation rate vs lookback time from observation (redshift) for 6 of the quiescent galaxies. }
    \label{fig:sfhs2}
\end{figure*}

\begin{figure*}
    
    \centering
    
    \includegraphics[width=\sfhwidth]{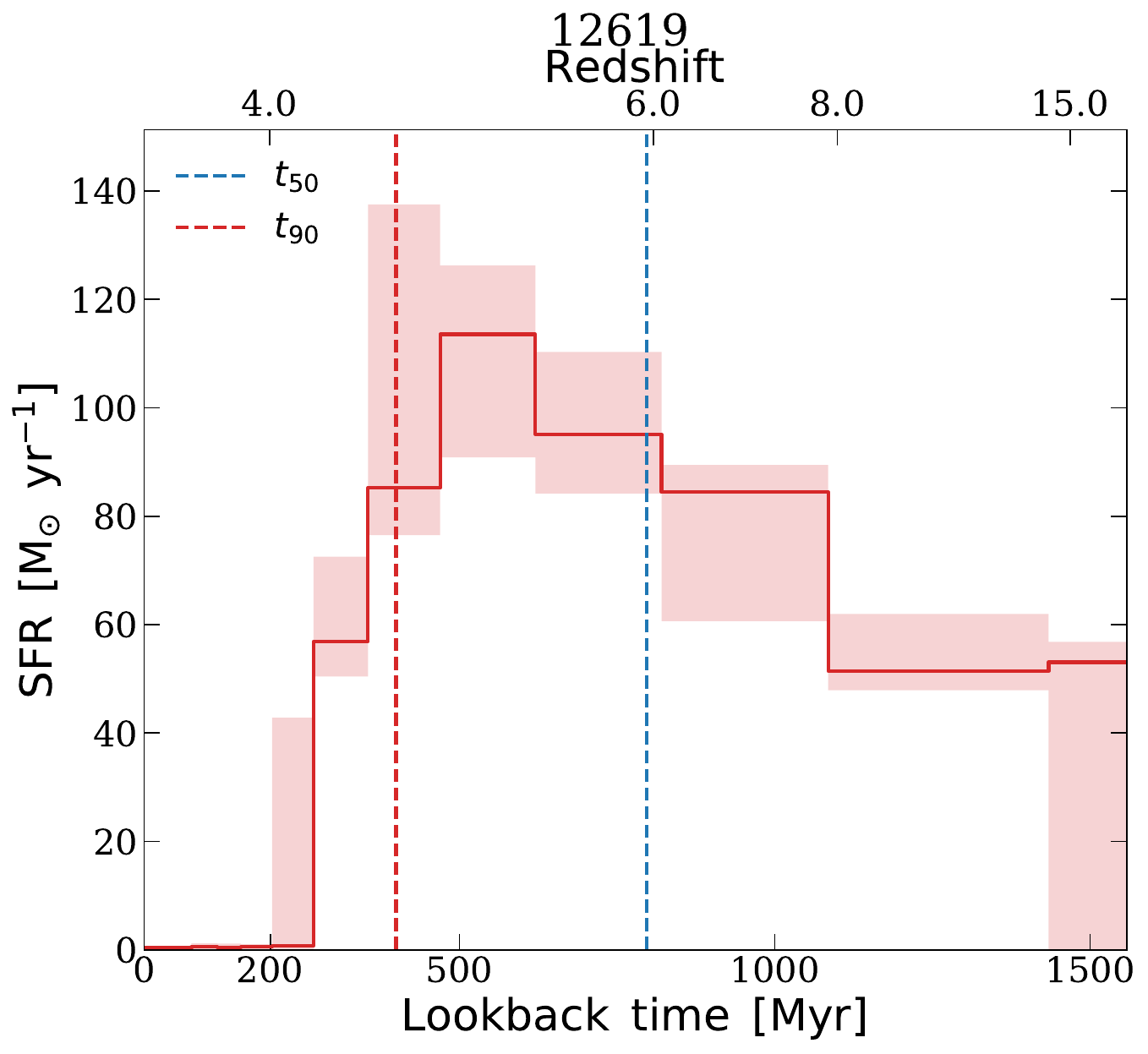}
    \includegraphics[width=\sfhwidth]{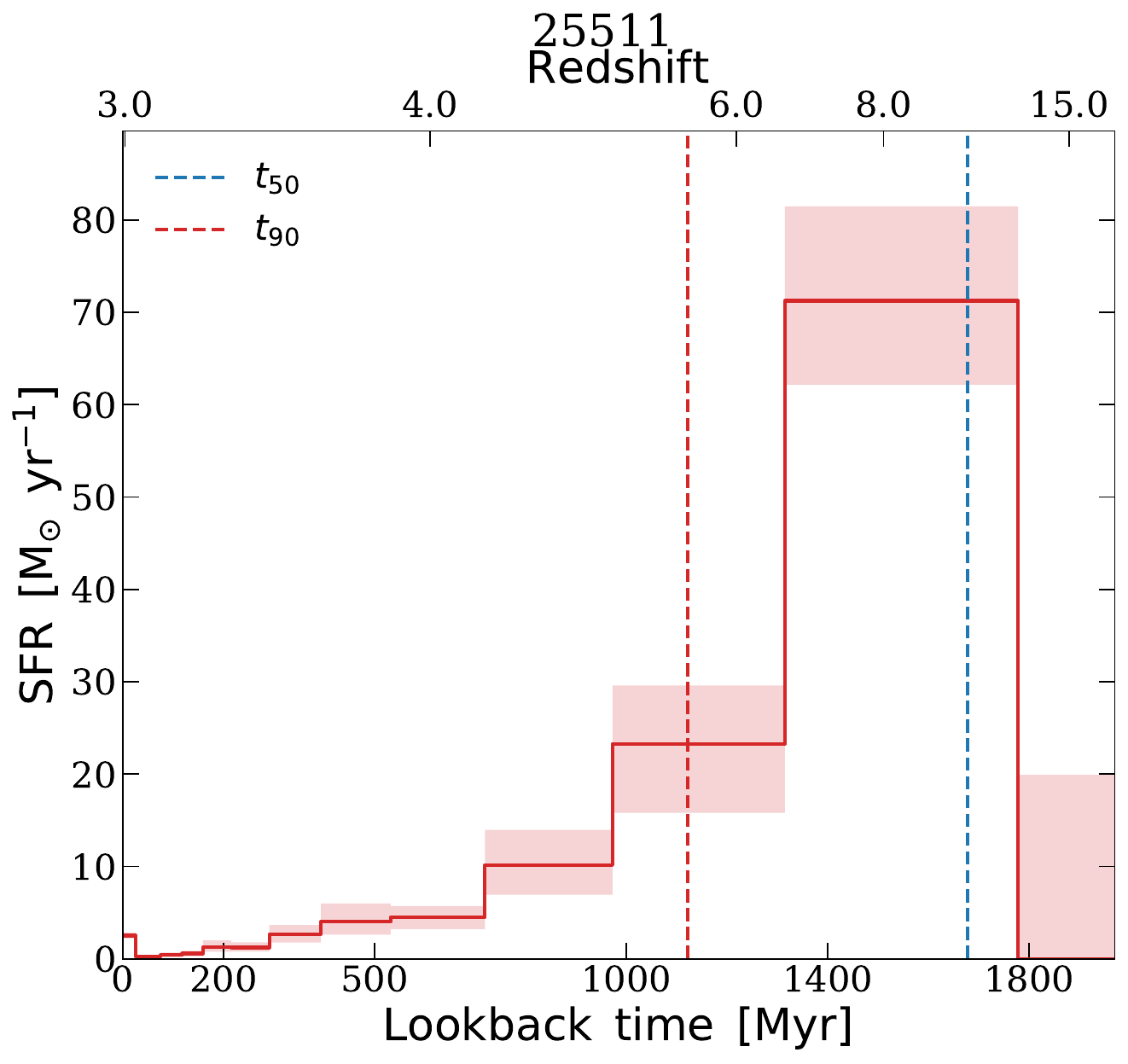}  
    \includegraphics[width=\sfhwidth]{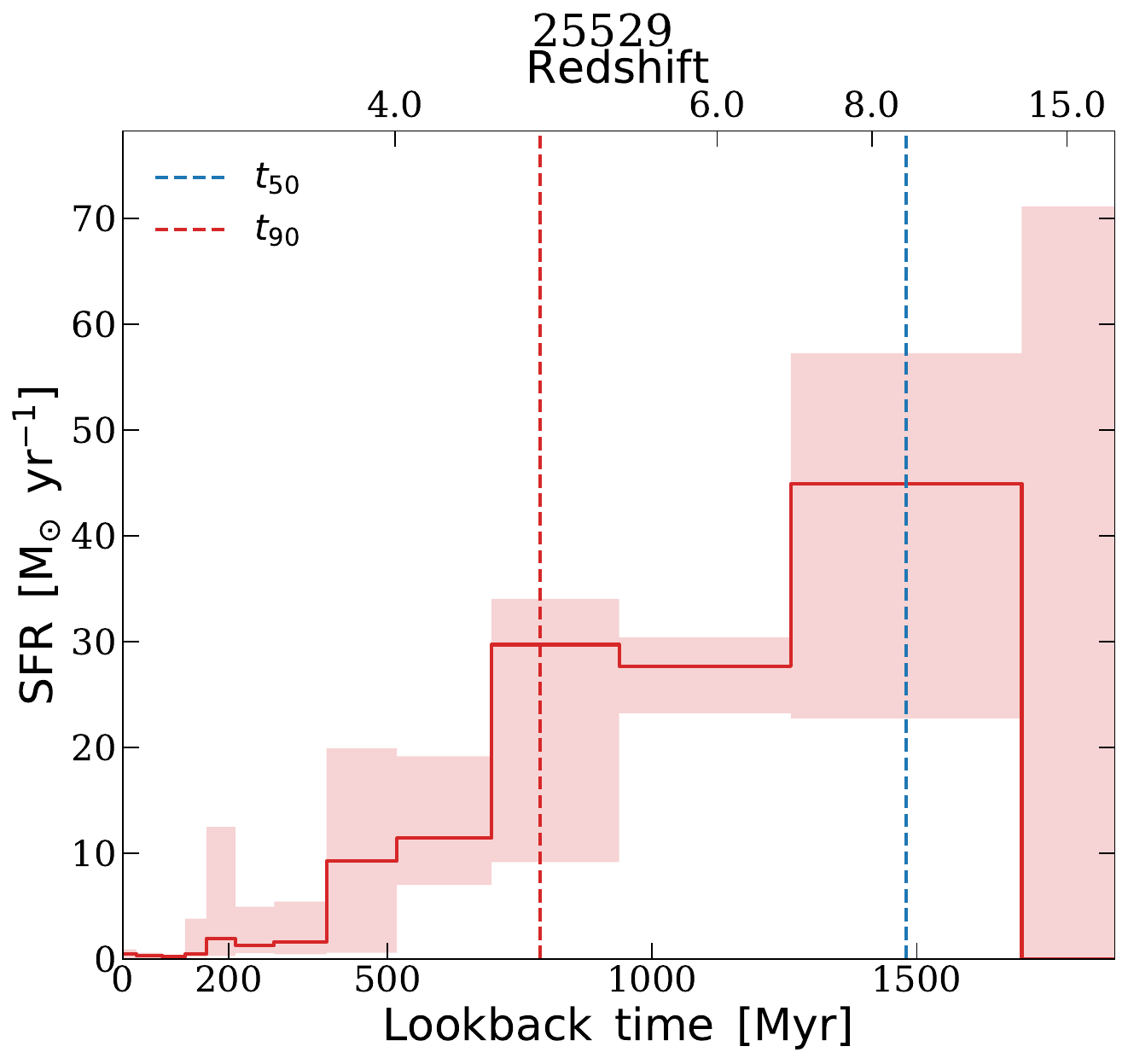}
    \includegraphics[width=\sfhwidth]{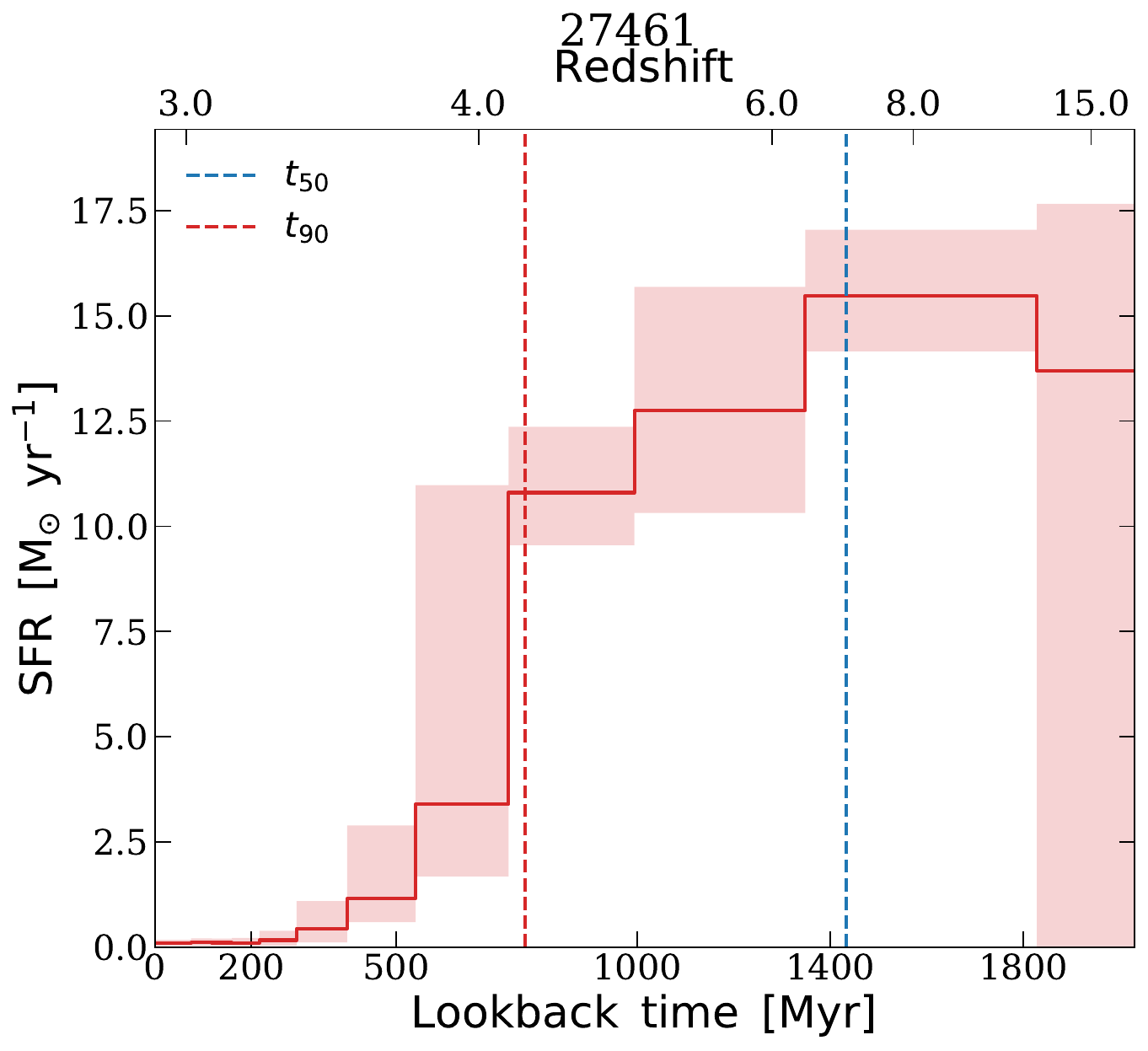}
    \includegraphics[width=\sfhwidth]{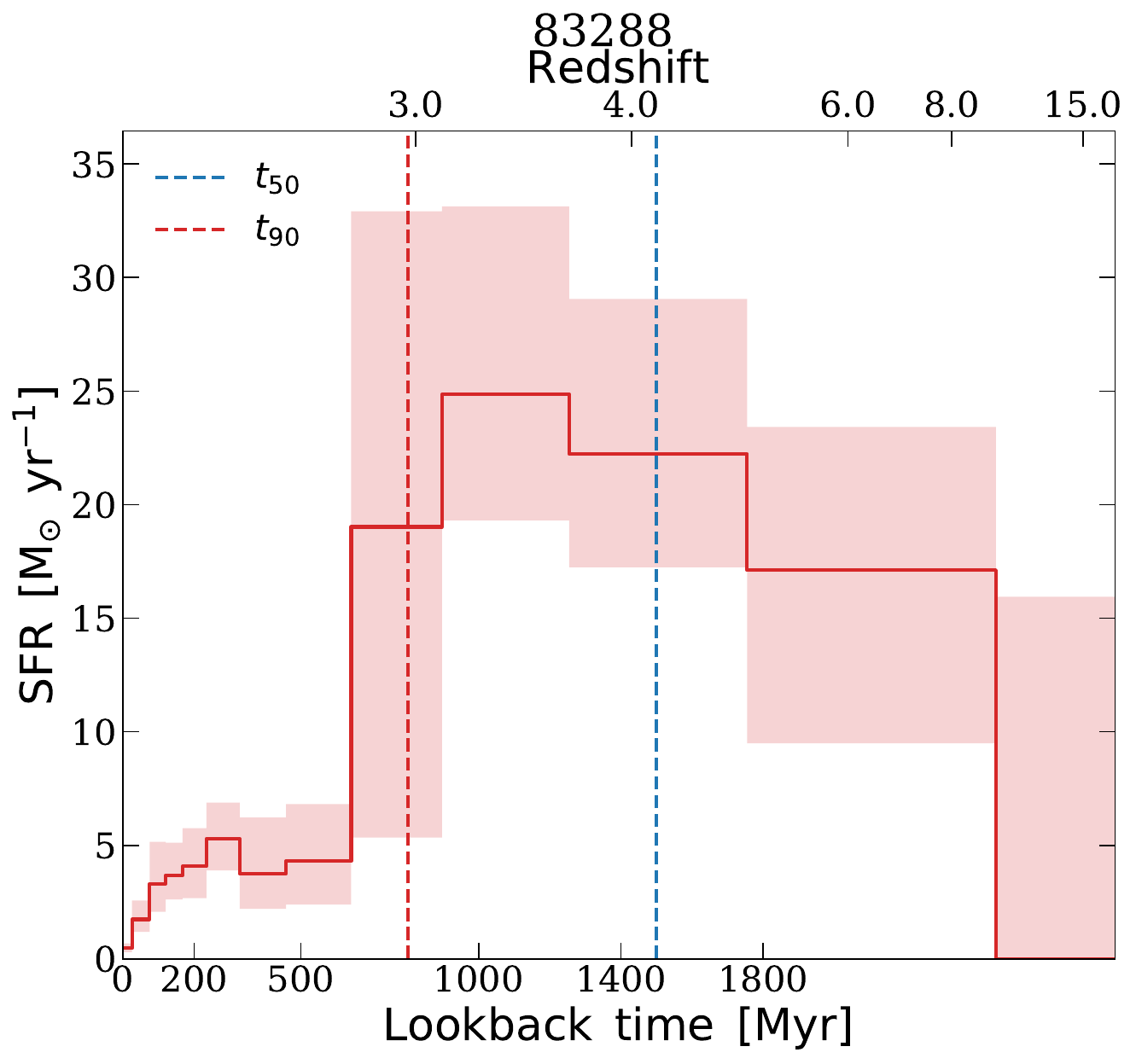}
    \includegraphics[width=\sfhwidth]{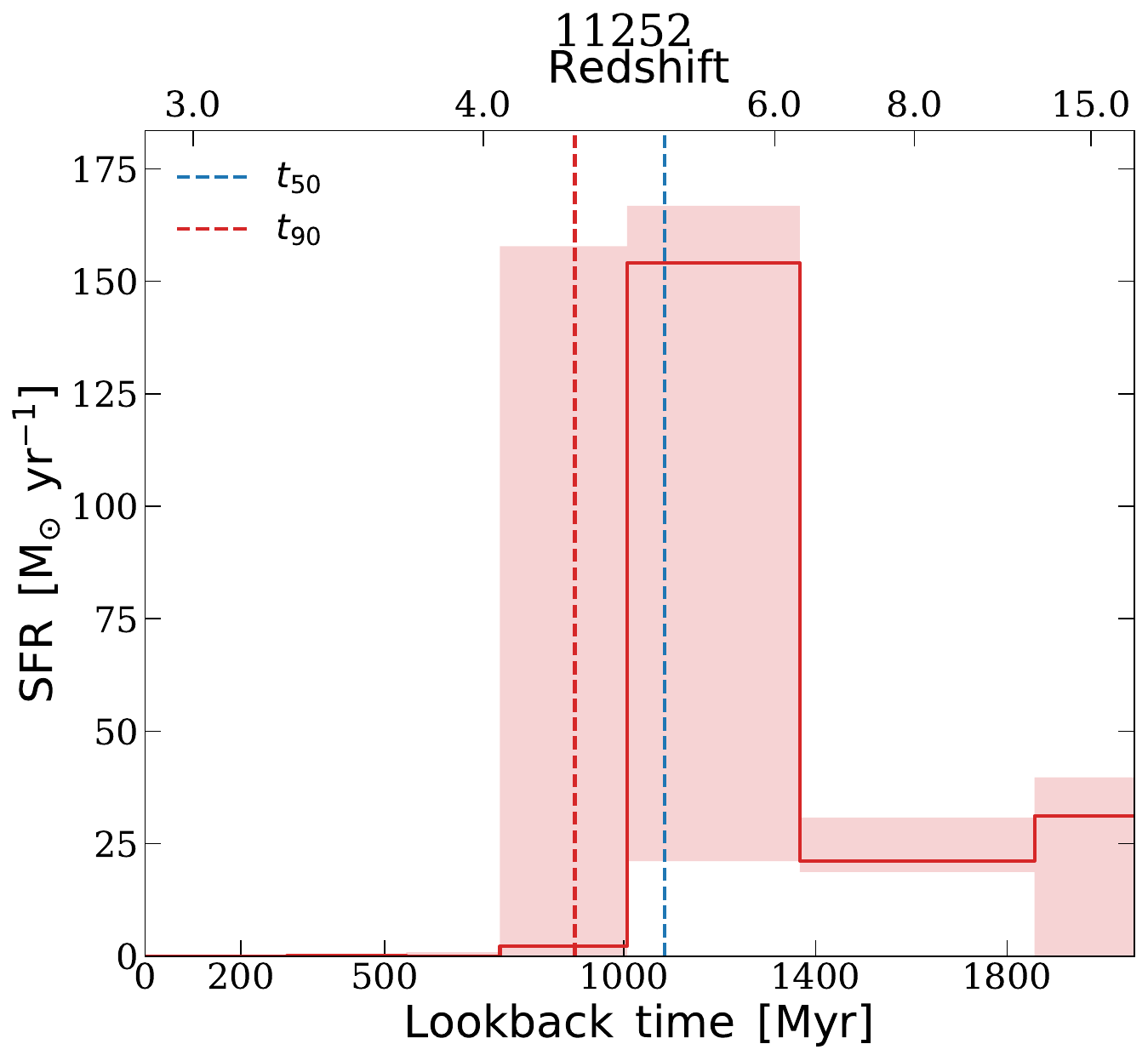} 
    
    \caption{Star-formation rate vs lookback time from observation (redshift) for 6 of the quiescent galaxies. }
    \label{fig:sfhs3}
\end{figure*}

Figs. \ref{fig:sfhs2} and \ref{fig:sfhs3} show the star-formation histories for the remaining 12 massive quiescent galaxies.
The vertical dashed lines correspond to the formation time $\rm t_{50}$ and the quenching time $\rm t_{90}$.

\section{Fitting both photometry and spectra when offset }
\label{app.s.offset}

\begin{figure}
    \centering
    \includegraphics[width=1\linewidth]{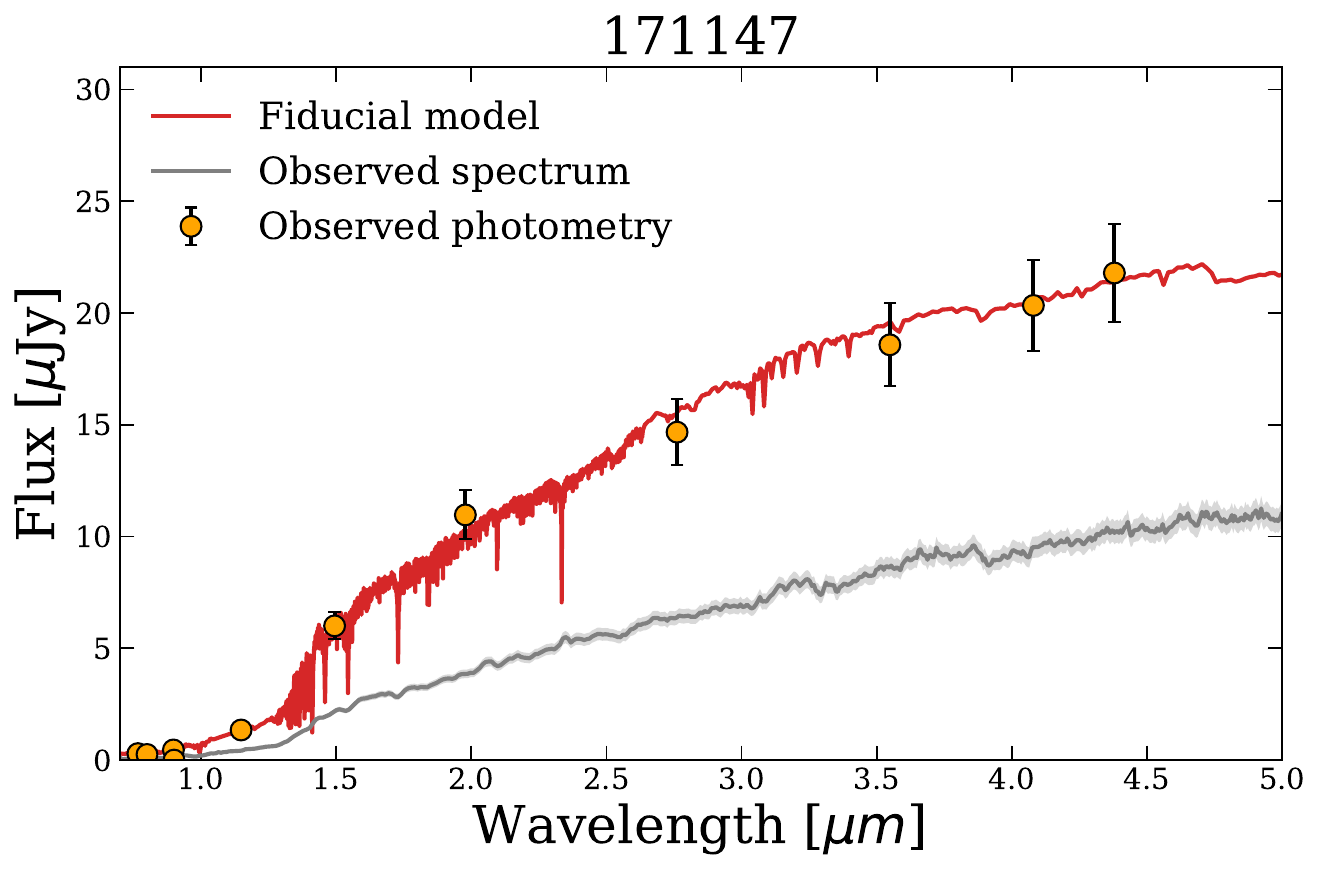}
    \caption{NIRSpec Prism spectrum (grey) and NIRCam photometry (yellow points) for the massive quiescent galaxy 141147. The red spectrum is our best-fit model to both the photometry and spectrum. This galaxy is significantly larger than the NIRSpec slit so we can see the best-fit model accurately traces the photometry, whilst still incorporating information from the spectrum. This figure demonstrates how prospector fits both spectroscopy and photometry and incorporates them both into the best-fit model for the overall galaxy. }
    \label{fig:141147}
\end{figure}

Fig. \ref{fig:141147} shows the NIRSpec Prism spectrum (grey) and NIRCam photometry (yellow points) for the massive quiescent galaxy 171147. This galaxy is significantly larger than the NIRSpec slit, as can be seen in Fig. \ref{fig:rgb}. Therefore, it provides a helpful example of how we fit both photometric and spectroscopic data at the same time. We want to include information from the spectrum, whilst also being aware that this traces a subregion of the galaxy, whereas the photometry traces the overall galaxy as a whole. 
The red spectrum in Fig. \ref{fig:141147} corresponds to our best-fit model of the spectra and the photometry. We see that our best-fit model accurately traces the shape and flux of the photometry (i.e. the galaxy as a whole) but is able to also take into account the superior information from the spectrum on quantities such as redshift, Balmer break shape, emission lines, and more. Clearly, this methodology is not a substitute for missing data; there could be scenarios where the spectrum differs substantially from the photometry for physical reasons (e.g., the spectrum captures only the quiescent bulge of an early spiral, while the star-forming arms are completely outside the NIRSpec shutter). However, the compact nature and smooth morphology of most of our targets suggest that such effects are not a concern in this work.

\section{Fitting the grating spectra}\label{app:grating.sfh}

\begin{figure*}
\centering
    \includegraphics[width=1.6\columnwidth]{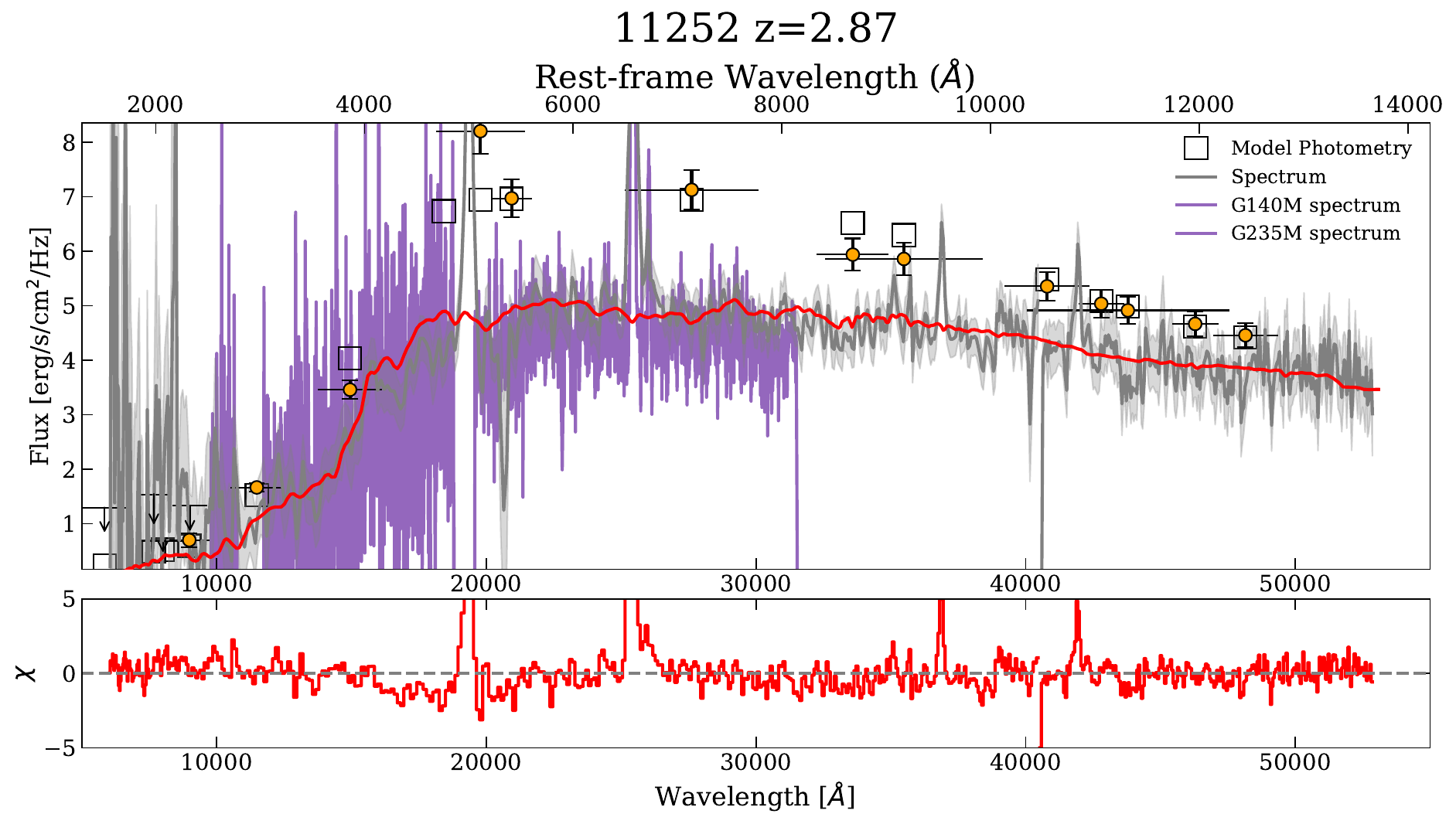}
    \includegraphics[width=1.4\columnwidth]{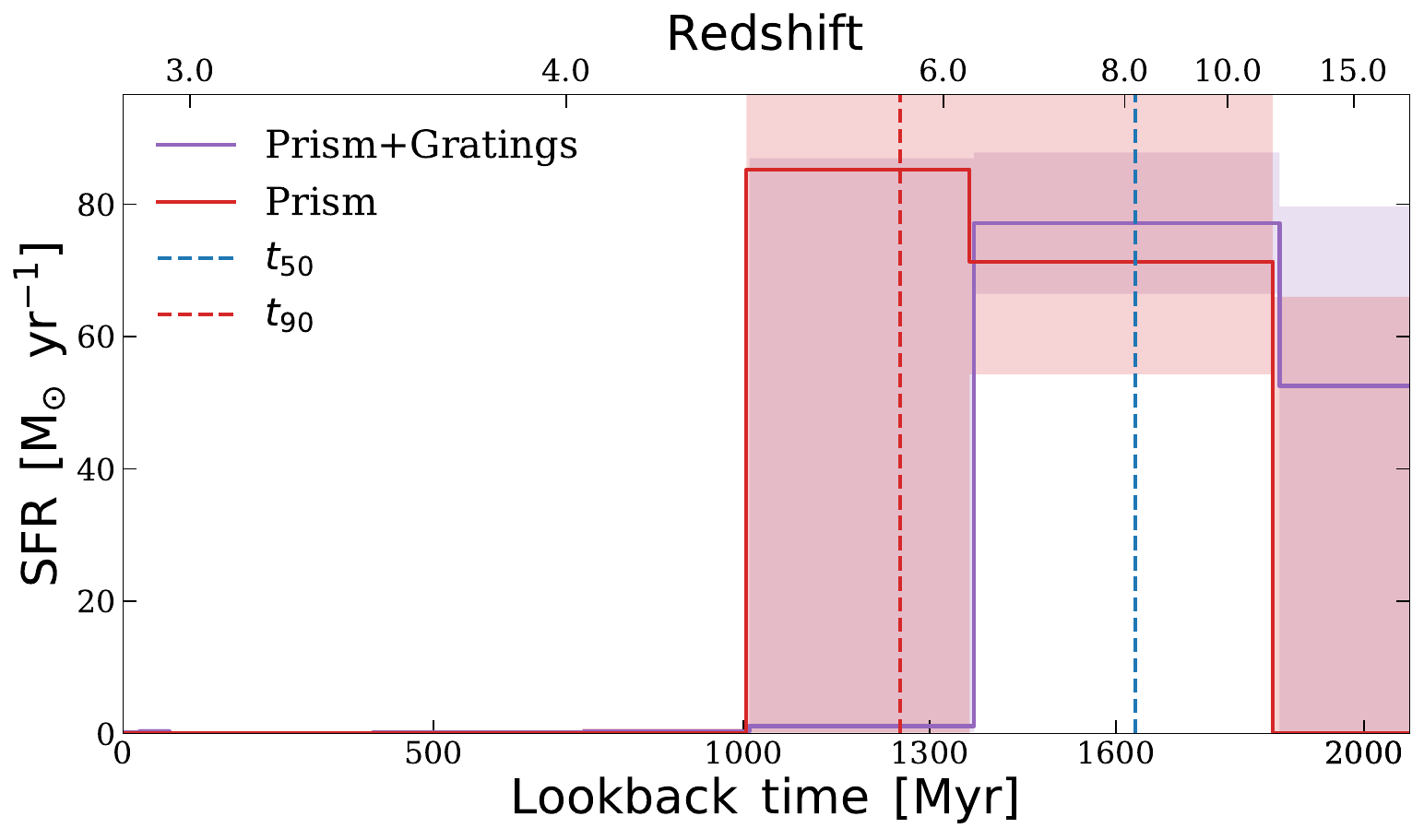}
    \caption{Upper panel: observed prism spectrum (grey), grating spectra (purple), and photometry (yellow points) for 11252. The red line is the best-fit model spectrum incorporating the data from the prism
    all the observed spectra and the black squares are the best-fit model photometry. Lower panel: Star-formation rate vs lookback time from observation (redshift) for 11252 comparing the fit without the grating spectra (red) and the fit including the grating spectra (purple). We see we obtain consistent star-formation histories in both cases (to within 1\sig).}
    \label{fig:grating_comp}
\end{figure*}

An interesting test is to explore what effects fitting the grating spectra in addition to the prism spectrum has on the star-formation history. This enables us to explore the effects of higher-resolution data which should be better able to fit absorption features within the quiescent galaxies spectrum. We test this by using galaxy 11252 for which we have gratings spectra. 
Fig. \ref{fig:grating_comp} shows the best-fit spectra to the prism, grating, and photometry and the difference between the SFHs for the grating + prism and prism only fits. We see that both star-formation histories are consistent to within 1\sig. We do obtain differences in the formation and quenching times, with them being roughly 30\% older with the inclusion of the grating data. This highlights the effects of additional grating data, particularly in the case of AGN with strong emission lines. 

\section{PySersic Fits}

\begin{figure*}
    \centering
    \includegraphics[width=2\columnwidth]{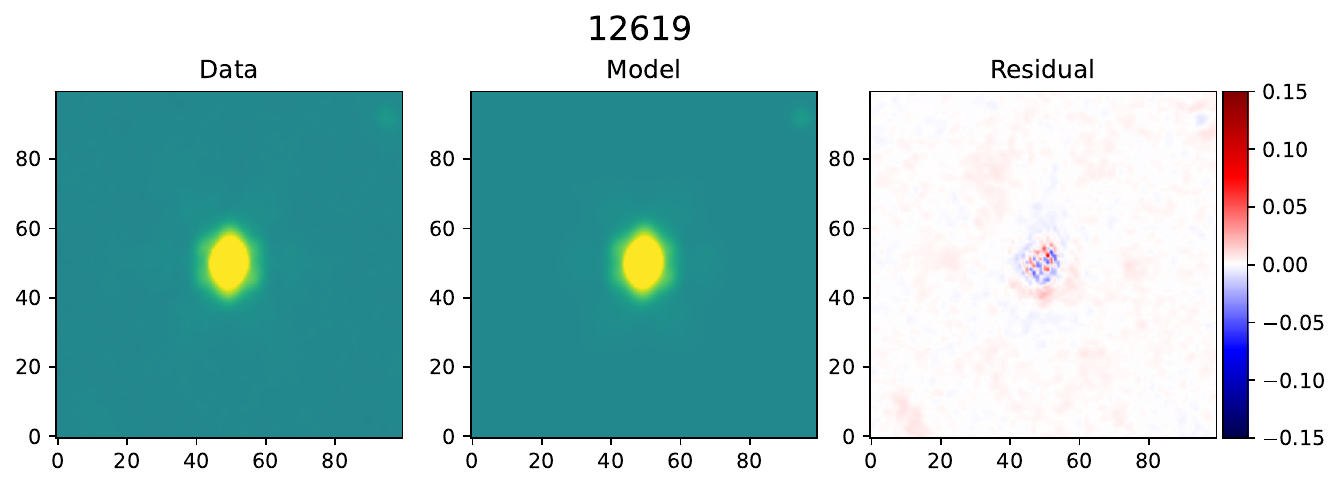}
    \includegraphics[width=2\columnwidth]{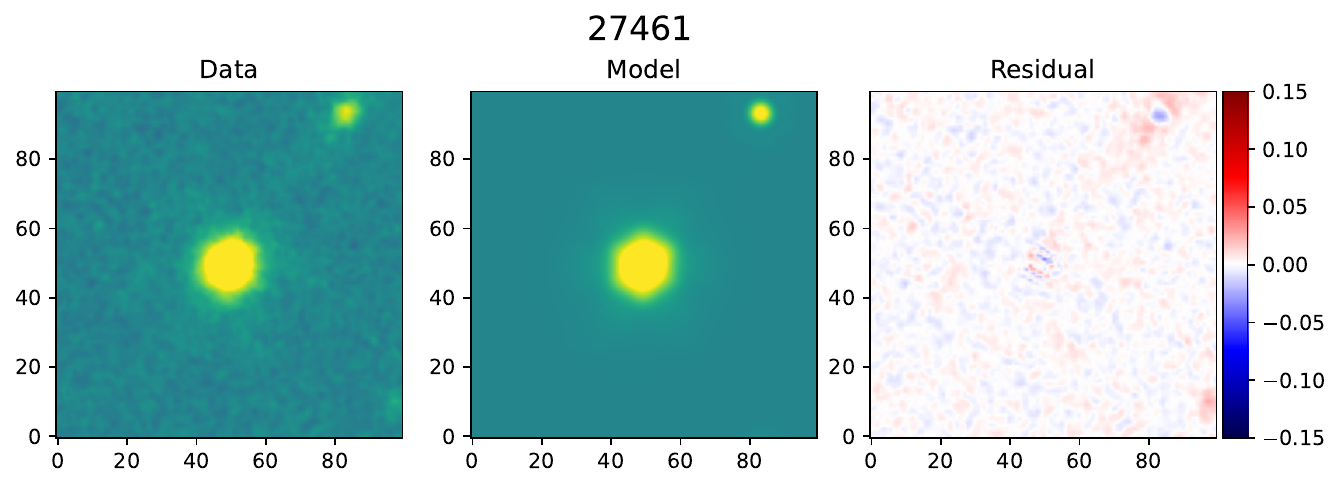}
    \includegraphics[width=2\columnwidth]{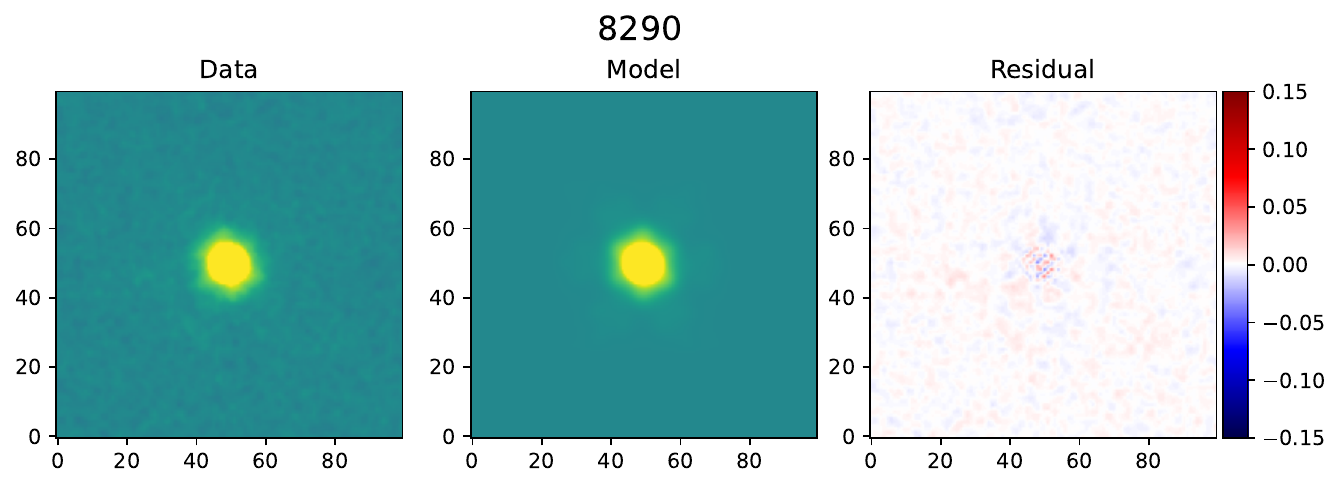}

    \caption{Example PySersic fits for three of the galaxies in the sample. Left is the data, middle is the model and right is the residual. We see that PySersic has modelled the galaxies' light distribution without significant residuals.}
    \label{fig:PySersic}
    
\end{figure*}

Fig. \ref{fig:PySersic} shows the data, model, and residual maps for the F200W morphological fits of three quiescent galaxies in our sample. The galaxies (and additional sources) appear to be well-modelled by the single \sersic profile fits without leaving significant residuals.

%%%%%%%%%%%%%%%%%%%%%%%%%%%%%%%%%%%%%%%%%%%%%%%%%%

% Don't change these lines
\bsp	% typesetting comment
\label{lastpage}
\end{document}

%% file: quiescent_tab_18_12Sep.tex
\begin{table*}
\centering
\begin{tabular}{|c|c|c|c|c|c|c|c|c|c|}
\hline
NIRSpec ID & R.A. & Dec. & Redshift & $\log(M_\star)$ & $z_{50}$ & $r_e$ & $n$ & Balmer Break & D$_n$4000 \\ 
\hline
 & deg & deg &  & $\Msun$ &  & kpc &  &  &  \\ 
 (1) & (2) & (3) & (4) & (5) & (6) & (7) & (8) & (9) & (10) \\
\hline
1397 & $53.06227$ & $-27.87504$ & $4.22$ & $10.46 \pm 0.02$ & $8.27 \pm 1.29$ & $0.22 \pm 0.00$ & $4.83 \pm 0.20$ & $3.33 \pm 0.05$ & $1.45 \pm 0.12$ \\ 
\hline
6620 & $53.07872$ & $-27.83960$ & $3.47$ & $10.40 \pm 0.01$ & $5.69 \pm 0.08$ & $1.31 \pm 0.03$ & $7.98 \pm 0.03$ & $3.26 \pm 0.13$ & $1.34 \pm 0.04$ \\ 
\hline
8290 & $53.08188$ & $-27.82880$ & $4.37$ & $10.51 \pm 0.02$ & $6.69 \pm 0.54$ & $0.33 \pm 0.00$ & $2.62 \pm 0.08$ & $3.43 \pm 0.11$ & $1.26 \pm 0.06$ \\ 
\hline
8777 & $53.10821$ & $-27.82519$ & $4.66$ & $10.81 \pm 0.01$ & $7.64 \pm 0.23$ & $0.21 \pm 0.00$ & $3.47 \pm 0.08$ & $2.75 \pm 0.06$ & $1.27 \pm 0.11$ \\ 
\hline
11252 & $53.12693$ & $-27.80468$ & $2.88$ & $10.71 \pm 0.05$ & $5.04 \pm 0.09$ & $0.65 \pm 0.00$ & $2.49 \pm 0.02$ & $3.44 \pm 0.16$ & $1.54 \pm 0.08$ \\ 
\hline
12214 & $53.19019$ & $-27.76921$ & $2.99$ & $10.69 \pm 0.01$ & $9.07 \pm 1.26$ & $0.53 \pm 0.00$ & $7.97 \pm 0.04$ & $3.41 \pm 0.11$ & $1.45 \pm 0.07$ \\ 
\hline
12619 & $53.19691$ & $-27.76053$ & $3.61$ & $10.87 \pm 0.01$ & $5.95 \pm 0.17$ & $0.49 \pm 0.00$ & $2.06 \pm 0.02$ & $3.24 \pm 0.10$ & $1.34 \pm 0.09$ \\ 
\hline
25511 & $189.18254$ & $62.23201$ & $2.99$ & $10.55 \pm 0.01$ & $10.03 \pm 0.63$ & $0.47 \pm 0.00$ & $3.74 \pm 0.11$ & $2.78 \pm 0.04$ & $1.34 \pm 0.06$ \\ 
\hline
25529 & $189.02575$ & $62.26050$ & $3.12$ & $10.54 \pm 0.02$ & $8.63 \pm 1.22$ & $0.81 \pm 0.19$ & $6.32 \pm 1.31$ & $3.12 \pm 0.06$ & $1.30 \pm 0.06$ \\ 
\hline
27461 & $189.22644$ & $62.24715$ & $2.92$ & $10.15 \pm 0.01$ & $6.89 \pm 0.54$ & $0.66 \pm 0.01$ & $3.40 \pm 0.09$ & $2.73 \pm 0.04$ & $1.37 \pm 0.09$ \\ 
\hline
72127 & $189.26572$ & $62.16839$ & $4.13$ & $10.55 \pm 0.01$ & $9.62 \pm 1.18$ & $0.30 \pm 0.00$ & $7.65 \pm 0.26$ & $2.86 \pm 0.06$ & $1.30 \pm 0.08$ \\ 
\hline
72396 & $189.26305$ & $62.17038$ & $2.78$ & $10.96 \pm 0.01$ & $8.98 \pm 0.86$ & $1.95 \pm 0.01$ & $7.98 \pm 0.02$ & $3.32 \pm 0.07$ & $1.38 \pm 0.06$ \\ 
\hline
80660 & $189.27545$ & $62.21414$ & $4.39$ & $10.37 \pm 0.02$ & $6.41 \pm 0.47$ & $1.62 \pm 0.06$ & $6.12 \pm 0.27$ & $2.77 \pm 0.02$ & $1.22 \pm 0.07$ \\ 
\hline
83288 & $189.26741$ & $62.24780$ & $2.21$ & $10.49 \pm 0.01$ & $4.19 \pm 0.34$ & $0.76 \pm 0.01$ & $7.99 \pm 0.01$ & $3.15 \pm 0.18$ & $1.56 \pm 0.08$ \\ 
\hline
171147 & $53.05563$ & $-27.87406$ & $2.55$ & $11.32 \pm 0.00$ & $5.72 \pm 0.12$ & $2.42 \pm 0.01$ & $2.82 \pm 0.03$ & $3.40 \pm 0.11$ & $1.39 \pm 0.10$ \\ 
\hline
197911 & $53.16531$ & $-27.81414$ & $3.07$ & $11.30 \pm 0.01$ & $5.05 \pm 0.08$ & $0.72 \pm 0.01$ & $5.06 \pm 2.90$ & $3.12 \pm 0.08$ & $1.25 \pm 0.06$ \\ 
\hline
199773 & $53.16324$ & $-27.80906$ & $2.81$ & $10.77 \pm 0.01$ & $6.53 \pm 0.58$ & $0.51 \pm 0.01$ & $3.63 \pm 0.49$ & $3.07 \pm 0.08$ & $1.31 \pm 0.08$ \\ 
\hline
200733 & $53.10714$ & $-27.80482$ & $2.87$ & $11.04 \pm 0.01$ & $4.77 \pm 0.16$ & $0.88 \pm 0.00$ & $2.74 \pm 0.01$ & $3.09 \pm 0.14$ & $1.39 \pm 0.05$ \\ 
\hline
\end{tabular}
\caption{Our massive quiescent galaxies and their physical properties. (1) NIRSpec ID, corresponding to the target name in the MSA configuration. (2, 3) ICRS equatorial coordinates. (4) Redshift from prism observations. (5) Stellar mass. (6) Formation (half-mass) redshift. (7) Half-light F444W radius of the best-fit \sersic model. (8) \sersic index. (9) Balmer break index strength. (10) D$_n$4000 index strength. The uncertainties are the median and 16--84 interpercentile range from the relevant posterior distribution. (5, 6) are from \textsc{prospector} (Section~\ref{s.analysis.ss.sedfit}). (7, 8) are from \textsc{PySersic} (Section~\ref{s.analysis.ss.morphology}).}
\label{tab:values_errors}
\end{table*}